\documentclass[a4paper,11pt]{article}
\pdfoutput=1 

\usepackage{jheppub} 

\usepackage[T1]{fontenc} 

\usepackage[T1]{fontenc} 
\usepackage[utf8]{inputenc} 
\usepackage{microtype} 
\usepackage{lmodern} 
\usepackage{booktabs} 
\usepackage{mdframed} 
\usepackage{graphicx}

\usepackage{tikz}
\usepackage{tikz-3dplot}

\usepackage{comment} 

\usepackage{hyperref}
 \usepackage{xcolor}
 \definecolor{dark-red}{rgb}{0.4,0.15,0.15}
 \definecolor{dark-blue}{rgb}{0.15,0.15,0.4}
 \definecolor{medium-blue}{rgb}{0,0,0.5}
 \hypersetup{colorlinks, linkcolor={dark-red}, citecolor={dark-blue}, urlcolor={medium-blue}}

\usepackage{enumitem} 

\usepackage{amsmath,amsfonts,amssymb}
\usepackage{eucal} 
\usepackage{mleftright} \mleftright
\usepackage{tensor} 
\usepackage{braket} 
\usepackage{bm} 
\usepackage{mathtools} 
\usepackage{amsthm} 
\usepackage{slashed}

\newtheoremstyle{break}
  {\topsep}{\topsep}%
  {\normalshape}{}%
  {\bfseries}{}%
  {\newline}{}%
  
\theoremstyle{break}

\theoremstyle{remark}

\usepackage{mathrsfs} 

\providecommand*{\pd}{\partial}
\renewcommand*{\pd}{\partial}
\providecommand*{\vd}{\delta}
\renewcommand*{\vd}{\delta}

\providecommand*{\RR}{{\mathbb{R}}}
\renewcommand*{\RR}{{\mathbb{R}}}
\providecommand*{\CC}{{\mathbb{C}}}
\renewcommand*{\CC}{{\mathbb{C}}}

\usepackage{trimclip} 

\makeatletter
\newcommand{\shortcancel}[1]{%
  \mathord{\mathpalette\shortcancel@{#1}}%
}
\newcommand{\shortcancel@}[2]{
  \clipbox{0pt 0pt 0pt 0pt}{$#1\cancel{#2}$}
}
\makeatother
\newcommand{\subm}{\shortcancel{m}}

\usepackage{orcidlink}
\usepackage{cancel}

\usepackage[normalem]{ulem}

\usepackage{tikz,pgfplots}
\usepgfplotslibrary{fillbetween}
\usetikzlibrary{calc,arrows,arrows.meta,cd,shapes.geometric,positioning,through,decorations.markings,decorations.text}

\title{\boldmath Carrollian quantum states and flat space holography}

\author{Stefan Fredenhagen \orcidlink{0000-0002-5744-8581}, Stefan Prohazka \orcidlink{0000-0002-3925-3983}, Robert Tiefenbacher \orcidlink{0009-0007-4667-2207}}

\affiliation{Mathematical Physics, Faculty of Physics, University of Vienna, Boltzmanngasse 5, 1090 Vienna, Austria}

\emailAdd{stefan.fredenhagen@univie.ac.at}
\emailAdd{stefan.prohazka@univie.ac.at}
\emailAdd{robert.tiefenbacher@univie.ac.at}

\abstract{We study free Carrollian quantum field theories from an
  algebraic perspective and explore their implications for flat space
  holography. As explicit examples, we construct the electric and
  magnetic Carrollian Weyl algebras obtained from Carroll limits of
  the relativistic scalar field and analyze their states, including
  vacuum and thermal configurations.

  For the massive electric theory, we find a regular Carroll-invariant
  vacuum state and a regular KMS state, yielding a consistent
  Carrollian thermodynamic system. By contrast, the massless electric
  and magnetic theories are more subtle: depending on the
  quantization, they admit either no regular distinguished vacuum or
  only nonregular Carroll-invariant ground states, while still
  supporting nonregular thermal states. We further analyze alternative
  classes of states in the massless electric theory, including
  spatially homogeneous pure regular quasifree states and Sorkin--Johnston
  states.

  Motivated by these results, we discuss consequences for flat space
  holography. We construct a well-defined state relevant for
  Carrollian holography whose Hilbert-space representation factorizes
  into a standard Fock sector and a nonseparable zero-mode sector,
  thereby highlighting the role of infrared degrees of freedom in the
  boundary theory.}

\begin{document} 
\maketitle
\flushbottom

\section{Introduction}
\label{sec:intro}

Carroll symmetries~\cite{Levy1965,SenGupta1966OnAA} have recently
attracted renewed interest due to their intriguing connection to flat
space holography at null
infinity~\cite{Duval:2014uva,Donnay:2022aba,Bagchi:2022emh}, as well
as at spatial and timelike
infinity~\cite{Gibbons:2019zfs,Figueroa-OFarrill:2021sxz,Borthwick:2023lye,Have:2024dff,Borthwick:2024wfn,Borthwick:2024skd}. Beyond
holographic contexts, Carroll symmetries have also been linked to a
variety of other systems, including cosmological
models~\cite{deBoer:2021jej} and condensed matter phenomena such as
flat bands~\cite{Bagchi:2022eui,Ara:2024fbr,Biswas:2025dte} and
fractons~\cite{Bidussi:2021nmp,Marsot:2022imf,Baig:2023yaz,Kasikci:2023tvs,Figueroa-OFarrill:2023vbj}
(see the recent
reviews~\cite{Bagchi:2025vri,Nguyen:2025zhg,Ruzziconi:2026bix,Bekaert:2026cvx}).

On the other hand, the framework of algebraic quantum field theory
(AQFT)~\cite{Haag:1996hvx,Araki:1999ar,Fewster:2019ixc} provides a
model-independent and mathematically precise formulation of quantum
field theory in terms of local algebras and their states, cleanly
isolating structural principles such as locality, causality, and the
spectrum condition from specific Lagrangian realizations. This
structural viewpoint has proven powerful in the analysis of thermal
(KMS) states, modular theory, and quantum field theory on curved
spacetimes. It has, furthermore, played a central role in recent
progress on the emergence of bulk geometric structures in holography
and, more broadly, in quantum gravity (see~\cite{Liu:2025krl} for a
review).

In this work, we take a first step toward unifying these developments
by formulating (conformal) Carrollian quantum field theories within
the algebraic framework. We focus in particular on their local
algebras of observables and on the structure of vacuum and thermal
states. A virtue of the algebraic approach to
quantum field theory is that it allows one to define and analyze
states without fixing a Hilbert space representation from the
outset. In particular, equilibrium states can be characterized
intrinsically via the KMS condition, without relying on a trace. This
flexibility is especially useful in the Carrollian and flat space
holography setting, where infrared effects and associated divergences
play a central role. Within AQFT, such features instead guide the
construction of physically meaningful states, naturally leading to
nonregular representations and even nonseparable Hilbert spaces
through the GNS construction. These structures lie beyond conventional
Fock space methods and are directly relevant for the zero-mode sector
in flat space holography. For clarity of exposition, we restrict our
discussion to scalar fields and examine their Carroll limit in detail,
using explicit examples as guiding benchmarks. Throughout, we restrict
attention to unitary theories.

This work is conceived as part of a systematic dialogue between the
Carrollian and algebraic QFT communities. To make the discussion
accessible to both audiences, we occasionally review structural
aspects that may be standard from one perspective but less familiar
from the other. The goal is to establish a common language in which
the structural properties of Carrollian theories can be analyzed using
algebraic methods.\footnote{Algebraic methods have previously been
  applied to aspects of flat space holography, see, e.g.,
  \cite{Ashtekar:1987tt,Ashtekar:2018lor,Dappiaggi:2005ci,Prabhu:2022zcr}
  and references therein.}

To illustrate the emergence of Carrollian dynamics and to motivate the 
subsequent algebraic formulation, we begin by revisiting the Carroll 
limit of the relativistic scalar field. For this purpose, we first 
present the theory in Hamiltonian form before implementing the Carroll 
limit at the level of the operator algebras. The starting point is 
the Poincaré-invariant massive scalar field
\begin{align}
  \label{eq:action}
  S[\phi,\pi] &= \int dt d^{d}x \,
  \left[
  \pi \pd_{t}\phi - \mathcal{H}
  \right]\,,
  &
    \mathcal{H} = \frac{1}{2}
    \left(
    \pi^{2} + \pd_{i}\phi \pd^{i}\phi + m^{2}\phi^{2}
    \right) \, .
\end{align}
This theory admits two Carroll
limits~\cite{Barnich:2012rz,Duval:2014uoa}, called electric and
magnetic~\cite{Henneaux:2021yzg,deBoer:2021jej}.\footnote{A more
  intrinsic way is to think about Carrollian theories as theories which
  create elementary Carroll
  particles~\cite{Figueroa-OFarrill:2023qty}.} For the electric theory,
we substitute $\phi(t,x)=\epsilon^{1/2}\phi_{\epsilon}(t/\epsilon,x)$
and $\pi(t,x)=\epsilon^{-1/2}\pi_{\epsilon}(t/\epsilon,x)$ and rescale
the time $t'=t/\epsilon$. The limit is well defined and given by
\begin{align}
  \label{eq:electr}
  S_{\mathrm{e}}[\phi_{0},\pi_{0}] &= \lim_{\epsilon \to 0}\int dt' d^{d}x \,
  \left[
  \pi_{\epsilon} \pd_{t'}\phi_{\epsilon} - \mathcal{H_{\epsilon}}
  \right]\,,
  &
    \mathcal{H_{\epsilon}} &= \frac{1}{2}
    \left(
    \pi_{\epsilon}^{2} +\epsilon^{2} \pd_{i}\phi_\epsilon \pd^{i}\phi_\epsilon + \epsilon^{2}m^{2}\phi_\epsilon^{2}
    \right)\,,
\end{align}
where all fields are now functions of $t'$. Inspired by a similar
analysis of the nonrelativistic, Galilean limit in~\cite[Appendix
B]{Brunetti:2019cax}, we here think of taking the Carroll limit not by
changing parameters of the theory, but by redefining the fields to
$\phi_\epsilon$, $\pi_\epsilon$, which means that we zoom into a short
time interval where the theory becomes effectively Carrollian. In this
way, the mass term in~\eqref{eq:electr} vanishes for $\epsilon\to
0$. To obtain a massive electric theory, we can use a different
rescaling,\footnote{A massive electric theory can be obtained by the
  rescaling
  $\phi_\epsilon(t,x):=\epsilon^{-\frac{d}{2}}\phi(t,x/\epsilon)$.} or
-- slightly against the advertised spirit of not changing parameters
-- rescale the mass by $m\mapsto m/\epsilon$.  The characteristic
feature of electric theories is the absence of spatial gradient terms
and their resulting ultralocal behavior, a property that has also been
studied using complementary methods by Klauder; see, e.g.,~\cite{Klauder:1970cs,Klauder:1971zz} and references therein.
Since we can integrate out the momenta, we can also write them as
\begin{align}
  \label{eq:electr-no-mom}
  S_{\mathrm{e}}[\phi_{0}] &= \frac{1}{2}\int dt d^{d}x 
                             \left[
                             (\pd_{t}\phi_{0})^{2}- m^{2} \phi_{0}^{2}
                             \right] \,.
\end{align}

For magnetic theories we rescale as
$\phi(t,x)=\epsilon^{-1/2}\tilde \phi_{\epsilon}(t/\epsilon,x)$ and
$\pi(t,x)=\epsilon^{1/2}\tilde \pi_{\epsilon}(t/\epsilon,x)$ to obtain
\begin{align}
  \label{eq:mag}
  S_{\mathrm{m}}[\tilde\phi_{0},\tilde\pi_{0}] &= \lim_{\epsilon \to 0}\int dt' d^{d}x \,
  \left[
  \tilde\pi_{\epsilon} \pd_{t'}\tilde \phi_{\epsilon} - \mathcal{H_{\epsilon}}
  \right]\,,
  &
    \mathcal{H_{\epsilon}} &= \frac{1}{2}
    \left(
    \epsilon^{2}\tilde\pi_{\epsilon}^{2} + \pd_{i}\tilde\phi \pd^{i}\tilde\phi + m^{2}\tilde\phi^{2}
    \right) \, .
\end{align}
In the limit, the momentum term of the Hamiltonian vanishes, and
therefore magnetic theories are naturally in first order form. 
The corresponding equations of motion are
\begin{align}\label{magneticEOM}
    \dot{\tilde{\phi}}_0&=0\,, & \dot{\tilde{\pi}}_0 &= (\Delta -m^2)\tilde{\phi}_0  \,.
\end{align}

\subsection*{Outline}
\label{sec:outline-summary}

In Section~\ref{sec:electric}, we analyze the electric contractions
of the free scalar from the algebraic quantum field theory perspective. 
We cover both the massive and massless cases and discuss their electric 
Weyl algebras together with their (conformal) Carroll symmetries, 
providing the algebra of observables.

Section~\ref{sec:states} then examines the corresponding states, where
we first review the definition of a state in the AQFT framework
(Section~\ref{sec:states-aqft}). We then focus on vacuum (Section
\ref{sec:massive-vacuum}) and thermal states
(Section~\ref{sec:thermal-states}), but also emphasize subtleties in
the massless limit for which we present a novel vacuum state
(Section~\ref{sec:electricstate}). In
Section~\ref{sec:classical-state}, we demonstrate that the commonly
used prescription of subtracting the divergent $1/m$ term to obtain a
massless two-point
function~\cite{Bagchi:2022emh,Chen:2023pqf,deBoer:2023fnj,Chen:2024voz}
does not yield a well-defined state on the Weyl algebra and misses infrared physics. For
the massless case we also analyze all spatially homogeneous pure regular quasifree
states (Section~\ref{sec:spat-homog-pure}) and a distinguished
state that can be constructed on generic (bounded) backgrounds, the
so-called Sorkin--Johnston state (Section~\ref{sec:sork-johnst-stat}).

The magnetic limit, along with its Weyl algebra, symmetries and vacuum
and thermal states, is discussed in
Section~\ref{sec:magnetic}.

In Section~\ref{sec:flat-space-hologr} we relate our results to flat
space holography. In Section~\ref{sec:quasifree-state} we construct a
well-defined nonregular quasifree vacuum state from a Carrollian
perspective~\cite{Donnay:2022wvx}. This construction is subtle due to
a divergence in the two-point function for operators identified with
boundary values of bulk fields, analogous to that encountered in the
massless electric scalar theory. Upon smearing and passing to the Weyl
algebra, this apparent singularity can be handled and leads to a
Hilbert space that factorizes into a standard Fock space and a
nonseparable sector describing the zero modes. In
Section~\ref{sec:relat-infr-phys}, we show that this AQFT-based
construction naturally connects to standard notions in infrared
physics. Finally, in Section~\ref{sec:conf-carr-symm} we
demonstrate that the vacuum state is invariant under the conformal
Carroll group, highlight additional symmetries, and analyze the
associated unitary operators and their continuity
properties.

We conclude in Section~\ref{sec:discussion} with a
short summary of our results and a discussion of open questions and
future directions.

Technical results are collected in the
appendices. Appendix~\ref{sec:positivity} analyzes the positivity
properties of the states, and Appendix~\ref{sec:stat-flat-hologr}
provides details on the Hilbert space and vacuum representation for
flat space holography.

\section{Electric contraction and Carroll symmetries}
\label{sec:electric}

In this section, we consider the electric Carroll rescaling of the
free scalar field and the resulting one-parameter family of electric
Weyl algebras. The massive and massless cases are treated separately,
as their commutators differ,
\begin{align}
  \label{eq:comm-fct-sum}
  [\phi_{0}(t,x),\phi_{0}(0,0)]&=\int \frac{d^d p}{(2\pi)^d} e^{i p \cdot x} \int dp^0 \, \mathrm{sgn}(p^0) \, \delta((p^0)^2 - m^2) \, e^{-i p^0 t} \\
                                 &=\delta(x) \cdot
  \begin{cases}
    -i \frac{\sin(mt)}{m} & m>0 \\
    -i t & m=0 \, .
  \end{cases} \label{eq:carroll-comm}
\end{align}
Compared to the Poincaré-invariant theory, the delta function
in~\eqref{eq:comm-fct-sum} is independent of the spatial momentum, so
the spatial and temporal integrations decouple. In both cases this
leads to spatial ultralocality, as reflected by the delta function in
\eqref{eq:carroll-comm}.

The massive case can be seen as a useful warm-up and serves, as we
will see, as the prototypical example of a Carrollian theory. The
massless case is additionally conformal Carroll invariant, which makes
it an interesting toy example for flat space holography.  We also
explicitly check the realisation of (conformal) Carroll
transformations as automorphisms.

\subsection{Massive electric theory}
\label{sec:mass-electr-theory}

Consider a free real massive scalar field in $d+1$ dimensional
Minkowski space. It satisfies the commutation relations
\begin{equation}
    [\phi(t,x),\phi(s,y)] = i\Delta(t-s,x-y)\, ,
\end{equation}
with the Jordan--Pauli commutator function
\begin{equation}
i\Delta(t, x) := \int \frac{d^d p}{(2\pi)^d} \int dp^0 \, \mathrm{sgn}(p^0) \, \delta((p^0)^2 - p^2 - m^2) \, e^{-i p^0 t + i p \cdot x}\,.
\end{equation}
We then introduce the electric Carrollian fields by an appropriate
rescaling of the time coordinate and the mass,
\begin{align}
  \label{eq:phieps}
\phi_\epsilon(t, x) &:= \epsilon^{-\frac{1}{2}} \phi(\epsilon t, x)\,, &  m&\mapsto m/\epsilon  \, ,
\end{align}
such that in the limit one obtains the action~\eqref{eq:electr-no-mom}
for the field $\phi_0$.  We introduce smeared fields
\begin{equation}\label{smearedfields}
    \phi_\epsilon(f) = \int dt\,d^dx\,f(t,x)\,\phi_\epsilon(t,x)
\end{equation}
for some real smooth test function
$f\in \mathcal{C}_c(\mathbb{R}^{d+1})$ with compact support.  The
commutator defines an antisymmetric bilinear form $\sigma^m_\epsilon$ on
the space $\mathcal{C}_c(\mathbb{R}^{d+1})$ of test functions,
\begin{align}
  [\phi_\epsilon(f), \phi_\epsilon(g)] &=
   \int \frac{d^d p}{(2\pi)^d \, (2\epsilon E_p)}\left( \hat{f}(-\epsilon E_p,- p)\, \hat{g}(\epsilon E_p, p) - \hat{f}(\epsilon E_p, p)\, \hat{g}(-\epsilon E_p, -p) \right)\\
& =: i\sigma^m_\epsilon(f,g)\, ,
\end{align}
where $\hat{f}$ denotes the Fourier transform,
\begin{equation}
    \hat{f}(p^0,p) = \int dt\,d^d x\, f(t,x)\,e^{ip^0\,t - ip\cdot x} \,,
\end{equation}
and 
\begin{equation}
    E_p=\sqrt{p^2+\frac{m^2}{\epsilon^2}}\,.
\end{equation}
$\sigma^m_\epsilon$ has a
well defined limit when $\epsilon$ goes to zero,
\begin{align}
\sigma^m_0(f,g) &= 
\lim_{\epsilon \to 0} \sigma^m_\epsilon(f,g) \\ 
&=
-i \int \frac{d^d p}{(2\pi)^d \, (2m)}
\left( \hat{f}(-m,- p)\, \hat{g}(m, p) - \hat{f}(m, p)\, \hat{g}(-m, -p) \right) \label{eq:massiv-Sym} \\
&=
-\frac{1}{m}\int dt\,d^d x \,\int ds\,f(t,x)\,g(s,x)\,\sin (m(t-s)) \, .
\label{sigma0-massive-position}
\end{align}
This result agrees with~\cite{deBoer:2023fnj}.

This anti-symmetric bilinear form defines a symplectic form
$\sigma^m_\epsilon$ on the quotient space
\begin{equation}
    V_\epsilon^m = \mathcal{C}_c(\mathbb{R}^{d+1}) / \mathrm{ker}\, \sigma^m_\epsilon\,,
\end{equation}
where the kernel of $\sigma^m_\epsilon$ is divided out.  To such a
symplectic vector space we can associate a Weyl algebra
$\mathcal{W}(V_\epsilon^m,\sigma_\epsilon^m)$ (sometimes called CCR
algebra, see e.g., \cite{BratteliOla1997Oaaq}): a $C^*$-algebra
generated by elements $W^{m}_\epsilon(f)$, $f\in V_\epsilon^m$, that
satisfy
\begin{equation}\label{Weylrelation}
    W_\epsilon^m(f)W_\epsilon^m(g) = e^{-i\sigma^m_\epsilon(f,g)/2}W_\epsilon^m(f+g)\,, 
  \end{equation}
as well as
\begin{align}
    W_\epsilon^m(0) &= \mathbf{1}\,,& W_\epsilon^m(-f)&=W_\epsilon^m(f)^*\,.
\end{align}
Formally we can think of the $W_\epsilon^m(f)$ as the exponentiated fields
\begin{equation}\label{Weyl-fields}
  W_\epsilon^m(f) = e^{i \phi_\epsilon(f)}\, ,
\end{equation}
which satisfy the Weyl relations~\eqref{Weylrelation} upon using the
Baker--Campbell--Hausdorff formula and the fact that
$\sigma^m_{\epsilon}(f,g)$ commutes with the fields.  The Weyl
relations are a rigorous way to express the canonical commutation
relations. In contrast to the fields themselves, the elements
$W_\epsilon^m(f)$ can be represented as bounded operators, and one
avoids subtleties due to unboundedness of operators.

The resulting Weyl algebra can be thought of as the algebra of
(exponentiated) on-shell fields. Indeed, the equivalence relation
$f\sim f+g$ with $g\in \mathrm{ker}\, \sigma^m_\epsilon$ can be
understood as $\phi_0(f)=\phi_0(f+g)$ or $\phi_0(g)=0$. The kernel of
$\sigma^m_0$ equals the image of $P_0^m=\partial_t^2+m^2$, see
\cite{Fredenhagen:2026xxx}. Hence, we have
\begin{equation}
  0=\phi_0(P^m_0 h) = \int dtd^dx\,\phi_0(t,x)\,(\partial_t^2+m^2)h(t,x) = \int dt d^dx\, \big((\partial_t^2+m^2)\phi_0(t,x)\big)h(t,x)
\end{equation}
for any $h \in \mathcal{C}_c(\mathbb{R}^{d+1})$, which means that the field satisfies the equation of motion.

We have obtained a family of Weyl algebras labelled by the parameter
$\epsilon$. As the form $\sigma^m_\epsilon$ has a well-defined limit for
$\epsilon\to 0$, we can also define a Weyl algebra that one may call
the \textsl{electric Carrollian Weyl algebra}
\begin{equation}
    W_0^m(f)W_0^m(g) = e^{-i\sigma^m_0(f,g)/2}W_0^m(f+g)\,.
\end{equation}
Here, we view $f$ as representing an equivalence class in $V_0^m=\mathcal{C}_c(\mathbb{R}^{d+1})/\ker \sigma^m_0$, where 
\begin{equation}\label{kersigma-m-0}
    \ker \sigma^m_0 = \big\{ f\in \mathcal{C}_c(\mathbb{R}^{d+1}):\, \hat f(\pm m,p)=0 \,, \forall \,  p \in \RR^{d}\big\}\,.
\end{equation}
We denote the electric Carrollian Weyl algebra with $\mathcal{W}(V_0^m,\sigma_0^m)$.

\subsection{Massless electric theory}
\label{sec:massl-electr-theory}

For the massless electric theory we again rescale~\eqref{eq:phieps},
but this time we do not rescale the mass, which effectively means that
$m \to 0$ in~\eqref{eq:comm-fct-sum}.  The commutator is now
\begin{align} 
[\phi_\epsilon(f), \phi_\epsilon(g)] &= \int \frac{d^d
    p}{(2\pi)^d \, (2\epsilon E_p)}
\left( \hat{f}(-\epsilon E_p,- p)\, \hat{g}(\epsilon E_p, p) - \hat{f}(\epsilon E_p, p)\, \hat{g}(-\epsilon E_p, -p) \right)\\
& =:
i\sigma_\epsilon(f,g)\, ,
\end{align}
with 
\begin{equation}
    E_p = \sqrt{p^2 + m^2}\,.
\end{equation}
$\sigma_\epsilon$ has a well-defined limit when $\epsilon$ goes to zero,
\begin{align}
\sigma_0(f,g) 
&=
-i\int \frac{d^d p}{(2\pi)^d} \left(  \hat{f}(0, -p) \,\partial_0 \hat{g}(0, p)
- \partial_0 \hat{f}(0, p) \, \hat{g}(0, -p) \right)
\label{exprsigma0} \\
&=
-\int dt\,d^d x \,\int ds\, (t-s)\,f(t,x)\,g(s,x) \,.\label{eq:sigma0position}
\end{align}
Here, we denote the derivative of the Fourier transform with respect
to $p^0$ as $\partial_0$. The result also coincides with the $m\to 0$ limit of~\eqref{sigma0-massive-position}.

As before, we can define a symplectic vector space $V_0=\mathcal{C}_c(\mathbb{R}^{d+1}) / \mathrm{ker}\, \sigma_0$, where the kernel of $\sigma_0$ is given by
\begin{equation}\label{kersigma0}
    \ker \sigma_0 = \big\{ f\in \mathcal{C}_c(\mathbb{R}^{d+1}):\, \hat{f}\big|_{p^0=0} = \partial_{0}\hat{f}\big|_{p^0=0}=0\big\}\,.
\end{equation}
The corresponding Weyl algebra $\mathcal{W}(V_0,\sigma_0)$ is generated by elements $W_0(f)$ with
\begin{equation}
  W_0(f) W_0(g) = e^{-i\sigma_0(f,g)/2}W_0(f+g)\, ,   
\end{equation}
where $f$ is understood as a representative of an equivalence class in $V_0$.

\subsection{(Conformal) Carroll symmetries as automorphisms}
\label{sec:Carrollsymmetries}

We now investigate symmetries of the electric Carrollian Weyl algebra. We
will first consider the extended Carroll group which includes
arbitrary supertranslations. It is generated by
\begin{subequations}\label{globalconformalCarroll}
\begin{itemize}
    \item rotations and spatial translations
    \begin{equation}\label{spatialtrafos}
        (t,x) \longmapsto \alpha_{R,a}(t,x)=(t,R\,x+a) \,,
    \end{equation}
    with a rotation matrix $R$ and a vector $a\in \mathbb{R}^d$,
    \item supertranslations 
    \begin{equation}\label{supertranslations}
        (t,x) \longmapsto \alpha_h(t,x)= (t+h(x),x)\,,
    \end{equation}
    with a smooth function $h$ (a constant $h$ describes time translation, a linear $h$ describes Carroll boosts).    
\end{itemize}
\end{subequations}
For any transformation $\alpha(t,x)=(t',x')$ given in equations~\eqref{globalconformalCarroll}, we define a map
\begin{equation}
   \begin{array}{rl}
       \rho_{\alpha}: \, \mathcal{C}_c(\mathbb{R}^{d+1}) &  \longrightarrow \mathcal{C}_c(\mathbb{R}^{d+1})\\
        f & \longmapsto \rho_\alpha(f)=f\circ \alpha^{-1}\,.
    \end{array}
\end{equation}
This defines a representation on the space of test functions on
$\mathbb{R}^{d+1}$. The action of $\rho_\alpha$ respects the bilinear
forms $\sigma_0^m$ and $\sigma_0$,
\begin{align}\label{invariancesigma0}
    \sigma_0^m (\rho_\alpha (f), \rho_\alpha ( g)) &= \sigma_0^m (f,g)\,, &
    \sigma_0 (\rho_\alpha (f), \rho_\alpha ( g)) &= \sigma_0 (f,g)\,.
\end{align}
This is obvious from the expressions \eqref{sigma0-massive-position}
for $\sigma_0^m$ and \eqref{eq:sigma0position} for $\sigma_0$ in
position space. Because of~\eqref{invariancesigma0}, $\rho_\alpha$
preserves the kernel of $\sigma_0^m$ and of $\sigma_0$ and descends to a map on the
quotient space $V_0^m$ and $V_0$, respectively, that leaves the symplectic forms invariant. It
thus induces an automorphism of the Weyl algebra, acting on the
generators as
\begin{equation}\label{automorphismelectric}
    \pi(\alpha): W_0^m(f) \mapsto W_0^m(\rho_\alpha (f))\,.
\end{equation}
Similarly for $W_0(f)$.
Additionally, the bilinear forms are invariant under any spatial diffeomorphism,
\begin{equation}\label{spatialdiffeos}
    (t,x) \longmapsto \alpha_\zeta(t,x)=(t,\zeta(x))\,,
\end{equation}
when the test functions transform as half-densities,
\begin{equation}
    \rho_{\alpha_\zeta}(f)(t,x') = \left| \frac{\partial x'}{\partial x}\right|^{-1/2} f(t,x)\quad \text{with}\ x'=\zeta(x)\,.
\end{equation}
This again is obvious from the explicit form of $\sigma_0^m$
(see~\eqref{sigma0-massive-position}) and of $\sigma_0$
(see~\eqref{eq:sigma0position}).

In the case of the electric massless scalar, we additionally have invariance under dilatations,\footnote{Together with the invariance under spatial diffeomorphisms~\eqref{spatialdiffeos}, this implies also invariance under separate dilatations of space and time directions.}
\begin{equation}
    \alpha(t,x)= (\lambda t, \lambda x)\,,
\end{equation}
where the action on test functions is
\begin{equation}
    \rho_\alpha(f) = \lambda^{-(d+3)/2} f\circ \alpha^{-1}\,.
\end{equation}
Indeed, from~\eqref{eq:sigma0position} we have
\begin{align}
    \sigma_0\big(\rho_\alpha(f),\rho_\alpha(g)\big) &=
    -\lambda^{-(d+3)} \int dt\, ds\,d^dx\,(t-s)\, f(\lambda^{-1}t,\lambda^{-1}x) \,g(\lambda^{-1}s,\lambda^{-1}x) \\
    &= \sigma_0(f,g)\,.
\end{align}
The temporal special conformal transformations are simply
supertranslations with a function $h(x)\propto x^2$, and they thus
constitute a symmetry of the electric scalar. The spatial special
conformal transformations are of the form
\begin{equation}
    \alpha(t,x)=(t',x')=\bigg( \frac{t}{1-2b\cdot x + b^2 \,x^2},\frac{x-b\,x^2}{1-2b\cdot x + b^2 \,x^2}\bigg)\,.
\end{equation}
Analogously to the relativistic case, they are -- for a noncompact background -- not defined everywhere, but only on
$\mathbb{R}\times \big(\mathbb{R}^d \setminus \{ b/b^2\}\big)$. Such transformations become well defined when the spatial part is replaced by a sphere $S^d$. The expected action on test functions
is
\begin{equation}
\rho_\alpha(f)(t',x') = (1-2b\cdot x + b^2\,x^2)^{(d+3)/2} \,f(t,x)\,.
\end{equation}
This still results in a well-defined function, but it does not have
compact support, unless $f$ vanishes around $x=b/b^2$. For a given
parameter $b$, $\rho_\alpha$ can only be defined on a subspace of the
space of test functions and not on the whole space. For those,
however, $\sigma_0$ is invariant:
\begin{align}
        \sigma_0 (\rho_\alpha(f),\rho_\alpha(g)) &= - \int dt'\,ds'\,d^dx'\,(t'-s')\,\rho_\alpha(f)(t',x')\,\rho_\alpha(g)(s',x')\\
        &= - \int dt'\,d's\,d^dx'\, (t'-s')(1-2b\cdot x + b^2\,x^2)^{(d+3)} \,f(t,x) \,g(s,x)\\
        &= - \int dt\,ds\,d^dx \, (t-s)\,f(t,x)\, g(s,x)\\
        &=\sigma_0 (f,g)\,, 
          \label{invariace-spatial-SCT}
\end{align}
where we used the standard Jacobian determinant of a special conformal
transformation $x'=(x-b\,x^2)/(1-2\,b\cdot x + b^2\,x^2)$. In summary,
the massless electric theory admits an action of the global conformal
Carroll group.

\section{States of the electric Carrollian Weyl algebra}
\label{sec:states}

In this section, we analyze the states of free massive and massless
electric Carrollian theories. Our discussion will include vacuum and
thermal states, as well as more general classes of states. For free
theories, the so-called quasifree or Gaussian states are completely
determined by their two-point functions, which will therefore serve as
a central object of our analysis.

Before we formulate these ideas in the language of AQFT let us
describe our results in distributional form. The starting point is the
vacuum (not time-ordered) two-point function for a massive scalar field
in Minkowski space
\begin{align}
  \langle \phi(t,x) \phi(s,y) \rangle &= \int \frac{d^{d}p}{(2\pi)^{d}} \frac{1}{2 E_{p}} e^{i \left(-E_{p}(t-s) + p \cdot (x-y)\right)}  \, .
\end{align}
A fast way to arrive at the massive electric limit is to use the
Carrollian energy dispersion relation $E_{p}=m$, which leads to
\begin{align}
  \langle \phi(t,x) \phi(s,y) \rangle  &= \frac{1}{2 m} e^{-i m(t-s)} \delta(x-y)\\
                                       &= \frac{1}{2m}\delta(x-y) - \frac{i}{2}(t-s)\delta(x-y) + O(m) \, . \label{eq:massless-sublte}
\end{align}
The limit is again ultralocal, and we will show that it leads for the
massive case to a well-defined regular vacuum state.

From \eqref{eq:massless-sublte} we also see that the massless limit
diverges and is thus more subtle. One could first subtract the
divergent term and then take the limit leading
to~\cite{Bagchi:2022emh,Chen:2023pqf,deBoer:2023fnj,Chen:2024voz}
\begin{align}
 \label{eq:isolated}
  \braket{\phi(t,x)\phi(s,y)}_{\subm}  = - \frac{i}{2}(t-s)\delta(x-y)  \, .
\end{align}
The limit is now finite, but this modified two-point function does not
correspond to a well-defined state and cannot be realized as the
expectation value with respect to a vector in a Hilbert space (see
Section~\ref{sec:classical-state}). There is a similar divergence in
the two-point function of boundary fields in flat space holography
(see Section~\ref{sec:flat-space-hologr}), which captures interesting
infrared physics.  We therefore propose a different and well-defined
vacuum state. The price we pay is that the state is not regular,
similarly to sharp momentum eigenstates.  Motivated by the possible
relation to flat space holography and these subtleties, we further
analyze states of the massless theory and construct the most general
spatially homogeneous pure regular quasifree state and so-called
Sorkin--Johnston states, both of which solidify our earlier result
that the vacuum is either a well-defined, but nonregular state, or
that there exists no distinguished regular vacuum.

Finally we look at the thermal states, which in many ways mirror the
discussed vacuum states in the sense that we find well-defined regular
and nonregular states.  We show that these states are
Kubo--Martin--Schwinger (KMS) states, i.e., they satisfy the KMS
condition. By importing this condition to the Carrollian world we
provide a notion of thermal equilibrium for quantum systems in the
large volume limit which circumvents problems related to density
matrices that might not be traceclass.  From this perspective the
massive electric Carrollian KMS state is a well-defined regular thermal
state.\footnote{It is actually better behaved than thermal states for
  the massless scalar in finite volume which has IR divergences. The
  KMS condition circumvents problems of ill-defined partition
  functions~\cite{deBoer:2023fnj} and shows that Carrollian theories can
  have well defined thermal states, see
  also~\cite{Figueroa-OFarrill:2023qty,deBoer:2023fnj,Ecker:2023uwm,Cotler:2024xhb}.}

\subsection{States in AQFT}
\label{sec:states-aqft}

Before we explicitly construct states for our Carrollian theories, we
first recall their definition within the framework of algebraic
quantum field theory. We begin by reviewing the notion of a general
state, and then turn to the subclass of quasifree states, of which
the vacuum ground state and thermal Gibbs states are particular instances
(see, e.g.,~\cite{Fewster:2019ixc,Haag:1996hvx,Kay:1988mu} for
more details).

A state $\omega$ on a unital $C^*$-algebra is a linear,
positive, normalized functional,
\begin{equation}
  \label{eq:state-definition}
    \omega:\,A \mapsto \omega(A)\in \mathbb{C}\ \text{linear}\,,\ \omega(A^*A)\geq 0\,, \ \omega(\mathbf{1})=1\, .
\end{equation}
It is called pure if it cannot be written as
$\omega=\lambda\,\omega_1 + (1-\lambda)\,\omega_2$ for $0<\lambda<1$
for two different states $\omega_1\not=\omega_2$.  When $A=A^{*}$,
where we may think of $A^{*}$ as the hermitian adjoint, we can
interpret $\omega(A)$ as the expectation value of $A$ in the state
$\omega$. For example, when we have an algebra of bounded operators,
then any density matrix $\rho$ on the Hilbert space induces a state
via $\omega_{\rho}(A)=\mathrm{tr}(\rho A)$. The definition in terms of
a functional satisfying~\eqref{eq:state-definition}, however, is more
general and does not make reference to a particular representation of
the $C^*$-algebra on a Hilbert space and a particular vector or
density matrix.

Given a state~\eqref{eq:state-definition}, the Gelfand--Naimark--Segal
(GNS) construction provides us with a Hilbert space (see,
e.g.,~\cite[III.2.2]{Haag:1996hvx}): a state $\omega$ on a unital
$C^{*}$ algebra $\mathcal{A}$ gives rise to a Hilbert space
$\mathcal{H}_{\omega}$, a representation
$\pi_{\omega}:\mathcal{A} \to \mathcal{B}(\mathcal{H}_{\omega})$ and a
cyclic vector $\ket {\Omega_{\omega}}$ such that
\begin{align}
  \omega(A) = \braket{\Omega_{\omega}|\pi_{\omega}(A)|\Omega_{\omega}} \, .
\end{align}
We stress that positivity of the state is essential to ensure that one obtains a Hilbert space.

\subsubsection{Quasifree states}
\label{sec:quasifree-states}

For free theories we can construct so called quasifree or Gaussian
states. These states are fully determined by their two-point function
(we assume vanishing one-point functions). We describe here the general situation for a Weyl algebra over a symplectic vector space $(V,\sigma)$.

Quasifree states are given by
\begin{align}
  \omega(W(f)) = e^{-\frac{1}{2}q(f)} \, , \label{eq:Gaussian-state}
\end{align}
and are specified by their covariance $q$, a map $q:V\to[0,\infty]$
with $q(0)=0$ satisfying $q(\lambda\,f)=\lambda^2\,q(f)$ for
$\lambda\neq0$ and the parallelogram law
$q(f+g)+q(f-g)=2\,q(f)+2\,q(g)$; we use the convention
$e^{-\infty}=0$.  The set $V_q=\{f\in V:\,q(f)<\infty\}$ is a linear
subspace,\footnote{If $q(f)$ and $q(g)$ are finite, the
    parallelogram law makes $q(f+g)+q(f-g)$ finite and hence both
    summands finite.} and the polarization
\begin{align}
  \mu(f,g)=\frac{1}{4}\big(q(f+g)-q(f-g)\big)
\end{align}
 defines a
real, symmetric, positive semi-definite bilinear form on $V_q$ with
$q(f)=\mu(f,f)$.  Note that only the quadratic form $q$ extends to all of
$V$; the bilinear form $\mu$ does not.

The functional~\eqref{eq:Gaussian-state} is positive, and hence a state,
if and only if $\mu$ satisfies the following inequality with the
symplectic form $\sigma$ (see, e.g.,~\cite{Kay:1988mu} and expanded upon
in Appendix~\ref{app:positivity_quasi_free})
\begin{align}
  \label{eq:indeterm}
  \mu(f,f)\,\mu(g,g) \geq \frac{1}{4}\,|\sigma(f,g)|^{2}
\end{align}
for all $f,g\in V_q$.  One may interpret~\eqref{eq:indeterm} as a
generalization of the uncertainty principle that any quasifree quantum
state needs to satisfy for consistency.

The state is called \emph{regular} if $\lambda\mapsto\omega(W(\lambda f))$
is continuous for all $f\in V$, which here is the case precisely if
$V_q=V$, i.e.\ if $q$ is finite; otherwise it is \emph{nonregular}, and
$\omega(W(\lambda f))$ fails to be continuous at $\lambda=0$ for
$f\notin V_q$.\footnote{In the literature the covariance is usually assumed to be
  finite (see, e.g.,~\cite{Kay:1988mu,Fewster:2019ixc}), so that
  quasifree states are automatically regular; this is the class we refer
  to as regular quasifree states.  Covariances that are allowed to take
  the value $+\infty$ are called generalized quasifree states
  in~\cite{Acerbi:1993pp}.}

On $V_q$, the covariance is the symmetric part of the two-point function
\begin{align}
  \label{eq:decomp}
  G(f,g)= \mu(f,g) + \frac{i}{2}\sigma(f,g) \, ,
\end{align}
which we can interpret as $``G(f,g)= \omega(\phi(f)\phi(g))"$ if we think of Weyl-elements as $``W(f)=e^{i\phi(f)}"$. This function contains all the information of the state, while the anti-symmetric symplectic form $\sigma$ encodes the dynamical information of the theory and is independent of the state. Conversely, any bilinear hermitian positive functional $G(f,g)$ with $\text{Im}\,G(f,g)=\frac{1}{2}\sigma(f,g)$ defines a regular state \eqref{eq:Gaussian-state} with $\mu(f,g)=\text{Re}\,G(f,g)$, provided \eqref{eq:indeterm} still holds. For example, this can happen if $G$ is the restriction of a positive sesquilinear form on complex valued functions. Then \eqref{eq:indeterm} holds by the Cauchy-Schwarz-inequality.

For regular quasifree states, the GNS Hilbert space is (unitarily equivalent
to) a Fock space~\cite{Kay:1988mu}.  Its construction becomes
particularly simple if the state is pure, which is equivalent to the
case when the inequality~\eqref{eq:indeterm} is saturated in the sense
that
 \begin{equation}
     \mu(f,f)= \frac{1}{4} \sup\limits_{g\neq 0} \frac{|\sigma(f,g) |^{2}}{\mu(g,g)}\quad\forall f\in V\,.
 \end{equation}
Briefly summarized, the construction first uses $\mu$ as a real scalar product to complete $V$ to a real Hilbert space $\mathcal{H}'$, on which one constructs an operator $J$ satisfying
\begin{align}
    \mu (f,Jg) &= -\frac{1}{2}\sigma (f,g)\,,& J^2 &= -\mathbf{1}\,,& J^\dagger=-J\,.\label{eq:Prop-J}
\end{align}
The operator $J$ can be used to define a complex structure on $\mathcal{H}'$ where multiplication by $i$ corresponds to acting with $J$. On this complex vector space one defines a scalar product by
\begin{equation}
    \langle f,g\rangle_\mathcal{H}:=\mu(f,g)+\frac{i}{2}\sigma(f,g)=G(f,g)\,,
\end{equation}
giving it a structure of a complex Hilbert space, which we denote by $\mathcal{H}$.\footnote{For nonpure states, the Hilbert space $\mathcal{H}$ and the associated embedding of $V$ become more complicated. However, the properties of $J$ and the inner product still hold, with the embedding inserted where needed.}

Given $\mathcal{H}$, we can construct the Fock-space $\mathcal{F}(\mathcal{H})$, where we denote with $a(f)$, $a^\dagger(g)$ and $\ket{0}$ the annihilation operator, the creation operator and the vacuum vector, respectively. The annihilation and creation operator are conjugate linear and linear, respectively, and satisfy
\begin{equation}
    \left[a(f),a^\dagger(g)\right]=\langle f,g\rangle_\mathcal{H}\,.
\end{equation}
Finally, we define a representation $\Pi$ of the Weyl algebra $\mathcal{W}(V,\sigma)$ on $\mathcal{F}(\mathcal{H})$ via\footnote{$\overline{\left(a(f)+a^\dagger(f)\right)}$ denotes the closure of $a(f)+a^\dagger(f)$.} 
\begin{equation}
\Pi(W(f)):=e^{i\overline{\left(a(f)+a^\dagger(f)\right)}}\quad\text{for}\  f\in V\,,\label{eq:Fock-space-rep}
\end{equation}
which implies
\begin{equation}
    \langle 0| \Pi(W(f)) |0\rangle = e^{-\frac{1}{2}G(f,f)}=e^{-\frac{1}{2}\mu(f,f)} = \omega(W(f))\,,
\end{equation}
where $\ket 0$ denotes the vacuum of the Fock space.

The simplest example of a regular quasifree state is the vacuum state of the canonically quantized free (relativistic) scalar on Minkowski spacetime. Its value on the generating
elements $W(f)=e^{i\phi(f)}$ is determined by the two-point (Wightman)
function
\begin{equation}
    \langle 0| \phi(f)\,\phi(g) |0\rangle =: G(f,g)\,,
\end{equation}
and is given by
\begin{equation}
  \omega\big( W(f)\big) = e^{-\frac{1}{2}G(f,f)}\, .
\end{equation}
The two-point function reads
\begin{equation}
    G(f,g) = \int \frac{d^dp}{(2\pi)^d (2E_p)} \hat{f}(-E_p,-p)\,\hat{g}(E_p,p)\,. 
\end{equation}

As stressed before, all information about the quasifree state is contained in the covariance $q$, whose associated bilinear form $\mu$ satisfies~\eqref{eq:indeterm}. Different classes of physically
relevant states must satisfy additional conditions that encode their
physical interpretation, such as representing a ``no-particle'' ground
state or a state in thermal equilibrium.

We shall remain agnostic for the moment about the precise definition
of a ground state, although several commonly imposed criteria are
standard in the literature. Ground states are typically taken to be
pure states that minimize the energy and are invariant
under (at least part of) the symmetries of the theory. In addition,
one often demands a spectrum condition on the generators of spacetime
translations, ensuring the positivity and stability of the
vacuum.

\subsection{Massive vacuum}
\label{sec:massive-vacuum}

When we perform the Carroll rescaling of the fields as in~\eqref{eq:phieps}, we obtain the two-point functions
\begin{align}
    G^m_\epsilon(f,g) = \langle 0| \phi_\epsilon (f)\,\phi_\epsilon (g) |0\rangle = \int \frac{d^dp}{(2\pi)^d (2\epsilon E_p)} \hat{f}(-\epsilon E_p,-p)\,\hat{g}(\epsilon E_p,p)\,.
\end{align}
In the limit $\epsilon\to 0$, we find
a well-defined two-point function for the massive electric
case
\begin{align}
    G^m_0(f,g) &= \lim_{\epsilon \to 0} G^m_\epsilon(f,g) \\
    &= 
    \int \frac{d^dp}{(2\pi)^d (2 m)} \hat{f}(-m,-p)\,\hat{g}(m,p) \\
    &=
    \frac{1}{2m}\int dt\,d^d x \,\int ds\,f(t,x)\,g(s,x)\,e^{-im(t-s)}\,. \label{eq:massive-2pt}
\end{align}
It is easily checked, that $G^m_0$ is a bilinear hermitian positive form and that $\text{Im}\,G^m_0(f,g)=\frac{1}{2}\sigma^m_0(f,g)$. Since $G^m_0$ can be viewed as the restriction of a positive sesquilinear form on the space of complex test functions via
\begin{equation}
    \tilde{G}^m_0(f,g) = \frac{1}{2m}\int dt\,d^d x \,\int ds\,\overline{f(t,x)}\,g(s,x)\,e^{-im(t-s)}\,,
\end{equation}
it satisfies the Cauchy-Schwarz inequality
\begin{equation}
    |G^m_0(f,g)|^2\leq G^m_0(f,f)\,G^m_0(g,g)\,.
\end{equation}
This implies \eqref{eq:indeterm} with $\mu^m_0(f,g):=\text{Re}\,G^m_0(f,g)$. By the discussion in Section \ref{sec:quasifree-states}, we conclude that the corresponding functional $\omega^m_0$ on the Weyl algebra, given by
\begin{equation}
    \omega^m_0(W^m_0(f))=e^{-\frac{1}{2}G^m_0(f,f)}\,,\label{eq:massive-vacu-state}
\end{equation}
is a regular quasifree state. Regularity is immediate here,
since the covariance $G^m_0$ is finite on all of $V^m_0$. In addition, $\omega^m_0$ is pure as will be shown in
Section~\ref{sec:symm-mass-vacu}.

Lastly, we point out that the state $\omega_0^m$ can be seen as a sort of limit of the states $\omega_\epsilon^m$, in the sense that 
\begin{equation}
    \omega^m_0(W^m_0(f))=e^{-\frac{1}{2}G^m_0(f,f)}=\lim_{\epsilon \to 0} e^{-\frac{1}{2}G^m_\epsilon(f,f)}=``\lim_{\epsilon \to 0} \omega^m_\epsilon(W^m_\epsilon(f))"\,.
\end{equation}
The quasifree vacuum states $\omega^m_\epsilon$, defined through their two-point function $G_\epsilon^m$, are states of the Weyl-algebras $\mathcal{W}(V^m_\epsilon,\sigma^m_\epsilon)$. As the underlying vector spaces $V_\epsilon^m=\mathcal{C}_c(\mathbb{R}^{d+1})/\ker \sigma_\epsilon^m$ of those Weyl-algebras depend on $\epsilon$, one needs to be careful when one tries to define a limit of states. In the above case, however, the procedure leads to a well-defined state. 
In the massless case, the same approach will again lead to a well-defined result (see Section~\ref{sec:electricstate}), even though the two-point function itself diverges. 

\subsubsection{Symmetries of the massive vacuum}
\label{sec:symm-mass-vacu}

Under a Carroll transformation
\begin{equation}
    \alpha(t,x) = (t+h(x),Rx+a)\,,
\end{equation}
including arbitrary supertranslations, the state $\omega_0^m$ is invariant with respect to the induced automorphism discussed in Section~\ref{sec:Carrollsymmetries}, i.e., $\omega_0^m(W_0^m(\rho_\alpha(f)))=\omega_0^m(W_0^m(f))$ for every $f$. Indeed, the two-point function is invariant,
\begin{equation}
    G^m_0\big(\rho_\alpha(f),\rho_\alpha(g)\big) = G^m_0(f\circ \alpha^{-1},g\circ \alpha^{-1}) = G^m_0(f,g)\,,
\end{equation}
which follows straightforwardly from the explicit expression~\eqref{eq:massive-2pt} for $G^m_0$. This implies invariance of the corresponding state $\omega^m_0$.

Via the GNS-construction, one can associate a Hilbert space representation of the Weyl algebra to any state. It is shown in \cite{Fewster:2019ixc}, that if a state is invariant under an automorphism, the said automorphism can be implemented as a unitary operator in the GNS-representation which leaves the GNS-vector invariant. This makes the interpretation of the GNS-vector as a vacuum vector or ground state more credible. 

Further, we want to check if such unitary implementations of symmetries are strongly continuous, which is necessary for the formulation of energy, momentum and the spectrum condition. We have already seen in Section \ref{sec:quasifree-states} that the GNS-representation of $\omega^m_0$ is a Fock space. To check strong continuity, we first want to make the representation more explicit. 

First, we write down the maps used in the construction \ref{sec:quasifree-states} explicitly for this case. The covariance $\mu^m_0$ is, as before, the real part of $G^m_0$.
\begin{equation}
    \mu^m_0(f,g):=\text{Re}\,G^m_0(f,g)=\frac{1}{2m}\int dt\,d^d x \,\int ds\,f(t,x)\,g(s,x)\,\cos (m(t-s)) \, .
\end{equation}
The well-defined operator 
\begin{equation}
    J:V^m_0\rightarrow V^m_0,\,f(t,x)\mapsto f\Big(t-\frac{\pi}{2m},x\Big)
\end{equation}
satisfies all the properties in \eqref{eq:Prop-J}. Indeed, if $f$ and $g$ represent the same element in $V^m_0$, they differ by a function $P^m_0h$. Since constant shifts commute with derivatives, we conclude that $Jf$ differs from $Jg$ by $P^m_0Jh$, hence $J$ is well-defined. This operator was constructed to satisfy 
\begin{align}
    \mu^m_0(f,Jg)&=\frac{1}{2m}\int dt\,d^d x \,\int ds\,f(t,x)\,g\Big(s-\frac{\pi}{2m},x\Big)\,\cos (m(t-s))\\
    &=\frac{1}{2m}\int dt\,d^d x \,\int ds\,f(t,x)\,g(s,x)\,\cos \Big(m(t-s)-\frac{\pi}{2}\Big)=-\sigma^m_0(f,g)/2\,.
    \label{rel-mu-sigma}
\end{align}
$J^\dagger$ is given by $f(t,x)\mapsto f(t+\frac{\pi}{2m},x)$ and satisfies $J^\dagger J=\mathbf{1}$. Moreover, one observes that $\mu^m_0(f,J^2 g)=-\mu^m_0(f,g)$ which implies $J^2=-\mathbf{1}$ by nondegeneracy.

Hence, $J$ satisfies \eqref{eq:Prop-J} on the inner product space $(V^m_0,\mu^m_0)$. This allows for the following: By the Cauchy-Schwarz-inequality for $\mu^m_0$ and the properties of $J$ we get
\begin{equation}
    \frac{|\sigma^m_0(f,g)|^2}{4}=\mu^m_0(f,Jg)^2\leq \mu^m_0(f,f)\mu^m_0(Jg,Jg)=\mu^m_0(f,f)\mu^m_0(g,g)
\end{equation}
for any $f,g \in V^m_0$. For fixed $f$, the Cauchy-Schwarz-inequality has equality, when $f\propto Jg$. Therefore, for any $f \in V^m_0$ there is a $\lambda$ such that $g=-\lambda Jf$ yields $\mu^m_0(f,f)=|\sigma^m_0(f,g)|^2 / 4\mu^m_0(g,g)$. In particular, this means that~\eqref{eq:indeterm} is saturated and hence $\omega^m_0$ is pure.

Furthermore, since $\omega^m_0$ is pure, the real one-particle Hilbert space $\mathcal{H}'$ is just the completion of $V^m_0$ with respect to $\mu^m_0$. $J$ can then be continuously extended to a bounded operator on $\mathcal{H}'$, which we also denote with $J$ and induces a complex structure, as outlined in Section \ref{sec:quasifree-states}. The resulting complex Hilbert space we denote with $\mathcal{H}$.

The representation of $\omega^m_0$ on the Fock space $\mathcal{F}(\mathcal{H})$ will be denoted by $\Pi^m_0$, and is given by \eqref{eq:Fock-space-rep}. Finally, any transformation $\alpha$, under which $\omega^m_0$ is invariant, is then unitarily implemented as
\begin{equation}
    U_\alpha \Pi^m_0(W_0(f))\ket{0}:=\Pi^m_0(W_0(f\circ\alpha^{-1}))\ket{0}=e^{i\overline{\left(a(f\circ\alpha^{-1})+a^\dagger(f\circ\alpha^{-1})\right)}}\ket{0}\,.
\end{equation}
This implementation is indeed strongly continuous, by
\cite[Prop.~5.2.4]{BratteliOla1997Oaaq}. The same proposition
shows that $\Pi^m_0(W_0(tf))$ is strongly continuous, in accordance
with the regularity of $\omega^m_0$
\cite[Section~5.2.3]{BratteliOla1997Oaaq}. 

\subsection{Massless analogue of vacuum state}
\label{sec:electricstate}

We will now focus on the Carroll limit of the massless theory. We can
again evaluate the two-point function of the scaled field
$\phi_\epsilon$ in the vacuum state,
\begin{align}
    G_\epsilon(f,g) &= \langle 0 | \phi_\epsilon(f)\phi_\epsilon(g) |0\rangle \\
    &= \int \frac{d^dp}{(2\pi)^d (2\epsilon E_p)} \hat{f}(-\epsilon E_p,-p)\,\hat{g}(\epsilon E_p,p)\,.
\end{align}
Although its imaginary part has a well-defined limit, the two-point
function diverges when $\epsilon \to 0$. However, motivated by
the interpretation of the massive vacuum as a limit of states, we observe
that we can still define a state on the corresponding Weyl algebra,
\begin{equation}
  \label{eq:state-electr-massless}
  \omega_0 \big( W_0(f) \big)=\lim_{\epsilon\to 0} e^{-\frac{1}{2}G_\epsilon(f,f)}
  = \left\{ \begin{array}{cl}
    1 & \text{for}\ \hat{f}\big|_{p^0=0} = 0\\
    0 & \text{for}\ \hat{f}\big|_{p^0=0} \neq0 
    \end{array}\right.
  = \left\{ \begin{array}{cl}
    1 & \text{for}\ f^{\phi} = 0\\
    0 & \text{for}\ f^{\phi} \neq 0  \, ,
    \end{array}\right.  
\end{equation}
where the last equality is expressed in position space, with
$f^\phi (x) = \int dt\, f(t,x)$.  This state minimizes the energy, is
invariant under Carroll transformations, supertranslation and
dilatations and is well-defined in the sense that it satisfies all
conditions of~\eqref{eq:state-definition}. However,
$\omega_0(W_0(\lambda f))$ is not continuous in $\lambda$ and it is
therefore not regular. This implies, e.g., that we can not even
formally differentiate $\omega_{0}$ to access the unexponentiated
fields and correlation functions.

One could also quantize this theory using canonical quantization. In
that case one does not find a distinguished vacuum vector, which
agrees with the statement~\cite{deBoer:2023fnj} that the vacuum is
not normalizable. Let us emphasize that this is nothing unusual: for theories with zero modes or theories on time-dependent spacetimes this is the typical situation, as discussed, e.g., in~\cite{Witten:2021jzq}.

We will now argue that~\eqref{eq:state-electr-massless} is a
well-defined state. Recalling that $W_0(f)$ depends only on the
equivalence class, $W_0(f+g)=W_0(f)$ for $g\in\ker \sigma_0$, one
notices that this map is well-defined because of the form of
$\ker \sigma_0$ given in~\eqref{kersigma0}. This map can be linearly
extended to the Weyl algebra to define a linear functional that is
normalised ($\omega_0(W_0(0))=1$), and one can explicitly check that
it is positive and defines a pure state \cite{Tiefenbacher:Carroll}. Actually, it is an instance of a state
$\omega_K$ of a Weyl algebra defined over a symplectic space
$(V,\sigma)$ that one can associate to any Lagrangian\footnote{A
  maximal subspace on which $\sigma$ is evaluated to zero.} subspace
$K\subset V$ by setting $\omega_K(W(x))=1$ for $x\in K$, and to zero
otherwise (see, e.g., \cite{Fredenhagen:lecturenotes}). In our case
\begin{equation}
    K = \{ f\in V_0\,:\ \hat{f}\big|_{p^0=0} =0\}\,.
\end{equation}
The simplest instances of such states arise in ordinary quantum
mechanics where they can be obtained by considering states in which
position is sharp and momentum is completely undetermined, or the
other way round. In our case, the state can be interpreted as a state
in which the value of the field is completely undetermined, and the
value of the field momentum is sharply zero.

\subsubsection{Symmetries of the massless state}
\label{sec:symm-massless-vacu}

The state $\omega_0$ is invariant under the Carroll transformations
given in~\eqref{globalconformalCarroll} including supertranslations,
and also under dilatations. This follows by inspecting all cases
discussed in Section~\ref{sec:Carrollsymmetries} and verifying that
for any such Carroll transformation $\rho_\alpha$, we have
\begin{equation}
    \hat{f}\big|_{p^0=0} =0\, \Longleftrightarrow \, \widehat{\rho_\alpha(f)}\big|_{p^0=0}=0\, ,
\end{equation}
or, equivalently,
\begin{equation}
    \int dt\, f(t,\_) = 0 \, \Longleftrightarrow \, \int dt\, \rho_\alpha(f)(t,\_)=0 \, .
\end{equation}
As remarked before, one can use the GNS construction to represent the Weyl algebra on a Hilbert space. The Carroll symmetries are realized as unitary operators that leave the unit vector, corresponding to the state $\omega_0$, invariant.

To prepare for the concrete realization of such a representation on a
Hilbert space, we introduce some notation. For a test function
$f\in \mathcal{C}_c(\mathbb{R}^{d+1})$, we define
\begin{align}
    f^\phi (x) &= \int dt\, f(t,x)\,, & f^\pi (x) = \int dt\, t\,f(t,x)\,.
\end{align}
This notation is motivated from the concrete expression for a smeared field $\phi_0(f)$ for a solution $\phi_0$ that satisfies
\begin{equation}
    \phi_0(t,x) = \phi_0(0,x) + t\,\pi_0(0,x)
\end{equation}
for the conjugate field momentum $\pi_0 = \dot{\phi}_0$. We have
\begin{equation}
    \phi_0(f) = \int dtd^dx f(t,x) \big(\phi_0(0,x)+t\,\pi_0(0,x) \big) = \int d^dx \big( f^\phi(x)\phi_0(0,x) + f^\pi(x) \pi_0(0,x)\big)\, .
\end{equation}
The bilinear form $\sigma_0$ can then be expressed as
\begin{equation}
    \sigma_0 (f,g) = - \int d^dx \, \big( f^\pi(x)g^\phi(x) - f^\phi(x) g^\pi(x) \big)\,.
\end{equation}
A Hilbert space representation that is unitarily equivalent to the GNS representation corresponding to the state $\omega_0$ can be realized as the space of square-summable (complex) functions over $\mathcal{C}_c(\mathbb{R}^d)$,
\begin{equation}
    \mathcal{H} = \ell^2\big(\mathcal{C}_c(\mathbb{R}^d)\big)\,.
\end{equation}
This Hilbert-space is nonseparable, and an orthonormal basis is given
by $\{ \delta_g \}_{g\in \mathcal{C}_c(\mathbb{R}^d)}$,
\begin{equation}
    \langle \delta_{g_1} | \delta_{g_2}\rangle = \delta_{g_1,g_2}\,.
\end{equation}
The representation of the Weyl algebra is given by
\begin{equation}
    \Pi\big(W(f)\big) \delta_g = e^{\frac{i}{2}\int d^dx \big(2\,f^\pi(x)g(x) + f^\pi(x)f^\phi(x)\big)} \delta_{f^\phi+g}\,.
\end{equation}
Spatial translations and rotations, $\alpha(t,x)=(t,Rx+a)$ are represented unitarily by
\begin{equation}
    U(\alpha)\delta_g = \delta_{g\circ \alpha^{-1}}\,.
\end{equation}
On the other hand, supertranslations $\alpha(t,x)=(t+h(x),x)$ are represented as
\begin{equation}
    U(\alpha)\delta_g = e^{\frac{i}{2}\int d^d x \,h(x)\,g(x)^2} \delta_g\,.
\end{equation}
Whereas spatial translations and rotations are not represented strongly continuously, the representation of the supertranslations is strongly continuous. In particular, for temporal translations, $h(x)=h_0$, we have
\begin{equation}
    U(\alpha)\delta_g = e^{ih_0 \frac{1}{2}\int d^d x \,g(x)^2} \delta_g\,,
\end{equation}
which shows that the spectrum of the generator $P^0$ of temporal translations is positive.

\subsection{Thermal states}
\label{sec:thermal-states}

Another important class of states we will consider consists of thermal
or KMS states (see, e.g., \cite[Section V]{Haag:1996hvx}). By
definition, these must satisfy the KMS condition: given a
$C^*$-algebra $\mathcal{A}$ and a group of $*$-automorphisms
$t\mapsto \tau_t$, a KMS-state $\omega$ is a $\tau_t$-invariant state on $\mathcal{A}$
such that $F(A,B,z):=\omega(A\tau_z(B))$ is analytic in $z$ on
$0<\mathrm{Im}\,z<\beta$, and continuous on $0\leq \mathrm{Im}\,z \leq \beta$, and it satisfies
\begin{equation}
    \omega(A\,\tau_{t+i\beta}(B))=\omega(\tau_t(B)A)\quad\forall t\in\RR\label{eq:KMS-condition}
\end{equation}
for any $A,B\in\mathcal{A}$.

The KMS condition generalizes the
notion of a Gibbs state on an operator algebra,
\begin{align}
\omega_{\rho}(A) &= \mathrm{tr}(A \rho),
&
\rho &= \frac{e^{-\beta H}}{\mathrm{tr}( e^{-\beta H})},
\end{align}
which is well-defined only if $e^{-\beta H}$ is trace-class, and
therefore often requires placing the theory in a finite volume.

For the relativistic real scalar, the two-point function in a thermal
state at inverse temperature $\beta$ is given by
\cite{Bellac:2011kqa}
\begin{equation}
    G^\beta(t,x; s,y)=\int \frac{d^dp}{(2\pi)^d} \frac{1}{2 E_p}\left(\frac{e^{i E_p (t-s)-ip(x-y)}}{e^{\beta E_p}-1}-\frac{e^{-i E_p (t-s)+ip(x-y)}}{e^{-\beta E_p}-1}\right)\,,
\end{equation}
which in terms of test functions reads
\begin{equation}
    G^\beta(f,g)= \left\langle\phi(f)\phi(g)\right\rangle^{\beta} = \int \frac{d^dp}{(2\pi)^d} \frac{1}{2 E_p}\left(\frac{\hat{f}(E_p,p)\hat{g}(-E_p,-p)}{e^{\beta E_p}-1}-\frac{\hat{f}(-E_p,-p)\hat{g}(E_p,p)}{e^{-\beta E_p}-1}\right)\,.
\end{equation}
The corresponding quasifree state of the Weyl algebra is defined by
\begin{equation}
    \omega^{\beta}\big(W(f)\big) = e^{-\frac{1}{2}G^\beta(f,f)}\,.\label{eq:Base-thermal-state}
\end{equation}
Because we rescale time in the Carroll limit, it is natural to consider a family of states with $\beta\mapsto\epsilon \beta$. The two-point function of the rescaled fields then becomes\footnote{Note that the same two-point functions can be directly obtained by the procedure in \cite{Bellac:2011kqa} using the Lagrangian density for the rescaled field 
\begin{equation}
    \mathcal{L}=\frac{1}{2}\left(\partial_t\phi_\epsilon)^2-\epsilon^2(\nabla_x\phi_\epsilon)^2-\epsilon^2m^2\phi_\epsilon^2\right)\,.
\end{equation}
The rescaling of $\beta$ is then implicit with the choice of integration limits in \cite{Bellac:2011kqa}. For the massive case, we rescale $m\mapsto m/\epsilon$.}
\begin{align}
G^{\beta}_\epsilon(f,g)&:=\left\langle\phi_\epsilon(f)\phi_\epsilon(g)\right\rangle^{\beta \epsilon}\\
    &=
    \int \frac{d^dp}{(2\pi)^d} \frac{1}{2 \epsilon E_p}\left(\frac{\hat{f}(\epsilon E_p,p)\hat{g}(-\epsilon E_p,-p)}{e^{\beta \epsilon E_p}-1}-\frac{\hat{f}(-\epsilon E_p,-p)\hat{g}(\epsilon E_p,p)}{e^{-\beta \epsilon E_p}-1}\right)\,.
    \label{thermal-two-point-function-electric}
\end{align}

To check the KMS-Condition in our case, we first need to figure out which group of $*$-automorphisms we want to use. On the level of fields, the usual $\tau_t$ one considers is $\tau_s(\phi(t,x)):=e^{isH}\phi(t,x)e^{-isH}=\phi(t+s,x)$, given a Hamiltonian $H$. Smeared out with a test function $f$ this reads
\begin{equation}
    \tau_s(\phi(f))=\int dtd^dx\, \tau_s(\phi(t,x))f(t,x)=\int dtd^dx\,\phi(t+s,x)f(t,x)=\phi(f_{s})
\end{equation}
where $f_s(t,x):=f(t-s,x)$. For an exponential of the form $e^{i\phi(f)}$ we can then compute
\begin{equation}
    \tau_s(e^{i\phi(f)})=e^{i\phi(f_s)}\,.
\end{equation}
This allows us to extend the definition of $\tau_t$ to elements of a Weyl-algebra via
\begin{equation}
    \tau_t(W(f)):=W(f_t)\,,
\end{equation}
which is independent of any Hamiltonian $H$ and is therefore suitable to be used for Carrollian Weyl-algebras. Moreover, since $f_s$ is a test function if $f$ is one, it suffices to check the KMS-condition \eqref{eq:KMS-condition} for $t=0$.

For a state of the form $\omega(W(f))=e^{-\frac{1}{2}G(f,f)}$ for an appropriate $G$,
like \eqref{eq:Base-thermal-state}, to be KMS with the above $\tau_t$, we need to check if
\begin{equation}
    t \mapsto e^{-\frac{1}{2}\left(G(f+g_{t},f+g_{t})+i\sigma(f,g_{t})\right)}\,,
\end{equation}
or equivalently
\begin{equation}
    t\mapsto -\frac{1}{2}\left(G(f+g_{t},f+g_{t})+i\sigma(f,g_{t})\right) \label{eq:Analytic-cond}
\end{equation}  
can be continued analytically on $0<\text{Im}\,z<\beta$, for all test functions $f,g$. The KMS-condition then reads\footnote{Note that expressions like $\sigma(f,g_{i\beta})$ make sense, even though $g_{i\beta}$ is nonsensical if we only consider test functions over the real numbers, since $\sigma$ can be seen as a distribution and $\sigma(f,g_{i\beta})$ makes sense after integration with an appropriate substitution.}
\begin{equation}
    e^{-\frac{1}{2}\left(G(f+g_{i\beta},f+g_{i\beta})+i\sigma(f,g_{i\beta})\right)}=e^{-\frac{1}{2}\left(G(g+f,g+f)+i\sigma(g,f)\right)}\,.\label{eq:KMS-massive}
\end{equation}
\subsubsection{Massive electric Carrollian thermal states}
\label{sec:mass-electr-carr}

In the massive limit we have seen that $\epsilon E_p \xrightarrow{}m$
as $\epsilon\xrightarrow{}0$. Using this we get the two-point function
\begin{equation}
    G^{m,\beta}_0(f,g)=\int \frac{d^dp}{(2\pi)^d} \frac{1}{2 m}\left(\frac{\hat{f}(m,p)\hat{g}(-m,-p)}{e^{\beta m}-1}-\frac{\hat{f}(-m,-p)\hat{g}(m,p)}{e^{-\beta m}-1}\right)\,.
\end{equation}
We immediately see that, although the two-point function still depends on $\beta$, we no longer have the characteristic Bose-Einstein-distribution present. The analog of \eqref{eq:Base-thermal-state},
\begin{equation}
    \omega^{m,\beta}_0\big(W_0(f)\big) = e^{-\frac{1}{2}G^{m,\beta}_0(f,f)}\,,
\end{equation}
then defines a state on the Weyl-algebra $\mathcal{W}(V^m_0,\sigma^m_0)$, since a small computation shows that $G^{m,\beta}_0$ indeed is bilinear, hermitian and positive (for $\beta,m \geq 0$) and satisfies $\text{Im}\,G^{m,\beta}_0(f,g)=\frac{1}{2}\sigma^m_0(f,g)$.

Another quick computation shows that $\widehat{g_{t}}(p_0,p)=\hat{g}(p_0,p)e^{i t p_0}$, which is easily continued analytically by replacing $t$ with $z$. Since $G^{m,\beta}_0$ and $\sigma^{m}_0$ are bilinear and we only deal with test functions, $G^{m,\beta}_0(f+g_{t},f+g_{t})$ and $\sigma^m_0(f,g_{t})$ extend analytically in the same way and hence so does \eqref{eq:Analytic-cond}. Moreover, it is easy to see that above relation implies $G^{m,\beta}_0(f_t,g_t)=G^{m,\beta}_0(f,g)$ and hence $\omega^{m,\beta}_0$ is $\tau_t$-invariant.

To show \eqref{eq:KMS-massive}, we first note that $G^{m,\beta}_0$ satisfies $G^{m,\beta}_0(f,g_{i\beta}) = G^{m,\beta}_0(g,f)$. This is indeed true by the above relation. 
\begin{align}
    G^{m,\beta}_0(f,g_{i\beta})
    &=
    \int \frac{d^dp}{(2\pi)^d} \frac{1}{2 m}\left(\frac{\hat{f}(m,p)\hat{g}(-m,-p)e^{\beta m}}{e^{\beta m}-1}-\frac{\hat{f}(-m,-p)\hat{g}(m,p)e^{-\beta m}}{e^{-\beta m}-1}\right)\\
    &=
    G^{m,\beta}_0(g,f)\,.
\end{align}
Second, we can compute 
\begin{align}
    G^{m,\beta}_0(f,g)-G^{m,\beta}_0(g,f)&=\int \frac{d^dp}{(2\pi)^d} \frac{1}{2 m}\left(\frac{\hat{f}(m,p)\hat{g}(-m,-p)}{e^{\beta m}-1}-\frac{\hat{f}(-m,-p)\hat{g}(m,p)}{e^{-\beta m}-1}\right.\\&\quad -\left. \frac{\hat{g}(m,p)\hat{f}(-m,-p)}{e^{\beta m}-1}+\frac{\hat{g}(-m,-p)\hat{f}(m,p)}{e^{-\beta m}-1}\right)\\
    &=
    -\int \frac{d^dp}{(2\pi)^d} \frac{1}{2 m}\left(\hat{f}(m,p)\hat{g}(-m,-p)-\hat{f}(-m,-p)\hat{g}(m,p)\right)\\
    &=
    i\sigma^m_0(f,g)\,.  \label{eq:G-antisym}
\end{align}
The difference to the relation $\text{Im}\,G^{m,\beta}_0(f,g)=\frac{1}{2}\sigma^m_0(f,g)$, which holds for real-valued test functions $f,g$, is that \eqref{eq:G-antisym} still holds if we replace $g$ with $g_{i\beta}$. Essentially, we lose that $G^{m,\beta}_0$ is hermitian when we continue $t\mapsto G^{m,\beta}_0(f,g_{t})$ analytically to complex values.\\
Lastly, by linearity of $G^{m,\beta}_0$ and \eqref{eq:G-antisym} we get
\begin{equation}
G^{m,\beta}_0(f+g,f+g)= G^{m,\beta}_0(f,f) +G^{m,\beta}_0(g,g)+ 2G^{m,\beta}_0(f,g) -i\sigma^m_0(f,g)\,.
\end{equation}
There are no issues in replacing $g$ with $g_{i\beta}$ in the above, so we conclude
\begin{equation}
G^{m,\beta}_0(f+g_{i\beta},f+g_{i\beta}) +i\sigma^m_0(f,g_{i\beta}) =G^{m,\beta}_0(g+f,g+f) +i\sigma^m_0(g,f)\,,
\end{equation}
using $G^{m,\beta}_0(f,g_{i\beta}) = G^{m,\beta}_0(g,f)$, implying \eqref{eq:KMS-massive}.

\subsubsection{Massless electric Carrollian thermal states}
\label{sec:massl-electr-carr}

In the massless case, the two-point function \eqref{thermal-two-point-function-electric} diverges in general because $\epsilon E_p \to 0$ as $\epsilon\to 0$. The leading piece becomes
\begin{align}
    G^\beta_\epsilon (f,f) &=\int \frac{d^dp}{(2\pi)^d} \,\frac{1}{\beta (\epsilon E_p)^2}  \big| \hat{f}(0,p)\hat{f}(0,-p)\big|^2 + \dots
\end{align}
On the other hand, if $\hat{f}(0,p)=0$ or equivalently $f^\phi=0$, the limit is finite,
\begin{align}
    G_0^\beta (f,f) \Big|_{f^\phi=0}&= \frac{1}{\beta} \int \frac{d^dp}{(2\pi)^d} \, \left( -\partial_0\hat{f}(0,p)\partial_0\hat{f}(0,-p)\right)\\
    &=\frac{1}{\beta} \int d^dx \, \big(f^\pi(x)\big)^2\,.
\end{align}
On the Weyl algebra, one then obtains a well-defined state in the limit:
\begin{equation}\label{masslessKMSstate}
    \omega_0^\beta \big( W_0(f)\big) = \left\{
    \begin{array}{ll}
        0 & \text{for}\ f^\phi\not=0  \\
        e^{-\frac{1}{2\beta}\int d^dx\,(f^\pi(x))^2}&  \text{for}\ f^\phi=0\,.
    \end{array}
    \right.
\end{equation}
Indeed, linearity and normalization are obvious, and positivity is proven in Appendix~\ref{app:positivity_thermal}. 

To show that $\omega_0^\beta$ is a KMS-state with $\tau_t$, we follow
similar arguments as in the previous subsection. We need to verify
that
\begin{equation}
    t\mapsto e^{-\frac{i}{2}\sigma_0(f,g_{t})}\omega^\beta_0(W_0(f+g_{t}))\label{eq:Analytic-massless}
\end{equation}
can be continued analytically to $0<\text{Im}\,z<\beta$ and that
\begin{equation}
    e^{-\frac{i}{2}\sigma_0(f,g_{i\beta})}\omega^\beta_0(W_0(f+g_{i\beta}))=e^{-\frac{i}{2}\sigma_0(g,f)}\omega^\beta_0(W_0(g+f)) \label{eq:KMS-cond-massless}
\end{equation}
holds for any $f,g$. Moreover, we need to show that $\omega_0^\beta$ is $\tau_t$-invariant. To do this, we first get
\begin{align}
    f_{t}^\phi=&f^\phi\\
    f_{t}^\pi=&f^\pi+tf^\phi
\end{align}
from a small computation. The $\tau_t$-invariance follows quickly, using the above two equations to compute $\omega_0^\beta \big( W_0(f_t)\big)$. Also, both of those equations continue analytically to the complex plane. Now two cases emerge.

On the one hand, if $g^\phi+f^\phi\neq0$, then so is $f^\phi+g_{t}^\phi=g^\phi+f^\phi\neq0$. In this case \eqref{eq:Analytic-massless} is just the $0$-function and extends trivially to $0<\text{Im}\,z<\beta$. Moreover, \eqref{eq:KMS-cond-massless} also holds, since both the right-hand and the left-hand side vanish by the same argument.

On the other hand, if $g^\phi+f^\phi=0$, then again $f^\phi+g_{t}^\phi=g^\phi+f^\phi=0$. In this case, \eqref{eq:Analytic-massless} becomes 
\begin{equation}
    t\mapsto e^{-\frac{1}{2}\left(\frac{1}{\beta}\int d^dx\,(f^\pi(x)+g^\pi_{t}(x))^2+i\sigma_0(f,g_{t})\right)}\,.
\end{equation}
Since $f_{t}^\phi$ and $f_{t}^\pi$ extend analytically, we can use the same argument as in the previous subsection to conclude that the above also continues analytically to $0<\text{Im}\,z<\beta$. Furthermore, \eqref{eq:KMS-cond-massless} reads as
\begin{equation}
    e^{-\frac{1}{2}\left(\frac{1}{\beta}\int d^dx\,(f^\pi(x)+g^\pi_{i\beta}(x))^2+i\sigma_0(f,g_{i\beta})\right)}=e^{-\frac{1}{2}\left(\frac{1}{\beta}\int d^dx\,(g^\pi(x)+f^\pi(x))^2+i\sigma_0(g,f)\right)}\,. \label{eq:KMS-massless}
\end{equation}
By some direct calculations we get
\begin{align}
    \sigma_0(f,g_{i\beta})&=- \int d^dx \, \big( f^\pi(x)g_{i\beta}^\phi(x) - f^\phi(x) g_{i\beta}^\pi(x) \big)\\
    &=
    -\sigma_0(g,f)+i\beta\int d^dx \, f^\phi(x) g^\phi(x)
\end{align}
and
\begin{align}
    \frac{1}{\beta}\int d^dx\,\big(f^\pi(x)+g^\pi_{i\beta}(x)\big)^2
    &=
    \frac{1}{\beta}\int d^dx\,\left[\big(f^\pi+g^\pi\big)^2+2i\beta g^\phi(f^\pi+g^\pi)-\beta^2 (g^\phi)^2\right]\\
    &=
    \frac{1}{\beta}\int d^dx\,\left[\big(f^\pi+g^\pi\big)^2+2i\beta (g^\phi f^\pi-f^\phi g^\pi)+\beta^2 f^\phi g^\phi\right]\\
    &=
    \frac{1}{\beta}\int d^dx\,\big(f^\pi+g^\pi\big)^2+2i\sigma_0(g,f)+\beta \int d^dx\,f^\phi g^\phi\,,
\end{align}
where we used $f^\phi=-g^\phi$.
Combining the results, we get
\begin{equation}
    \frac{1}{\beta}\int d^dx\,\big(f^\pi(x)+g^\pi_{i\beta}(x)\big)^2+i\sigma_0(f,g_{i\beta})=\frac{1}{\beta}\int d^dx\,\big(f^\pi+g^\pi\big)^2+i\sigma_0(g,f)
\end{equation}
and hence \eqref{eq:KMS-massless} holds.

\subsection{Other states of the massless theory}
\label{sec:other-stat-massl}

The nonregularity of the massless vacuum state proposed in
Section~\ref{sec:electricstate} is unconventional and motivates us to
examine alternative candidate states.  An alternative proposal is to
first subtract the divergent contribution in the two-point function
and subsequently take the massless
limit~\cite{Bagchi:2022emh,Chen:2023pqf,deBoer:2023fnj,Chen:2024voz}. However,
as we demonstrate in Section~\ref{sec:classical-state}, this procedure does not yield a well-defined state on the Weyl algebra and
discards the infrared information carried by the divergence (cf., Section~\ref{sec:flat-space-hologr}).
In Section~\ref{sec:spat-homog-pure}, we construct the most general spatially homogeneous
pure regular quasifree state and show that none of them is invariant under time translations. Finally, in
Section~\ref{sec:sork-johnst-stat}, we discuss Sorkin--Johnston
states, a distinguished class of covariantly defined quantum states
that are determined solely by the Pauli--Jordan function.

\subsubsection{``State'' without fluctuations}
\label{sec:classical-state}

We will analyze states for the two-point function of the massless
electric theory that was proposed
in~\cite{Bagchi:2022emh,Chen:2023pqf,deBoer:2023fnj,Chen:2024voz}
\begin{align}
  \label{eq:iso-cl}
  \braket{\Omega|\phi(t,x)\phi(s,y)|\Omega}_{\subm}  = - \frac{i}{2}(t-s)\delta(x-y)  \, ,
\end{align}
where the divergent term was first subtracted before taking the
limit. While this prescription leads to a finite quantity, one might
lose relevant information: a similar divergence appears in flat space
holography in the two-point function of operators corresponding to
boundary values of bulk fields (see
Section~\ref{sec:flat-space-hologr}). This divergence captures
infrared physics and should therefore be retained.

  Also, this two-point function cannot actually derive from a
  well-defined state on a Hilbert space. To show this, we first
  integrate \eqref{eq:iso-cl} against some test functions
\begin{align}
  \label{eq:nonzero}
  \braket{\Omega|\phi(f)\phi(g)|\Omega}_{\subm}   = -\frac{i}{2} \int dt \,d^{d}x \int ds  \, (t-s)\,f(t,x)\,g(s,x) \, .
\end{align}
Equating the test functions we find that the norm is vanishing 
\begin{align}
  \braket{\Omega|\phi(f)\phi(f)|\Omega}_{\subm}   = | \phi(f)\ket{\Omega}_{\subm} |^{2} =0
\end{align}
and, by unitarity, $\phi(f)\ket{\Omega}_{\subm} =0$ for any test
function $f$. Therefore,
$\braket{\Omega|\phi(f)\phi(g)|\Omega}_{\subm}=0$ which is in
tension with~\eqref{eq:nonzero} that is generically nonzero.

Another way to see that~\eqref{eq:iso-cl} is not a well-defined state
is to recall that we can decompose the two-point function into
symmetric and antisymmetric
parts~\eqref{eq:decomp}. For~\eqref{eq:nonzero} the symmetric part
vanishes and therefore $\mu_{\subm}(f,g)=0$. This implies that
this state does not satisfy the uncertainty
relation~\eqref{eq:indeterm} and hence positivity of the
state. Intuitively, we can understand this as follows: We have fixed
$\mu_{\subm}(f,g)$ to be strictly zero, which is analogous to
fixing both position and momentum to zero, which is clearly in tension
with the uncertainty relation.

If one attempts to define a quasifree state with the two point
function~\eqref{eq:iso-cl} above, one would arrive at a functional
$\omega$ satisfying $\omega(W(f))=1$. Such a functional is not a state
because positivity is violated: Take $g_1$ and $g_2$ from the
symplectic space underlying the Weyl-algebra such that
$\sigma(g_1,g_2)\neq 0$. This is always possible by nondegeneracy of
$\sigma$. Define $f_1:=g_1$, $f_2:=\frac{2\pi}{\sigma(g_1,g_2)}g_2$
and $f_3:=f_1+f_2$ and consider $C:=W(f_1)+W(f_2)+W(f_3)$. Then,
$C^*C$ is positive but
\begin{align}
\omega(C^*C)&=\omega\left(W(0)+e^{i\sigma(f_1,f_2)/2}W(-f_1+f_2)+e^{i\sigma(f_1,f_3)/2}W(-f_1+f_3)\right.\nonumber\\&\left.+e^{i\sigma(f_2,f_1)/2}W(-f_2+f_1)+W(0)+e^{i\sigma(f_2,f_3)/2}W(-f_2+f_3)\right.\nonumber\\&\left.+e^{i\sigma(f_3,f_1)/2}W(-f_3+f_1)+e^{i\sigma(f_3,f_2)/2}W(-f_3+f_2)+W(0)\right)\\
    &=
    3+3e^{i\sigma(f_1,f_2)/2}+3e^{i\sigma(f_2,f_1)/2}\,.
\end{align}
By construction we have $\sigma(f_1,f_2)=\sigma\Big(g_1,\frac{2\pi}{\sigma(g_1,g_2)}g_2\Big)=2\pi$ and hence
\begin{equation}
    \omega(C^*C)=3+3e^{i\pi}+3e^{-i\pi}=-3\,.
\end{equation}
In particular, this is consistent with the argument above that
\eqref{eq:iso-cl} cannot be realized on a Hilbert space.

Let us stress that this is not a general prohibition against infrared subtractions or against restricting to a suitable subspace of test functions.  Whenever a symmetric form $\mu'$ on a subspace $V'\subset V$ satisfies the inequality~\eqref{eq:indeterm} with respect to $\sigma|_{V'}$, it defines a quasifree state on the Weyl subalgebra $\mathcal{W}(V',\sigma|_{V'})$, and nothing so far excludes that such a $\mu'$ arises after a subtraction.  What is special about~\eqref{eq:iso-cl} is that the subtraction leaves nothing behind: the symmetric part vanishes identically, $\mu_{\subm}\equiv 0$, so that~\eqref{eq:indeterm} degenerates into the condition $\sigma(f,g)=0$.  The only subspaces on which~\eqref{eq:iso-cl} defines a state are therefore those on which the symplectic form vanishes, and on such a subspace the state carries no fluctuations at all. This is the sense in which we speak of a ``state'' without fluctuations.  On the full algebra, where $\sigma$ is nondegenerate, no such state exists.

The obstruction is thus tied to the degeneracy of the remaining covariance, not to the act of subtracting.  In Section~\ref{sec:quasifree-state} we will meet a case in which the corresponding remainder is a nontrivial positive covariance, and where the situation is accordingly different.

Let us stress that in both the massless electric case (Section~\ref{sec:electricstate})
and flat-space holography (Section~\ref{sec:quasifree-state}) we do
not subtract anything: we use exponentiated fields and remove the
regulator. This suppresses the Weyl generators for test functions with nonvanishing zero modes and leads to well-defined nonregular quasifree states.

\subsubsection{Spatially homogeneous pure regular quasifree states}
\label{sec:spat-homog-pure}

The state $\omega_0$ that we obtained from the vacuum state of the
relativistic scalar in the Carroll limit is not a regular state of the
Weyl algebra. In this subsection we explore other, regular states of
the electric scalar, and construct the most general spatially
homogeneous pure regular quasifree state.

A regular quasifree state is determined by its two-point function,
\begin{equation}
    G(f,g) = \mu(f,g)+i\sigma_0(f,g)/2\,,
\end{equation}
whose anti-symmetric part is given by the symplectic form. It is
convenient to collect the Fourier transforms of a test function into
$F=(\widehat{f^\phi},\widehat{f^\pi})^{T}$. The most general ansatz for
the symmetric part that is invariant under spatial translations is then
encoded in a matrix-valued symbol $M(p)$,
\begin{align}\label{eq:homogeneous-ansatz}
  \mu(f,g) &= \frac{1}{2}\int\frac{d^dp}{(2\pi)^d}\,F(-p)^{T}\,M(p)\,G(p)\,,
  & M(p) &=\begin{pmatrix} A_1(p) & B(p)\\ \overline{B(p)} & A_2(p)\end{pmatrix}\,,
\end{align}
so that $A_1$ and $A_2$ multiply
$\widehat{f^\phi}(-p)\widehat{g^\phi}(p)$ and
$\widehat{f^\pi}(-p)\widehat{g^\pi}(p)$, while $B$ and $\overline{B}$
multiply the two mixed terms. Symmetry of $\mu$, and the requirement
that $\mu$ take real values on real test functions, translate into two
conditions on the symbol,
\begin{equation}\label{eq:homogeneous-parities}
  M(p)^{\dagger}=M(p)\,,\qquad M(-p)=\overline{M(p)}\,.
\end{equation}
In components, $A_1$ and $A_2$ are real and even, whereas $B$ is
constrained only by $B(-p)=\overline{B(p)}$, so that its real part is
even and its imaginary part is odd.

Positivity of $\mu$ amounts to positivity of the hermitian matrix
$M(p)$ for every $p$, that is
\begin{equation}\label{eq:homogeneous-positivity}
  A_1\geq 0\,,\qquad A_2\geq 0\,,\qquad \det M = A_1\,A_2-|B|^2\geq 0\,.
\end{equation}

We are looking for two-point functions that lead to
a well-defined pure regular quasifree state. For such a state, there needs to
exist an operator $J:V_0\to V_0$ satisfying $J^2=-\mathbf{1}$ with
\begin{align}
    \mu(f,Jg) &= -\sigma_0(f,g)/2 \,,&
    \sigma_0(f,Jg)/2 &= \mu(f,g)\,.
\end{align}
From the second relation we can read off $J$, which acts on $F$ by
multiplication with the matrix
\begin{equation}\label{eq:homogeneous-J}
  \mathcal{J}(p) = \begin{pmatrix} \overline{B} & A_2\\ -A_1 & -B\end{pmatrix}\,,
  \qquad \operatorname{tr}\mathcal{J} = \overline{B}-B\,,
  \qquad \det\mathcal{J} = \det M\,,
\end{equation}
the first relation being implied by this together with
$J^2=-\mathbf{1}$. By the Cayley--Hamilton theorem, $J^2=-\mathbf{1}$ is
equivalent to
$(\operatorname{tr}\mathcal{J})\,\mathcal{J}=(\det\mathcal{J}-1)\,\mathbf{1}$.
A nonvanishing trace would force $\mathcal{J}$ to be proportional to the
identity and hence $A_1=A_2=0$, which is excluded
by~\eqref{eq:homogeneous-positivity}. Purity therefore requires that the
symbol be real and of unit determinant,
\begin{equation}\label{eq:homogeneous-purity}
  \operatorname{Im}B = 0\,,\qquad \det M(p) = A_1(p)\,A_2(p)-B(p)^2 = 1\,.
\end{equation}
In particular the odd imaginary part of $B$, which a general
spatially homogeneous covariance allows, is absent for pure states, so
that $B$ is real and even.

The most general solution consistent with positivity is given by 
\begin{align}
    A_1(p)&= \gamma(p)^{-2}+\delta(p)^2\gamma(p)^2\,,& A_2(p)&= \gamma(p)^2\,, & B(p)&= \delta(p)\,\gamma(p)^2
\end{align}
with a real even function $\delta$ and a nonzero even function $\gamma$ that we
choose to be positive, $\gamma >0$. The full
two-point function is then 
\begin{multline}
    G^{(\gamma,\delta)}(f,g) = \\
    \frac{1}{2}\int \frac{d^d p}{(2\pi)^d} \bigg(\big(\gamma(p)^{-2}+ \delta(p)^2\, \gamma(p)^2 \big) \, \widehat{f^\phi}(-p)\,\widehat{g^\phi}(p) 
    + \gamma(p)^2 \,\widehat{f^\pi}(-p)\,\widehat{g^\pi}(p) \\ + \big(i+\delta(p)\, \gamma(p)^2\big) \,\widehat{f^\phi}(-p)\,\widehat{g^\pi}(p) + \big(-i +\delta(p)\, \gamma(p)^2\big) \,\widehat{f^\pi}(-p)\,\widehat{g^\phi}(p)\bigg)\,.
    \label{alternativestates}
\end{multline}
It defines a state for the electric Carrollian Weyl algebra by
\begin{equation}
    \omega^{(\gamma,\delta)}\big( W_0(f)\big) = e^{-\frac{1}{2}G^{(\gamma,\delta)}(f,f)}\,.
\end{equation}
The state $\omega_0$ is recovered from $\omega^{(\gamma,\delta)}$ by taking the limit $\gamma\to 0$.

All correlation functions of the fields are then determined in terms of the 2-point function, as in a free theory (via Wick's theorem). Given such a state, one can construct a representation of the Weyl algebra on a Hilbert space. The vector space $V_0$ can be interpreted as a complex vector space using $J$ as the complex structure with a scalar product given by the two-point function. The Fock space of its completion defines a Hilbert space on which the Weyl algebra is represented, and the state corresponds to taking expectation values with respect to the Fock space vacuum.  

The states $\omega^{(\gamma,\delta)}$ are by construction invariant under spatial translations; they are in addition invariant under spatial rotations precisely if $\gamma$ and $\delta$ depend on $p$ only through $|p|$. Under time translations  $\alpha(t,x)=(t+h,x)$ with a constant $h$, the function $f$ transforms as
\begin{align}
     \rho_\alpha(f)^\phi&= f^\phi\,,&\rho_\alpha(f)^\pi &= f^\pi + h\,f^\phi \,.
\end{align}
For the two-point function we then find
\begin{align}
    G^{(\gamma,\delta)}(\rho_\alpha(f),\rho_\alpha(g)) &= G^{(\gamma,\delta+h)}(f,g)\,,
\end{align}
hence, the states $\omega^{(\gamma,\delta)}$ are not invariant under time translations, but transform into each other. This implies that no pure regular quasifree state that is invariant under spatial and time translations exists.

Under a time dilatation, $\alpha(t,x)=(\lambda_t\,t,x)$, the states transform as
\begin{equation}\label{transf-genstates-time-dilatations}
    \omega^{(\gamma,\delta)} \longmapsto \omega^{(\lambda_t^{1/2}\gamma, \lambda_t^{-1}\delta)}\,.
\end{equation}
Under Carroll boosts or more general supertranslations, the states $G^{(\gamma,\delta)}$ do not transform into each other.

Under general spatial diffeomorphisms, the states are invariant for constant parameters $\gamma$ and $\delta$. The explicit two-point function in this case reads
\begin{align}\label{two-point-function-family}
    G^{(\gamma,\delta)}(t,x;s,y) = \frac{1}{2}\bigg(\gamma^{-2}\Big(1+\frac{i}{2}\gamma^2(s-t)\Big)^2+ \gamma^2\Big(\delta+\frac{1}{2}(s+t)\Big)^2\bigg)\delta^{(d)}(x-y)\,.
\end{align}

\subsubsection{Sorkin--Johnston states}
\label{sec:sork-johnst-stat}

For a relativistic free field, one can construct a distinguished
state, called Sorkin--Johnston state \cite{Afshordi:2012jf} (see also
\cite{Fewster:2012ew,Fewster:2013lqa,Brum:2013bia,Wingham:2018pxx} for further details), if the
underlying spacetime satisfies certain boundedness properties. Based
on a construction of a state in causal set theory
\cite{Johnston:2009fr}, it starts from the observation that $i\sigma$
can be interpreted as a self-adjoint operator on the space of
square-integrable functions on space-time. Its positive part can then
be used to define the two-point function of a regular quasifree state.

A similar construction can be used in the Carrollian setting. We consider a Carrollian spacetime $\mathcal{M}_{a,T}=\big(a-\frac{T}{2},a+\frac{T}{2}\big)\times \mathbb{R}^d$ where the time is restricted to an interval. We then consider an operator $\Sigma$ on the space $L_2(\mathcal{M}_{a,T})$ of square-integrable functions by
\begin{equation}
    \Sigma (f) (t,x) = -i \int_{a-\frac{T}{2}}^{a+\frac{T}{2}} dt'\,(t-t')\,f(t',x)\,,
\end{equation}
which is bounded and self-adjoint. Its square is
\begin{align}
    \Sigma^2 (f)(t,x) &= - \iint (t-t')(t'-t'') f(t'',x)\,dt'dt''\\
    &= - \iint \big((t-a)-(t'-a)\big)\big((t'-a)-(t''-a)\big) f(t'',x)\,dt'dt''\\
    &= \int \bigg(T(t-a)(t''-a)+ \frac{T^3}{12}\bigg) f(t'',x)\,dt''\,.
\end{align}
Its positive square-root is given by 
\begin{align}
    |\Sigma|(f)(t,x) &= \int \bigg(\frac{2\sqrt{3}}{T}(t-a)(t'-a)+ \frac{T}{2\sqrt{3}}\bigg) f(t',x)\,dt'\,.
\end{align}
Hence, the positive part of $\Sigma$ is given by
\begin{align}
    \Sigma^+(f)(t,x) &= \frac{1}{2} (\Sigma + |\Sigma|)(f)(t,x)\\&= \frac{1}{2}\int  \bigg( -i(t-t') + \frac{2\sqrt{3}}{T} (t-a)(t'-a) + \frac{T}{2\sqrt{3}}\bigg)\,dt'\,.
\end{align}
It can be used to define a two-point function
\begin{equation}
    G_{a,T}(f,g) = \frac{1}{2}\int  \bigg( -i(t-t') + \frac{2\sqrt{3}}{T} (t-a)(t'-a) + \frac{T}{2\sqrt{3}}\bigg)f(t,x)g(t',x)\,dt\,dt'\,d^dx\,.
\end{equation}
This two-point function diverges for $T\to \infty$, and one observes that one obtains the state $\omega_0$ in the limit. Interpreting the result for finite $T$ as the two-point function on the full $\mathbb{R}\times \mathbb{R}^{d}$, we observe that we recover the two-point function that we identified in~\eqref{two-point-function-family} with $\delta=-a$ and $\gamma^2=\frac{2\sqrt{3}}{T}$. From this interpretation, the scaling of the parameters $\gamma$ and $\delta$ under time dilatations in~\eqref{transf-genstates-time-dilatations} is natural.

\section{The magnetic case}
\label{sec:magnetic}

This section analyses the magnetic Carrollian scalar. To study this limit, we need to realize the Weyl algebra by keeping track of the field momenta to obtain a magnetic contraction. We show that Carroll symmetries act as automorphisms on the resulting algebra. The vacuum state of the free scalar induces a Carroll-invariant state of the Weyl algebra of the magnetic Carrollian scalar that -- similarly to the massless electric scalar -- is not regular. The thermal Gibbs state gives rise to a well-defined (but nonregular) thermal (KMS-)state in the magnetic limit.

\subsection{Magnetic contraction of the Weyl algebra}
\label{sec:magn-contr-weyl}

The magnetic Carroll limit involves a different rescaling of the fields,
\begin{equation}
  \tilde{\phi}_\epsilon (t,x) = \epsilon \phi_\epsilon (t,x)=\epsilon^{\frac{1}{2}}\phi(\epsilon t, x)\,.
\end{equation}
Due to the extra factor of $\epsilon$, the commutator of two fields
vanishes in the limit,
\begin{equation}
    [\tilde{\phi}_\epsilon (f),\tilde{\phi}_\epsilon (g)] \xrightarrow{\epsilon\to 0} 0\,.
\end{equation}
On the other hand, the conjugate momenta $\pi(t,x)$ need to be rescaled as 
\begin{equation}
    \tilde{\pi}_\epsilon (t,x) = \epsilon^{-\frac{1}{2}}\pi (\epsilon t,x)\,,
\end{equation}
in the magnetic limit\footnote{In the electric case, they scale as $\pi_\epsilon(t,x)=\epsilon^{\frac{1}{2}}\pi(\epsilon t,x)$.} to satisfy the equal time commutation relations
\begin{equation}
    [\tilde{\phi}_\epsilon (t,x),\tilde{\pi}_\epsilon (t,y)] = i\,\delta^{(d)}(x-y)\,.
\end{equation}
Then the commutator of two field momenta does not vanish in the limit, but we find
\begin{align}
    [\tilde{\pi}_\epsilon (f),\tilde{\pi}_\epsilon (g)] &= 
    \int \frac{d^dp}{(2\pi)^d} \frac{E_p}{2\,\epsilon} \big( \hat{f}(-\epsilon E_p,-p)\,\hat{g}(\epsilon E_p,p) - \hat{f}(\epsilon E_p,p)\,\hat{g}(-\epsilon E_p,-p)\big)\\
    &\xrightarrow{\epsilon\to 0} \int \frac{d^dp}{(2\pi)^d} E_p^2\big( \hat{f}(0,-p)\,\partial_0 \hat{g}(0,p) - \partial_0 \hat{f}(0,p)\,\hat{g}(0,-p) \big)\,.
\end{align}
To investigate the Weyl algebra in the magnetic limit, we have to be aware that the Weyl algebra should contain fields and momenta. For the relativistic scalar as well as for the electric limit, we do not lose information if we only concentrate on the fields as long as we consider fields smeared over time with a test function. This is because, due to the equations of motion, we have
\begin{align}
    \dot{\phi}(t,x) &= \pi (t,x)\, , & \dot{\phi}_\epsilon (t,x) &= \pi_\epsilon (t,x)\,,
\end{align}
such that
\begin{align}
    \pi (f) &= \phi(-\dot{f})\, , &  \pi_\epsilon (f) &= \phi_\epsilon(-\dot{f})\,.
\end{align}
Hence, we do not need to include the field momenta explicitly in the electric limit. In the magnetic rescaling, however, we have
\begin{equation}
    \dot{\tilde{\phi}}_\epsilon = \epsilon^2\,\tilde{\pi}_\epsilon \,.
\end{equation}
In the limit $\epsilon\to 0$, the time derivative of the field $\tilde{\phi}_\epsilon$ goes to zero, and the field momentum cannot be expressed in terms of the field. We therefore consider generators of the form
\begin{equation}
    \tilde{W}_\epsilon (F)=\tilde{W}_\epsilon \big((f_1,f_2)\big) = e^{i\big(\tilde{\phi}_\epsilon (f_1) + \tilde{\pi}_\epsilon (f_2) \big)}
\end{equation}
for pairs of test functions, $F=(f_1,f_2)\in D_2=\mathcal{C}_c(\mathbb{R}^{d+1})\times \mathcal{C}_c(\mathbb{R}^{d+1})$.
The commutator of the exponents then defines an antisymmetric bilinear form on $D_2$:
\begin{align}
    \tilde{\sigma}_\epsilon \big( F,G) \big)&=
    -i\big[ \tilde{\phi}_\epsilon(f_1) +\tilde{\pi}_\epsilon  (f_2),\tilde{\phi}_\epsilon(g_1)+\tilde{\pi}_\epsilon (g_2) \big] \\
    & = -i\big[ \tilde{\phi}_\epsilon (f_1 -\epsilon^{-2}\,\dot{f}_2) , \tilde{\phi}_\epsilon (g_1 - \epsilon^{-2}\dot{g}_2)  \big]\\
    &= -i\big[\phi_\epsilon ( \epsilon\,f_1 - \epsilon^{-1}\,\dot{f}_2) , \phi_\epsilon (\epsilon \,g_1 - \epsilon^{-1}\,\dot{g}_2) \big]\\
    &= \sigma_\epsilon \big(\epsilon\,f_1 - \epsilon^{-1}\,\dot{f}_2,\epsilon \,g_1 - \epsilon^{-1}\,\dot{g}_2\big)\,.
\end{align}
In the limit $\epsilon\to 0$, we obtain
\begingroup
\allowdisplaybreaks
\begin{align}
    \tilde{\sigma}_0 \big( F,G \big)&=-i\int \frac{d^dp}{(2\pi)^d}\bigg( i\big( \hat{f}_1(0,-p)\,\hat{g}_2 (0,p)-\hat{f}_2(0,p)\,\hat{g}_1(0,-p)\big)\nonumber\\
    &\qquad\qquad\quad +E_p^2 \Big(\hat{f}_2(0,-p)\,\partial_0 \hat{g}_2(0,p) - \partial_0 \hat{f}_2 (0,p)\,\hat{g}_2(0,-p)\Big)\bigg) \label{tildesigma} \\
    &=
    \int dt\,d^d x\,\int ds\,\bigg(f_1(t,x)\, g_2(s,x)-f_2(t,x)\,g_1(s,x)\nonumber\\
    &\qquad\qquad\quad + t(\Delta-m^2)f_2(t,x)\, g_2(s,x)-f_2(t,x)\,s\,(\Delta-m^2)g_2(s,x)\bigg)\,.
    \label{eq:mag-position}
\end{align}
\endgroup
We introduce the combinations
\begin{align}
    F^\phi(x) &= \int dt \, \big(f_1(t,x) + t(\Delta -m^2)f_2(t,x)\big) \,,\\
    F^\pi (x) &= \int dt\, f_2(t,x)\,.
\end{align}
This notation is motivated by the expression for the smeared fields when $\tilde{\phi}_0$ and $\tilde{\pi}_0$ are solutions of the equations of motion~\eqref{magneticEOM}:
\begin{align}
    \tilde{\phi}_0(f_1)+\tilde{\pi}_0(f_2) & = \int d^dx\, dt\, \big( f_1(t,x) \tilde{\phi}_0(t,x) + f_2(t,x)\tilde{\pi}_0(t,x) \big)\\
    &= \int d^dx\, dt\, \big( f_1(t,x) \tilde{\phi}_0(0,x) + f_2(t,x)\big(\tilde{\pi}_0(0,x)+t (\Delta-m^2)\tilde{\phi}_0(0,x) \big)\big)\\
    &= \int d^dx\, \big( F^\phi(x)\tilde{\phi}_0(0,x) + F^\pi(x)\tilde{\pi}_0(0,x) \big)\,.
\end{align}
The bilinear form $\tilde{\sigma}_0$ can then be expressed as
\begin{equation}\label{mag-sigma0-Fphi-Fpi}
    \tilde{\sigma}_0(F,G) = - \int d^dx\, \big( F^\pi(x) G^\phi(x)-F^\phi(x)G^\pi(x)\big)\,.
\end{equation}
The kernel of $\tilde{\sigma}_0$ is given by 
\begin{equation}
    \ker \tilde{\sigma}_0 = \big\{ F\in D_2\,|\ F^\phi=0 \ \text{and}\ F^\pi=0\big\}\,.
\end{equation}
$\tilde{\sigma}_0$ induces a symplectic form on the quotient $\tilde{V}_0=D_2/\ker \tilde{\sigma}_0$, and one can define the corresponding Weyl algebra.

Let us note that the massless electric and massless magnetic theories
are related by duality (at least when the spatial manifold is
compact)~\cite{Bekaert:2022oeh,Bekaert:2024itn}.

\subsection{Carroll symmetries}
\label{sec:carroll-symmetries-mag}

As in the electric case (see Section~\ref{sec:Carrollsymmetries}), we consider the Carroll transformations including supertranslations,
\begin{equation}
\label{Carrollmagnetic}
    \alpha(t,x)= (t+h(x),Rx+a)\,.
\end{equation}
They are represented on the space $D_2$ of test functions $F=(f_1,f_2)$ as
\begin{equation}
\label{rhoalphamagnetic}
     \rho_\alpha (F) = \big(f_1- \Delta h\,f_2 - \partial h\cdot \partial f_2 +\tfrac{1}{2}(\partial h)^2 \dot{f}_2, f_2\big)\circ \alpha^{-1}\,.
\end{equation}
We now show that the antisymmetric form $\tilde{\sigma}_0$ is invariant under these maps, which implies that the maps $\rho_\alpha$ induce automorphisms of the corresponding Weyl algebra.

For the rigid spatial transformations, $x\mapsto Rx+a$, the transformation of $f_1$ and $f_2$ is as in the electric case, and hence the combinations $F^\phi,F^\pi$ transform as
\begin{align}\label{mag:Fpi-Fphi-spatialtrans}
    \tilde{F}^\phi (Rx+a) &= F^\phi(x)\,,& \tilde{F}^\pi(Rx+a) &= F^\pi(x)\,.
\end{align}
The invariance of $\tilde{\sigma}_0$ is obvious from its expression~\eqref{mag-sigma0-Fphi-Fpi}.

For a supertranslation, $\alpha_h(t,x)=(t+h(x),x)$, the test functions transform to $(\tilde{f}_1,\tilde{f}_2)=\rho_{\alpha_h}(f_1,f_2)$. To evaluate $\tilde{\sigma}_0$, we need the transformation of the combinations $F^\phi,F^\pi$:
\begin{align}
    \tilde{F}^\phi (x)&= F^\phi(x) +\partial h(x)\cdot \partial F^\pi(x) + h(x)\,(\Delta -m^2)F^\pi(x)\,,\\
    \tilde{F}^\pi (x) &= F^\pi(x) \label{mag:transf-Fpi-supertrans}\,.
\end{align}
Using the expression~\eqref{mag-sigma0-Fphi-Fpi} for $\tilde{\sigma}_0$, we find
\begin{align}
    \tilde{\sigma}_0(\tilde{F},\tilde{G}) &= - \int d^dx\,\big( \tilde{F}^\pi \tilde{G}^\phi - \tilde{F}^\phi \tilde{G}^\pi \big)\\
    &= - \int d^dx\, \Big( F^\pi \big(G^\phi + \partial h\cdot \partial G^\pi + h\,(\Delta-m^2)G^\pi\big) \nonumber\\
    &\qquad \qquad \quad - \big(F^\phi + \partial h\cdot \partial F^\pi + h\,(\Delta-m^2)F^\pi\big) G^\pi
    \Big)\\
    &= -\int d^dx\, \big(F^\pi G^\phi  - F^\phi G^\pi + F^\pi \partial \cdot (h \partial G^\pi)- G^\pi \partial\cdot (h \partial F^\pi)\big)\\
    &= \tilde{\sigma}_0 (F,G)\,
\end{align}
after partial integration. 
We conclude that $\tilde{\sigma}_0$ is invariant under Carroll transformations of the form~\eqref{Carrollmagnetic}. The representation~\eqref{rhoalphamagnetic} on the function space $D_2$ via $\rho_\alpha$ descends to a representation on the quotient $\tilde{V}_0$ as symplectomorphisms, which induces a representation on the corresponding Weyl algebra as automorphisms acting on the generators as
\begin{equation}\label{automorphismmagnetic}
    \tilde{\pi}(\alpha): \tilde{W}_0\big(F \big)\mapsto \tilde{W}_0 \big( \rho_\alpha (F)\big)\,.
\end{equation}
Because of the appearance of the mass, we do not have invariance under scale transformations. There is, however, invariance under rescalings of the time variable only,
\begin{equation}\label{timedilatations}
    \alpha_{\lambda_t}(t,x) = (\lambda_t t,x)\, ,
\end{equation}
which we represent on $D_2$ as
\begin{equation}
    \tilde{F}=\rho_{\alpha_{\lambda_t}} (F)= \big(\lambda_t^{-1/2}f_1,\lambda_t^{-3/2}f_2\big)\circ \alpha_{\lambda_t}^{-1}\,.
\end{equation}
The combinations $F^\phi,F^\pi$ then transform as
\begin{align}\label{transfundertimedilatations}
    \tilde{F}^\phi &= \lambda_t^{1/2}F^\phi\,,& \tilde{F}^\pi &= \lambda_t^{-1/2}F^\pi\,,
\end{align}
and hence the form $\tilde{\sigma}_0$ is invariant,
\begin{equation}
    \tilde{\sigma}_0(\tilde{F},\tilde{G}) = \tilde{\sigma}_0(F,G)\,.
\end{equation}

\subsection{A Carroll-invariant state}
\label{sec:magnetic-vacuum}

Analogously to the discussion in Section~\ref{sec:electricstate}, we consider the behaviour of the two-point function under the magnetic rescaling,
\begin{align}
    \tilde{G}_\epsilon (F,G ) &= \langle 0| \big(\tilde{\phi}_\epsilon (f_1)+\tilde{\pi}_\epsilon (f_2)\big)\big(\tilde{\phi}_\epsilon (g_1)+\tilde{\pi}_\epsilon (g_2)\big)|0\rangle\\
    &= G_\epsilon \big( \epsilon f_1 -\epsilon^{-1}\dot{f}_2,\epsilon g_1 - \epsilon^{-1}\dot{g}_2 \big)\\
    &=\epsilon \int \frac{d^dp}{(2\pi)^d (2E_p)} \hat{f}_1(-\epsilon E_p,-p)\,\hat{g}_1(\epsilon E_p,p)\nonumber \\
    &\quad +\frac{i}{2} \int \frac{d^dp}{(2\pi)^d} \big( \hat{f}_1 (-\epsilon E_p,-p)\,\hat{g}_2(\epsilon E_p,p) - \hat{f}_2 (-\epsilon E_p,-p)\,\hat{g}_1(\epsilon E_p,p)\big)\nonumber \\
    &\quad + \int \frac{d^dp}{(2\pi)^d (2\epsilon E_p)}E_p^2 \,\hat{f}_2(-\epsilon E_p,-p) \,\hat{g}_2 (\epsilon E_p,p)\,.
\end{align}
The last term shows that the two-point function diverges when $\epsilon$ goes to zero. On the other hand, the expectation value
\begin{equation}
\tilde{\omega}_\epsilon \big( \tilde{W}_\epsilon ( F)\big)  = e^{-\frac{1}{2} \tilde{G}_\epsilon (F,F)}
\end{equation}
has a well-defined limit,
\begin{equation}\label{state-magnetic}
    \tilde{\omega}_0 \big( \tilde{W}_0 (F)\big) = \left\{ \begin{array}{cl}
         1 & \text{for}\ F^\pi = 0\\
0 & \text{for}\ F^\pi \not=0 \,.
    \end{array}\right.
\end{equation}
This looks similar to the state~\eqref{eq:state-electr-massless} that we identified for the massless electric scalar (in particular, it provides a well-defined pure state), but its interpretation is reversed: whereas the vacuum state in the massless electric limit gives rise to a state of sharp field momentum $\pi_0=0$, in the magnetic case the limit produces a state of sharp field $\tilde{\phi}_0=0$.

The state~\eqref{state-magnetic} is invariant under the Carroll transformations~\eqref{Carrollmagnetic} and time dilatations~\eqref{timedilatations} because due to~\eqref{mag:Fpi-Fphi-spatialtrans} and~\eqref{mag:transf-Fpi-supertrans} as well as~\eqref{transfundertimedilatations}, we have
\begin{equation}
    \tilde{F}^\pi  =0 \ \Longleftrightarrow \ F^\pi = 0 \, .
\end{equation}
Using the GNS construction, one can define a corresponding Hilbert space which is unitarily equivalent to
\begin{equation}
    \mathcal{H} = \ell^2 \big( \mathcal{C}_c(\mathbb{R}^d) \big)
\end{equation}
with basis $\{\delta_k\}_{k\in \mathcal{C}_c(\mathbb{R}^d)}$. The Weyl algebra is then represented as
\begin{equation}
    \Pi\big( \tilde{W}_0(F)\big) \delta_k = e^{-\frac{i}{2}\int d^dx\,F^\phi(x) (F^\pi(x)+2k(x))}\delta_{k+F^\pi}\,.
\end{equation}
Spatial transformations $\alpha_{R,a}(t,x)=(t,Rx+a)$ are represented as unitary operators
\begin{equation}
    U(\alpha_{R,a})\delta_k = \delta_{k\circ \alpha_{R,a}^{-1}}\,,
\end{equation}
and we observe that this representation is not strongly continuous. On the other hand, supertranslations $\alpha_h(t,x)=(t+h(x),x)$ are represented strongly continuously as unitary operators
\begin{equation}
    U(\alpha_h) \delta_k = e^{\frac{i}{2}\int d^dx\, h(x) \big((\partial k(x))^2 + m^2k(x)^2\big)} \delta_k\,.
\end{equation}
We also see that the spectrum of the temporal translation operator $P^0$ is nonnegative.

Similarly to the electric case, one can consider other states which are invariant under rotations and spatial translations, but not under all Carroll transformations, by considering the two-point functions
\begin{multline}
    \tilde{G}^{(\gamma,\delta)}(F,G) = \\
    \frac{1}{2}\int \frac{d^d p}{(2\pi)^d} \bigg(\big(\gamma(|p|)^{-2}+ \delta(|p|)^2\, \gamma(|p|)^2 \big) \, \widehat{F^\phi}(-p)\,\widehat{G^\phi}(p) 
    + \gamma(|p|)^2 \,\widehat{F^\pi}(-p)\,\widehat{G^\pi}(p) \\ 
    + \big(i+\delta(|p|)\, \gamma(|p|)^2\big) \,\widehat{F^\phi}(-p)\,\widehat{G^\pi}(p) + \big(-i+\delta(|p|)\, \gamma(|p|)^2\big) \,\widehat{F^\pi}(-p)\,\widehat{G^\phi}(p)\bigg)\,.
    \label{alternativestates_magnetic}
\end{multline}

\subsection{Thermal states}
\label{sec:thermal-states-mag}

One can obtain a thermal KMS state by considering -- similarly to the discussion in Section~\ref{sec:thermal-states} -- a thermal Gibbs state where the inverse temperature is scaled as $\beta \to \epsilon \beta$. As in the computation of the vacuum two-point function in Section~\ref{sec:magnetic-vacuum}, we can express the two-point function of the magnetically rescaled fields by the two-point function of the electrically rescaled fields as long as $\epsilon$ is finite,
\begin{equation}
    \tilde{G}_\epsilon^\beta (F,G) = G_\epsilon^\beta \big(\epsilon f_1 - \epsilon^{-1}\dot{f}_2,\epsilon g_1 - \epsilon^{-1} \dot{g}_2\big)\,.
\end{equation}
This expression diverges in general, but the expression
\begin{equation}
    \tilde{\omega}_\beta \big(\tilde{W}_\epsilon(F)\big) =e^{-\frac{1}{2}\tilde{G}_\epsilon^\beta(F,F)}
\end{equation}
has a well-defined limit as we show in the following. Using the expression~\eqref{thermal-two-point-function-electric} for $G_\epsilon^\beta$, we find
\begin{align}
    \tilde{G}_\epsilon^\beta(F,F) &= \frac{1}{\epsilon^2}\frac{1}{\beta} \int \frac{d^dp}{(2\pi)^d} \hat{f}_2(0,p)\hat{f}_2(0,-p) + \dots 
\end{align}
as the leading contribution, hence $\tilde{G}_\epsilon^\beta(F,F)$ diverges for $\hat{f}_2(0,p)\not=0$, or in other words for $F^\pi\not=0$. On the other hand, for $F^\pi=0$, we obtain a finite limit,
\begin{align}
    \tilde{G}_0^\beta(F,F) \Big|_{F^\pi=0}&= \frac{1}{\beta} \int \frac{d^dp}{(2\pi)^d} \frac{1}{E_p^2} \hat{F}^\phi(p)\hat{F}^\phi(-p) \,. 
\end{align}
Finally, the resulting state on the Weyl algebra is
\begin{equation}
    \tilde{\omega}_0^\beta \big( \tilde{W}_0(F)\big) = \left\{ \begin{array}{ll}
        e^{-\frac{1}{2\beta}\int \frac{d^dp}{(2\pi)^d}\,\frac{1}{E_p^2}\hat{F}^\phi(p)\hat{F}^\phi(-p)}&  \text{for}\ F^\pi=0\,,\\
         0 & \text{for}\ F^\pi\not=0  \,.
    \end{array}\right.
\end{equation}
Following similar arguments as in Section \ref{sec:thermal-states},
using $\tilde{\tau}_t(\tilde{W}_0(F)):=\tilde{W}_0(F_t)$ with
$F_t:=(f_{1,t},f_{2,t})$, these states indeed satisfy the KMS
condition.

\section{Flat space holography}
\label{sec:flat-space-hologr}

In this section, we draw several lessons from our results for flat
space holography. We analyze the conformal Carrollian two-point
function for operators identified with boundary values of bulk
fields~\cite{Donnay:2022wvx}. This correlator shares features with
both the massless electric field studied here and massless scalar
fields in $1+1$ dimensions. Restricting attention to the free theory,
we discuss subtle issues that arise in quantization and in the
definition of the vacuum. 

We also relate these algebraic quantum field theory–motivated
constructions to standard notions of infrared (IR) physics, such as
the news, memory, soft charges, and their associated superselection
sectors.

Finally, we show that the vacuum state is invariant under the
conformal Carroll group, highlight additional symmetries, and discuss
the corresponding unitary operators and their continuity properties.

\subsection{Quasifree state for Carroll holography}
\label{sec:quasifree-state}

One general lesson for conformal Carrollian theories and consequently
for flat space holography is that their states and Hilbert spaces may
be nonregular, nonunique, or at least subtle to define. This
observation is consistent with recent discussions of the intricacies
surrounding the Hilbert space and partition functions of conformal
Carrollian theories;
see~\cite{deBoer:2023fnj,Cotler:2024cia,Aggarwal:2025hji} and
references therein.

We will now focus on the conformal Carrollian two-point function for
operators which are identified with boundary values of scalar bulk
fields~\cite{Donnay:2022wvx}
\begin{align}
&\braket{\Omega|  \phi(u_{1},z_{1},\bar z_{1})\phi(u_{2},z_{2},\bar z_{2})|\Omega} \nonumber \\
&\qquad  \overset{\beta\to 0}{\sim}\frac{1}{2\pi}\left[\frac{1}{\beta}-\left(\gamma+\log{|u_1-u_2|}+\frac{i\pi}{2}\mathrm{sign}(u_1-u_2)\right)\right]\delta^{(2)}(z_1-z_2)+\mathcal{O}(\beta)
\label{eq:2pt-ccft}
\end{align}
where $\gamma$ is the Euler--Mascheroni constant. The parameter $\beta>0$ is the infrared regulator of~\cite{Donnay:2022wvx}: the Carrollian correlator arises from a frequency integral that diverges logarithmically at the soft end $\omega\to0$, and is defined by analytic continuation in the exponent of $\omega$ (see~\cite[Sec.~2.4.4]{Donnay:2022wvx}).  It is therefore a dimensionless exponent rather than an energy or time scale, and in particular unrelated to the inverse temperature $\beta$ of Section~\ref{sec:thermal-states}. As stressed in~\cite{Donnay:2022wvx}, the pole in $1/\beta$ should be
retained, since it is required for boost invariance.

\begin{figure}
  \centering
  \includegraphics[width=0.3 \textwidth]{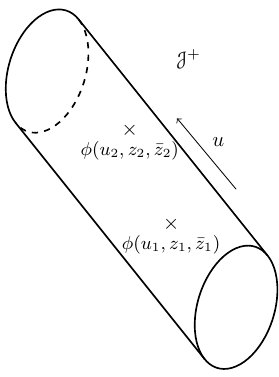}
  \caption{Operators on $\mathcal{J}^{+}=\mathbb{R}\times S^{2}$,
    which appear in the conformal Carrollian two-point function
    \eqref{eq:2pt-ccft} and are identified with boundary values of a
    bulk scalar field.}
    \label{fig:scri}
\end{figure}

In the equation above, $u\in\mathbb{R}$ denotes retarded time, and the
complex coordinates $z\in\mathbb{C}$ should be understood as
parameterizing the celestial Riemann sphere $S^2$ (see
Figure~\ref{fig:scri}). We see that similarly to the massless limit of
the electric scalar, there is a divergent term when $\beta\to
0$. Additionally, there is a logarithmic term reminiscent of massless
scalar fields in $1+1$ dimensions. Without additional input, this
two-point function is not a sensible correlation function. We now show
how to construct a well-defined vacuum state and its associated
Hilbert space.

In terms of smeared fields
\begin{equation}
    \phi(f) = \int du\,d^2z\, f(u,z,\bar{z})\,\phi(u,z,\bar{z})\,,
\end{equation}
the two point function becomes 
\begin{equation}\label{eq:2pt-from-flat-holography}
    \braket{\Omega|  \phi(f)\phi(g)|\Omega} = \mu_\beta (f,g) +\frac{i}{2} \sigma_\beta(f,g)\,,
\end{equation}
where the symmetric and the antisymmetric part of the two-point function are
\begin{align}
    \mu_\beta(f,g)
    &= \frac{1}{2\pi} \left[  \bigg(\frac{1}{\beta} -\gamma\right)\int d^2z\, f_0(z,\bar{z})\,g_0 (z,\bar{z}) \nonumber  \\
     & \qquad \qquad- \int du_1\,du_2\,d^2z\, \log |u_1-u_2|\,f(u_1,z,\bar{z})\,g(u_2,z,\bar{z})\bigg] + \mathcal{O}(\beta)\,, \label{eq:mu-flatholo-beta}\\
     \sigma_\beta(f,g) &= \sigma(f,g) + \mathcal{O}(\beta)\,,\qquad  
\end{align}
with 
\begin{equation}
    \sigma(f,g)  =-\frac{1}{2} \int du_1\,du_2\,d^2z\, \mathrm{sign}(u_1-u_2)\,f(u_1,z,\bar{z})\,g(u_2,z,\bar{z}) \,.
     \label{eq:sigma-flatholo}
\end{equation}
As further discussed in Section~\ref{sec:states-aqft}, for free
theories the symmetric part $\mu_{\beta}$ encodes all information
about quasifree states, while the antisymmetric part $\sigma_\beta$ captures
the dynamical content of the theory and is state-independent. Since the $\beta \to 0$ limit of $\sigma_{\beta}$ is well-defined, the IR divergence is a property of the particular vacuum state.  One
observes that the divergence is governed by the integral $f_0$ of the
test function over $u$,
\begin{equation}\label{f0-integral}
    f_0(z,\bar{z}) = \int du\,f(u,z,\bar{z})\,.
\end{equation}
If $f_0=0$, then there exists a test function $F$ with compact support such that $f=\partial_u F$, defined by
\begin{equation}
F(u,z,\bar{z}) = \int_{-\infty}^u du' \,f(u',z,\bar{z})\,.
\end{equation}
The field corresponding to a test function with $f_0=0$, 
\begin{equation}
    \phi(f) = \phi(\partial_u F) = -\int du\,d^2z\,\partial_u\phi(u,z,\bar{z})\,F(u,z,\bar{z})\,,
\end{equation}
therefore only depends on the derivative $\partial_u\phi$ of the
field, and not on a possible zero mode, more precisely on a zero-mode
function $\phi_{0}(z,\bar z)$ on the celestial sphere. In this sense,
$f_0$ measures the contribution of the zero mode to $\phi(f)$.

The limit $\beta\to 0$ of the two-point function does not lead to a
well-defined correlation function of the fields unless $f_0=0$ or
$g_0=0$.\footnote{\label{fn:restrict} One option is to restrict the test functions
    accordingly, which effectively amounts to restricting to
    derivatives of the field, $\pd_{u}\phi$. However, this leads to a
    Fock space construction that omits the zero-mode sector.}  On the
other hand, similarly to our discussion of the Carrollian scalars, one
might still be able to define expectation values of exponentiated
fields which we formally write as
\begin{equation}
    W(f) = e^{i\phi(f)}\,.
\end{equation}
These generate a Weyl algebra $\mathcal{W}(V,\sigma)$ with underlying
vector space $V=\mathcal{C}_c(\mathbb{R}\times S^2)$ and
symplectic form $\sigma$ (see the discussion
before~\eqref{Weyl-fields} in
Section~\ref{sec:mass-electr-theory}). We assume in the following that
we are in the quasifree situation (see
Section~\ref{sec:quasifree-states}), where the two-point function
determines all higher correlation functions, such that the expectation
value of the exponentiated operators (in the $\beta$-regulated vacuum
denoted by a subscript $\beta$) is given by the exponentiated
two-point function,
\begin{equation}
    \omega_\beta (W(f)) = \braket{\Omega|e^{i\phi(f)}|\Omega}_\beta = e^{-\frac{1}{2} \braket{\Omega| \phi(f)\phi(f)|\Omega}_\beta} = e^{-\frac{1}{2}\mu_\beta(f,f)}\,.\label{eq:omega_beta_evalv}
\end{equation}
The limit $\beta\to 0$ of these expectation values exists and yields the
functional
\begin{equation}
    \omega(W(f)):=\lim_{\beta\xrightarrow{}0}e^{-\frac{1}{2}\mu_\beta(f,f)}=\left\{ \begin{array}{cl}
    e^{-\frac{1}{2}\tilde{\mu}(f,f)} & \text{for}\ f_0 = 0\\
    0 & \text{for}\ f_0 \not=0 \,,
    \end{array}\right.\label{eq:state_evalv}
\end{equation}
with 
\begin{equation}\label{mu-flatholo}
\tilde{\mu}(f_1,f_2) =    -\frac{1}{2\pi}\int du_1\,du_2\,d^2z\,\log{|u_1-u_2|}\,f_1(u_1,z,\bar{z})\,f_2(u_2,z,\bar{z}) \,.
\end{equation}
This functional defines a state on the Weyl algebra: linearity and normalization are immediate, and positivity
follows from the explicit representation $\Pi_\ell$ constructed
below in~\eqref{eq:Pi-ell}, which satisfies
$\omega(A)=\braket{\Omega|\Pi_\ell(A)|\Omega}$ for all
$A\in\mathcal{W}(V,\sigma)$ and hence
$\omega(A^*A)=\big\|\Pi_\ell(A)\ket{\Omega}\big\|^2\ge0$
(see~\eqref{eq:state_exp_val} and
Appendix~\ref{sec:vac-rep-flatholo}). Note that the conclusion that $\omega$ defines a state does not rely on any assumption that the regulated functionals $\omega_\beta$ define states.

The state~\eqref{eq:state_evalv} is nonregular, in the sense that $\omega(W(\lambda\,f))$ is
not continuous in the parameter $\lambda$ when $f_0\not=0$. It can be
interpreted as the state where the momentum conjugate to the zero mode
is strictly zero, and hence the zero mode itself is completely
undetermined.

It is instructive to compare this with the subtraction prescription of Section~\ref{sec:classical-state}.  The divergent term in $\mu_\beta$, see~\eqref{eq:mu-flatholo-beta}, is proportional to $f_0\,g_0$ and hence absent on $V_0=\{f\in V:f_0=0\}$: there is nothing to subtract on $V_0$, the limit $\beta\to0$ exists by itself and leaves the nontrivial positive covariance $\tilde\mu$.  Restricting to $V_0$ -- as anticipated in footnote~\ref{fn:restrict} -- therefore does define a regular quasifree state, that of the radiative modes, on the Weyl subalgebra $\mathcal{W}(V_0,\sigma|_{V_0})$ -- in contrast to the massless electric case, where the remaining covariance vanishes identically and no such subspace exists. The restriction does not determine a state on the full algebra $\mathcal{W}(V,\sigma)$: a state on $\mathcal{W}(V_0,\sigma|_{V_0})$ admits many inequivalent extensions, differing in their treatment of the zero-mode sector.  The state~\eqref{eq:state_evalv} is one such extension, namely the one in which the momentum conjugate to the zero mode vanishes, and it is obtained without any subtraction, by working with exponentiated fields and removing the regulator.

One can obtain a realization of this state on a Hilbert space. First,
one uses the standard Fock space construction (see
Section~\ref{sec:quasifree-states}) for the part where $f_0=0$: based
on $V_0 = \{f\in \mathcal{C}_c(\mathbb{R}\times S^2)|\,f_0=0\}$, one
constructs a one-particle Hilbert space $\mathcal{H}_0$ with scalar
product (see Appendix~\ref{sec:positivity-non-zero-mode} for details)
\begin{equation}\label{scalarproduct-flatholo}
    \langle f,g\rangle_{\mathcal{H}_0}=\tilde{\mu}(f,g)+\frac{i}{2}\sigma(f,g)\,.
\end{equation}
Then one defines creation and annihilation operators on the Fock space $\mathcal{F}(\mathcal{H}_0)$ satisfying
\begin{equation}
  \left[a(f),a^\dagger(g)\right]=\langle f,g\rangle_{\mathcal{H}_0}\,.
\end{equation}
For the zero-mode part, we introduce the nonseparable Hilbert space
$\ell^2(\mathcal{C}(S^2))$ with the uncountable orthonormal
basis $\delta_h$, $h\in \mathcal{C}(S^2)$, i.e.,
$h(z,\bar z)$ is a smooth function on the Riemann sphere
$S^2$ and
\begin{align}
  \langle\delta_h,\delta_{h'}\rangle  = \delta_{h,h'} =
  \begin{cases}
    1 & \mathrm{if}\quad h'=h \\
    0 & \mathrm{otherwise}\, .
  \end{cases}
\end{align}
A generic element of this Hilbert space is then given by
$\Psi=\sum_{i=1}^{\infty} c_{i}\delta_{h_{i}}$ with
$\sum_{i}|c_{i}|^{2}< \infty$, so it is a superposition of different
sharply defined zero-mode profiles.

We now choose a test function $\ell\in \mathcal{C}_c(\mathbb{R})$ such
that its integral $\int du\,\ell(u)=1$. This allows us to define a
projection from the space
$V=\mathcal{C}_c(\mathbb{R}\times S^2)$ of test functions to the
space $V_0$ of test functions with vanishing $u$-integral:
\begin{equation}
    \big( P_\ell f\big) (u,z,\bar{z}) = f(u,z,\bar{z}) - \ell(u)\,f_0(z,\bar{z})\,.
\end{equation}
For ease of notation, we introduce (the overline denotes closure)
\begin{equation}
\phi_{\mathrm{Fock}}(f):=\overline{a(f)+a^\dagger(f)}\label{eq:phi_0}
\end{equation}
for $f\in V_0$. It follows that $\phi_{\mathrm{Fock}}(f)$ satisfies
\begin{equation}
\left[\phi_{\mathrm{Fock}}(f),\phi_{\mathrm{Fock}}(g)\right]=2i\,\text{Im}\langle f,g\rangle_{\mathcal{H}_0}=i\,\sigma(f,g)\,,
\end{equation}
and
\begin{equation}
    e^{i\phi_{\mathrm{Fock}}(f)}e^{i\phi_{\mathrm{Fock}}(g)}= e^{i\phi_{\mathrm{Fock}}(f+g)-\frac{i}{2}\sigma(f,g)}\,.
\end{equation}
Then we define a representation of the Weyl algebra on the Hilbert space
\begin{equation}
  \label{eq:hilbert-fac}
    \mathcal{H}=\mathcal{F}(\mathcal{H}_0)\otimes \ell^2(\mathcal{C}(S^2))
\end{equation}
in terms of operators
\begin{equation}\label{eq:Pi-ell}
\Pi_\ell(W(f))(v\otimes \delta_h):=e^{i\left(\phi_{\mathrm{Fock}}(P_\ell f) -\sigma(f,\ell h)-\frac{1}{2}\sigma(f,\ell f_0)\right)}v\otimes \delta_{f_0+h}\,.
\end{equation}
One can explicitly check that
$\Pi_\ell\big(W(f)W(g)\big)=\Pi_\ell(W(f))\,\Pi_\ell(W(g))$ holds (see
Appendix~\ref{sec:vac-rep-flatholo}).  The construction depends on the
choice of the function $\ell$, but another choice $\ell'$ leads to a
unitarily equivalent representation $\Pi_{\ell'}$ (see
Appendix~\ref{sec:vac-rep-flatholo}). The state $\omega$ can be
interpreted as expectation values in the vacuum defined by the tensor
product
\begin{equation}
    \ket\Omega = |0\rangle \otimes\delta_0\label{eq:flat_Omega}
\end{equation}
of the Fock vacuum $|0\rangle$ in $\mathcal{F}(\mathcal{H}_0)$ and the
basis vector $\delta_{0}$ corresponding to the zero configuration on
the celestial sphere. Indeed, one can verify that the expectation
value reproduces the state~\eqref{eq:state_evalv},
\begin{align}
    \langle\Omega|\Pi_\ell(W(f))|\Omega\rangle=&\langle 0|e^{i\left(\phi_{\mathrm{Fock}}(P_\ell f)-\frac{1}{2}\sigma(f,\ell f_0)\right)}|0\rangle\langle\delta_0,\delta_{f_0}\rangle\\
    =&e^{-\frac{1}{2}\tilde{\mu}(P_\ell f,P_\ell f)}\,\delta_{0,f_0}=\omega(W(f))\,.\label{eq:state_exp_val}
\end{align}
It follows, that $\Pi_\ell$ is unitarily equivalent to the GNS
representation of $\omega$ (see Appendix~\ref{sec:vac-rep-flatholo}).

The zero-mode factor $\ell^2(\mathcal{C}(S^2))$
in~\eqref{eq:hilbert-fac} is nonseparable, and $\mathcal{H}$ decomposes into a direct
sum of mutually orthogonal zero-mode sectors labelled by the profile
$h(z,\bar z)$. For the full Weyl algebra the labels are not superselected, since by \eqref{eq:Pi-ell} the generators $W(f)$ with $f_0\neq0$ shift $h\mapsto h+f_0$. If instead one restricts to the radiative algebra, they are: radiative observables do not change them. In many
physical applications one may therefore choose a sector and work
within it. We keep the full GNS representation because it
displays the zero-mode algebra and the operators that shift between
different zero-mode profiles explicitly.

The nonseparable zero-mode sector labelled by functions
$h\in \mathcal{C}(S^2)$ can be probed by a continuum of charges,
\begin{equation}
  \label{eq:soft-exp}
    U_\epsilon (v\otimes \delta_h) = e^{i\int d^2z\, \epsilon(z,\bar{z})\,h(z,\bar{z})} (v\otimes \delta_h)\,,
\end{equation}
where $\epsilon\in\mathcal{C}(S^2)$ is any test function.\footnote{As
  $\epsilon$ is smeared with the test function $h$, one could in
  principle generalize this to distributions $\epsilon$, but for
  simplicity we take $\epsilon \in \mathcal{C}(S^2)$.}  They act on
the fields as
\begin{equation}\label{charge-on-fields}
    U_\epsilon\, \Pi_\ell(W(f))\,U_\epsilon^{-1} = e^{i\int d^2z\, \epsilon(z,\bar{z})\,f_0(z,\bar{z})}\,\Pi_\ell(W(f))\,,
\end{equation}  
and can be understood
as charges arising from the derivatives of the field, formally
written as $U_\epsilon=e^{iQ_\epsilon}$ with 
\begin{equation}
  \label{eq:soft-charge}
    Q_\epsilon = \int du\,d^2z\, \epsilon(z,\bar{z}) \, \partial_u \phi(u,z,\bar{z}) \,.
\end{equation}
More precisely, one chooses a test function $\chi$ on $\mathbb{R}$
with compact support such that $\chi(u)=1$ in a neighborhood of
$u=0$. Then $\chi_\delta(u)=\chi(\delta u)$ converges pointwise to the constant function with value $1$ for $\delta\to 0$. Furthermore, we introduce its logarithmically average
\begin{equation}\label{defofXdelta}
X_\delta(u) := \frac{1}{\log\delta^{-1}}\int_0^{\log \delta^{-1}} ds\,\chi_\delta(e^{-s}u) \,,
\end{equation}
which is again a compactly supported test function equal to 1 in an interval around $u=0$ that scales with $1/\delta$. Then $Q_\epsilon$ formally can be approximated by the field $\phi(-X_\delta'\,\epsilon)$ smeared with the test function $-X_\delta'\,\epsilon$. Indeed, one can show (see Appendix~\ref{sec:chargeoperator}) that
\begin{equation}\label{Ueps-as-limit}
    \lim_{\delta\to 0} \,\Pi_\ell(W(-X_\delta'\,\epsilon)) = U_\epsilon
\end{equation}
in the strong operator topology (see~\cite{Bahns:2017mrt} for a similar discussion for the massless two-dimensional scalar).\footnote{The squared norm of $-X'_\delta \epsilon$ in $\mathcal{H}_0$ is of the order $\mathcal{O}(1/\log(1/\delta))$ (see~Appendix~\ref{sec:chargeoperator}).}

In summary, the vacuum of the conformal Carrollian theory relevant for 
flat space holography is subtle to define due to the zero-mode divergence. 
In the quasifree setting considered here, however, the expectation values 
of exponentiated fields admit a well-defined limit, yielding a nonregular 
state on the Weyl algebra. Its Hilbert-space representation factorizes 
into a standard Fock sector for the nonzero modes and a nonseparable 
zero-mode sector, reflecting the role of infrared degrees of freedom 
in the holographic setting.

The situation is reminiscent of the two-dimensional massless
scalar~\cite{Acerbi:1993pp,Acerbi:1993yu,Derezinski:2004,Bahns:2017mrt}. In
fact, the $u$-part of the two-point function has the same form as a
chiral massless scalar. In two dimensions, the massive two-point
function diverges when the mass goes to zero, and to obtain a
well-defined expression one can exclude the zero mode (in the above
case, this would mean to restrict to fields $\phi(f)$ with $f_0=0$),
or one can consider exponentiated fields for which the expectation
value has a well-defined limit. Instead, one can also look for other
states than the vacuum, by dropping for example the requirement that
the state be translation invariant. 

In our case -- similarly to~\cite{Derezinski:2004,Bahns:2017mrt} --
one can construct other representations, including representations on
separable Hilbert spaces, as follows. We first introduce a zero mode
$q(z,\bar{z})$ on the sphere, together with a conjugate momentum
$p(z,\bar{z})$ with commutator
\begin{equation}\label{CCRzeromode}
    [q(z_1,\bar{z}_1),p(z_2,\bar{z}_2)] = i\,\delta^{(2)} (z_1-z_2)\,.
\end{equation}
Now, given any representation of $q$ and $p$ on a Hilbert space\footnote{Or, more generally, a representation of the Weyl algebra corresponding to taking exponentiated and smeared operators $e^{i(q(h_q)+p(h_p))}$.} $\mathcal{K}$, we can build a representation $\Pi_{\ell,\mathcal{K}}$ of $\mathcal{W}(V,\sigma)$ on $\mathcal{F}(\mathcal{H}_0)\otimes \mathcal{K}$ by
\begin{equation}
  \label{eq:generic}
    \Pi_{\ell,\mathcal{K}}\big(W(f)\big) = e^{i\phi_{\mathrm{Fock}}(P_\ell f)} \otimes e^{i\,\big(q(f_0)-\sigma (f, \ell p)\big)}\,,
\end{equation}
where $q(f_0)=\int d^2z \,q(z,\bar{z})\,f_0(z,\bar{z})$, and $\sigma(f,\ell p)$ means the operator one obtains when smearing $p(z,\bar{z})$ with $-\frac{1}{2}\int du_1\,du_2\,\mathrm{sign}\,(u_1-u_2)\,f(u_1,z,\bar{z})\,\ell(u_2)$. The zero modes $q$ and $p$ can be written as
\begin{align}\label{qfromfield}
    1\otimes e^{iq(h)} &= \Pi_{\ell,\mathcal{K}}(W(\ell h))\,,\\
    1\otimes e^{ip(\epsilon)} &= \lim_{\delta\to 0} \Pi_{\ell,\mathcal{K}}(W(-X_\delta' \epsilon)) \,,
\end{align}
so $p(\epsilon)$ plays the role of the charge $Q_\epsilon$ (see~\eqref{Ueps-as-limit}).

The vacuum state we are considering above is a special case of this construction with $e^{i(q(h_q)+p(h_p))}$ represented on the nonseparable Hilbert space $\ell^2({\mathcal{C}(S^2}))$ as
\begin{equation}
  \label{eq:vac-rest}
    e^{i(q(h_q)+p(h_p))} \delta_h = e^{\frac{i}{2}\int d^2z\,(h_q(z,\bar{z})+2\,h(z,\bar{z}))\,h_p(z,\bar{z})}\delta_{h+h_q}\,.
\end{equation}
For a regular representation~\eqref{eq:generic} on a separable Hilbert
space, for which not only the exponentiated fields but also $q$ and
$p$ can be defined as operators, there is no distinguished vacuum
vector. Whereas $q(f_0)$ is invariant under a translation of the test
function $f$ in $u$, $\sigma (f, \ell p)$ changes because $\ell$ is
not invariant under shifts in $u$. For vacuum expectation values to be
invariant under translations in $u$, $p$ has to be represented as
$p=0$ on the vacuum vector, which would be in conflict with the
canonical commutation
relations~\eqref{CCRzeromode}. For~\eqref{eq:vac-rest}, we have such a
state, because $e^{i\,p(h_p)}\delta_0= \delta_0$, but the price we pay
is that the Hilbert space is nonseparable. This dichotomy is analogous
to the case of the massless scalar in $1+1$ dimensions, as, e.g.,
reviewed in~\cite{Bahns:2017mrt}.

\subsection{Relation to infrared physics}
\label{sec:relat-infr-phys}

The nonregular quasifree state constructed in Section~\ref{sec:quasifree-state}
captures the infrared (IR) structure expected of massless fields at
null infinity. In particular, the separation between radiative
(finite-frequency) and soft (zero-frequency) modes, the emergence of
superselection sectors, and the associated soft Ward identities arise
naturally within this framework. While these aspects of infrared
structure have been extensively explored in the literature, to our
knowledge there is no derivation directly based on the Carrollian
two-point function. We therefore do not claim originality in this
section; rather, our aim is to place our construction in the context
of standard terminology and established results (see,
e.g.,~\cite{Strominger:2017zoo,Campiglia:2017dpg}).

To set the stage, we introduce standard notions of infrared (IR)
physics adapted to the scalar case. The scalar news
$N(u,z,\bar z)=\partial_u\phi(u,z,\bar z)$ encodes the energy flux
through null infinity. The associated memory
$M(z,\bar z)=\int_{-\infty}^{+\infty}du \,N(u,z,\bar
z)=\phi_{+}(z,\bar z)-\phi_{-}(z,\bar z)$ can be interpreted as the
permanent shift of the boundary field between early and late retarded
times, where we introduced the (formal) limits
$\phi_\pm(z,\bar z)=\lim_{u\to\pm\infty}\phi(u,z,\bar z)$. There is
some freedom in choosing the canonical conjugate of the memory
(Goldstone mode), which can be encoded by a compactly supported
profile $\ell(u)$ with $\int du\,\ell(u)=1$,
\begin{equation}
    K_\ell (z,\bar{z}) = \int_{-\infty}^{\infty} du\, \ell(u)\,\phi(u,z,\bar{z})\,.
\end{equation}
Using the commutator
\begin{equation}
  \label{eq:phi-comm}
[\phi(u,z,\bar{z}),\phi(u',z',\bar{z}')]=-\frac{i}{2}\mathrm{sign}(u-u')\,\delta^{(2)}(z-z') \, ,
\end{equation}
one obtains
\begin{equation}
  \label{commutator-phi-M}
    [\phi(u,z,\bar{z}),M(z',\bar{z}')] = i \,\delta^{(2)}(z-z')\,,
\end{equation}
and therefore
\begin{align}
    [K_\ell(z,\bar{z}),M(z',\bar{z}')] &= \int du \,\ell(u) [\phi(u,z,\bar{z}),M(z',\bar{z}')]  \\
    &= \int du\, \ell(u) \,i\,\delta^{(2)}(z-z')\\
    &=i\,\delta^{(2)}(z-z')\,.
    \label{KMcommutator}
\end{align}
For the decomposition of the field, we define the radiative part with
respect to the choice of $\ell$ as
\begin{equation}
  \label{def-of-phirad}
    \phi_\ell^\mathrm{rad} (u,z,\bar{z}) = \frac{1}{2}\int_{-\infty}^\infty du'\, \mathrm{sign}(u-u') \, \partial_{u'}\phi(u',z,\bar{z}) - \frac{1}{2} s_\ell(u) M (z,\bar{z})\,.
\end{equation}
Here, 
\begin{equation}
    s_\ell(u) = \int_{-\infty}^{\infty} du'\, \ell(u')\,\mathrm{sign}(u-u') = 2 \int_{-\infty}^u du'\,\ell(u') -1
\end{equation}
smoothly interpolates between $s_\ell (-\infty)=-1$ and $s_\ell (+\infty)=+1$, and satisfies
\begin{equation}
    \int_{-\infty}^{\infty} du\,s_\ell(u)\,\ell(u) = 0\,.
\end{equation}
The radiative part only depends on the news $N=\partial_u \phi$ and (formally) vanishes at $u\to \pm\infty$. It commutes with the memory $M$ (because the memory commutes with the news due to~\eqref{commutator-phi-M}), as well as with the Goldstone mode $K_\ell$:
\begin{align}
    &[K_\ell(z,\bar{z}),\phi_\ell^\mathrm{rad}(u,z',\bar{z}')] \nonumber\\
    &\quad= \frac{1}{2}\int du' \int du''\,\ell(u')\, \mathrm{sign}(u-u'')\,[\phi(u',z,\bar{z}),\partial_{u''}\phi(u'',z',\bar{z}')] - \frac{i}{2} s_\ell(u) \,\delta^{(2)}(z-z')  \\
    &\quad= \frac{1}{2}\int du'  \int du''\,\ell(u')\, \mathrm{sign}(u-u'')\,i\,\delta(u'-u'')  \,\delta^{(2)}(z-z') - \frac{i}{2} s_\ell(u) \,\delta^{(2)}(z-z') \\
    &\quad= 0\,.
\end{align}
The commutator of $\phi^\mathrm{rad}_\ell$ with itself is formally given by
\begin{align}
    &[\phi^\mathrm{rad}_\ell(u,z,\bar{z}),\phi^\mathrm{rad}_\ell(u',z',\bar{z}')] \nonumber \\
    &\quad = \frac{1}{4} \int du_1 \int du_2\,\mathrm{sign}(u-u_1)\,\mathrm{sign}(u'-u_2)\,[\partial_{u_1}\phi(u_1,z,\bar{z}),\partial_{u_2}\phi(u_2,z',\bar{z}')]\\
    &\quad = \frac{1}{4} \int du_1 \int du_2\,\mathrm{sign}(u-u_1)\,\mathrm{sign}(u'-u_2)\,i\,\partial_{u_1}\delta(u_1-u_2)\,\delta^{(2)}(z-z') \\
    &\quad = -\frac{i}{2} \,\mathrm{sign}(u-u')\,\delta^{(2)}(z-z')\, ,
\end{align}
in agreement with~\eqref{eq:phi-comm}.  Notice that the above
computation is rather a formal manipulation, and the resulting
expression should be understood as a distribution defined on test
functions of vanishing $u$-integral.\footnote{Adding a constant to the
  $\mathrm{sign}$ term in the integrand produces a term proportional
  to $M$ that can be compensated by a change of the coefficient in
  front of the second term. Such a constant would formally alter the
  computation of the commutator, but this ambiguity goes away when
  $\phi^\mathrm{rad}_\ell$ is integrated against a test function whose
  $u$-integral vanishes.}

The field can then be decomposed as
\begin{equation}\label{decompositionflatholo}
    \phi (u,z,\bar{z}) = K_\ell(z,\bar{z}) + \frac{1}{2} s_\ell(u) M(z,\bar{z}) + \phi^{\mathrm{rad}}_\ell (u,z,\bar{z}) - \int_{-\infty}^{\infty} du'\, \ell(u')\,\phi_{\ell}^{\mathrm{rad}} (u',z,\bar{z})\,. 
  \end{equation}
  The decomposition can be compared
  to~\eqref{eq:generic}. Smearing~\eqref{decompositionflatholo} with a
  test function $f$, we obtain
\begin{align}
  \int du\,d^2z\, f(u,z,\bar{z}) \,\phi(u,z,\bar{z}) &= \int d^2z\, K_\ell(z,\bar{z})\,f_0(z,\bar{z}) \nonumber \\
    &\quad +\frac{1}{2} \int du\,du'\,d^2z\,\mathrm{sign}(u-u') \,f(u,z,\bar{z})\,\ell(u')\,M(z,\bar{z}) \nonumber \\
    &\quad + \int du\,d^2z\, \phi^\mathrm{rad}_\ell (u,z,\bar{z})\big( f(u,z,\bar{z})-\ell(u)f_0(z,\bar{z})\big)\,.
\end{align}
Notice that $\phi^\mathrm{rad}_\ell$ is smeared with the combination
$P_{\ell}f=f-\ell f_0$ whose $u$-integral vanishes. We can make contact
to~\eqref{eq:generic} when we represent $K_\ell(z,\bar{z})$ by
$q(z,\bar{z})$, $M(z,\bar{z})$ by $p(z,\bar{z})$, and
$\phi^\mathrm{rad}_\ell$ by $\phi_\mathrm{Fock}$. This means that we
represent the radiative part on a usual Fock space, and tensor it by
any representation of the zero mode part generated by $M$ and
$K_\ell$. Although the concrete representation depends on the choice
of $\ell$, different choices of $\ell$ lead to unitarily equivalent
representations.

\begin{figure}
    \centering
    \includegraphics[scale=0.3]{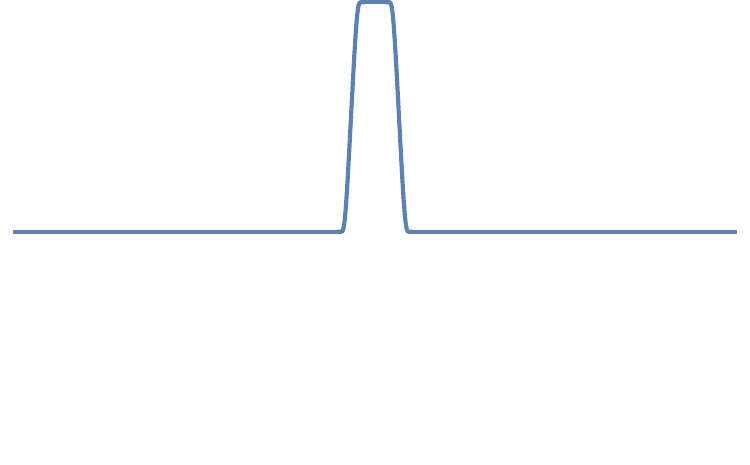} \hspace*{1cm}
    \includegraphics[scale=0.3]{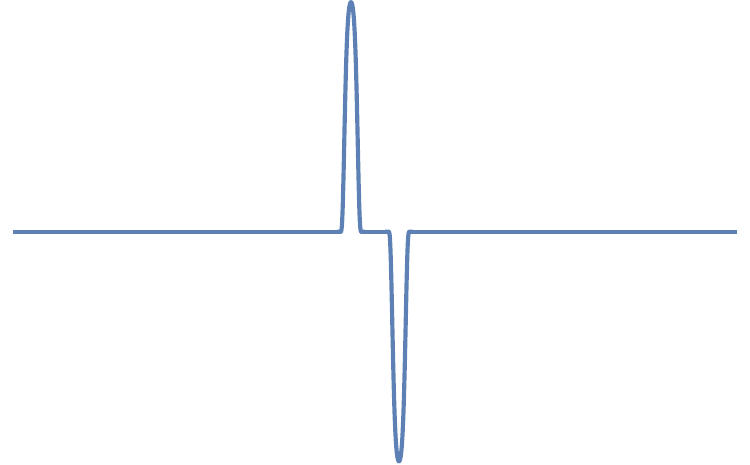}  \\\includegraphics[scale=0.3]{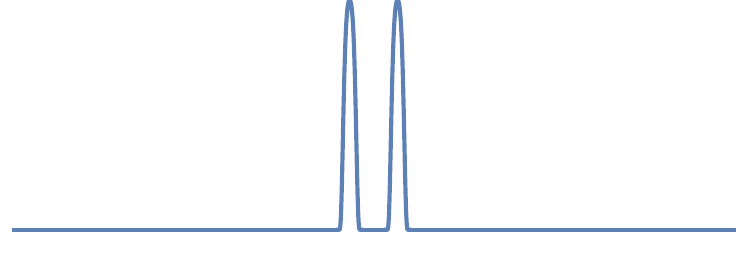}
    \includegraphics[scale=0.3]{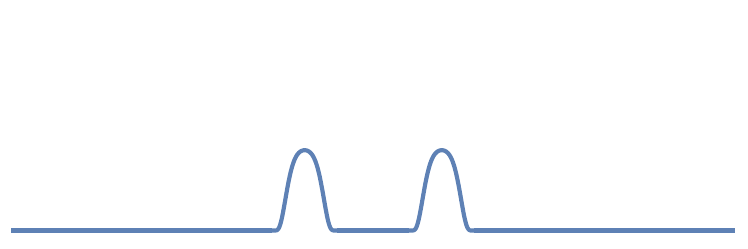}
    \includegraphics[scale=0.3]{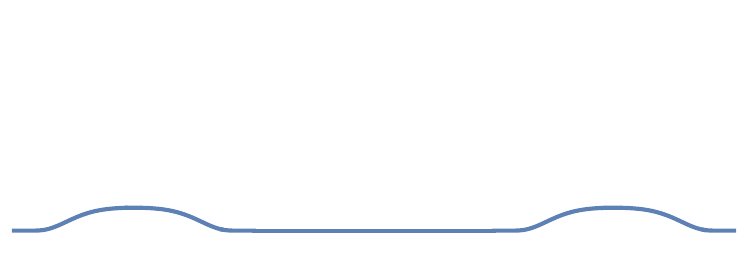}  
    \caption{The upper images show a typical compactly supported test
      function $\chi$ which has value $1$ in a neighborhood of $u=0$,
      as well as its derivative $\chi'$. The profiles
      $\ell_\delta = -\frac{1}{2}\chi_\delta'\,\mathrm{sign}$ are
      displayed in the lower line for decreasing values of $\delta$:
      the profile has two peaks of weight $\frac{1}{2}$ which are
      shifted towards larger values of $|u|$ and become flatter for
      smaller values of $\delta$.}
    \label{fig:ell-delta}
\end{figure}

There is no distinguished choice of the profile $\ell$, it can be any
compactly supported function normalized to unit integral. To proceed
further and to make contact to some expressions in the literature
(see, e.g.,
\cite{Chen:2024kuq,Bekaert:2024uuy,Araujo-Regado:2024dpr,AndradeeSilva:2026mpa}
and references therein) we now consider a specific example of profile
functions. Concretely, we set
$\ell_\delta (u)=-\frac{1}{2} \chi_\delta'(u)\,\mathrm{sign}(u)$
(where $\chi_\delta$ was introduced before~\eqref{Ueps-as-limit}) and
analyze the behaviour for $\delta\to 0$. These describe profiles with
two peaks that become flatter and centered around increasing values of
$|u|$ when $\delta$ goes to $0$ (see~Figure~\ref{fig:ell-delta}).
Then
\begin{equation}
  \label{Goldstone-limiting-value}
    K_{\ell_\delta} \xrightarrow{\delta\to 0} \frac{1}{2}(\phi_+ + \phi_-)\,.
\end{equation}
The interpolating function $s_{l_\delta}$ does not have a good limit; one obtains
\begin{equation}
    s_{\ell_\delta}(u) = \mathrm{sign}(u)\big( 1-\chi(\delta\,u)\big)\,.
\end{equation}
This is a function that goes to zero pointwise for $\delta\to 0$, but
for any $\delta$ satisfies
$s_{\ell_\delta}\xrightarrow{u\to \pm \infty}\pm 1$. The decomposition
then reads
\begin{equation}
    \phi(u) \sim \frac{1}{2} (\phi_+ + \phi_-) +\frac{1}{2} \mathrm{sign}(u)\big( 1 - \chi(\delta \,u)\big) M + \phi_{\ell_\delta}^\mathrm{rad}(u)\,.
\end{equation}
The choice to consider the sequence $\ell_\delta$ above is mainly
motivated by the resulting simple
expression~\eqref{Goldstone-limiting-value} for the Goldstone
mode. Notice, however, that this limiting expression
is not unproblematic when considering commutators: for any finite
$\delta$, $K_{\ell_\delta}$ and $M$ satisfy the canonical commutation
relations~\eqref{KMcommutator}, whereas naively (assuming $[\phi_{\pm},\phi_{\pm}]=0$) the limiting
expression leads to
\begin{equation}
    \Big[\frac{1}{2}(\phi_++\phi_-)(z,\bar{z}),(\phi_+-\phi_-)(z',\bar{z}')\Big] \overset{?}{=} \frac{i}{2} \delta^{(2)}(z-z')
\end{equation}
which is off by a factor of $2$. The reason is that the $\phi_+$ appearing in the limiting expression for $K_{\ell_\delta}$ should still be thought of as being at smaller $u$-values than the $\phi_+$ appearing in $M$ (which means that the naive commutator $[\phi_+,\phi_+]$ cannot be treated as zero) -- therefore considering $\phi_\pm$ as proper limiting values is problematic. In the following, we rather want to stay with a generic compactly supported profile function $\ell$ where these problems are absent. 

To continue the comparison with the previous subsection, we consider
the soft charge~\eqref{eq:soft-charge}. It can be expressed as
\begin{align}
Q_\epsilon=\int d^2z \,\epsilon(z,\bar z) \, M(z,\bar z) \, .
\end{align}
Since we are considering a free theory, there is no corresponding hard
contribution to the charge, such as one arising from a current to
which the scalar field would couple. The soft charge generates
asymptotic shift symmetries
\begin{align}
  \label{eq:large-gauge}
  \delta_\epsilon \phi(u,z,\bar z) = i[Q_\epsilon,\phi(u,z,\bar z)]= \epsilon(z,\bar z)
\end{align}
which act as shifts of the zero mode and realize the infinitesimal
version of~\eqref{charge-on-fields}. This
angle-dependent shift should not be confused with conformal Carroll or
BMS supertranslations, which act as spacetime transformations
$u\mapsto u+b(z,\bar z)$. The transformation generated by $Q_\epsilon$
acts at fixed retarded time and shifts the field by a function on the
celestial sphere. In the present scalar toy model it is therefore best
regarded as an internal shift symmetry associated
with the soft sector, rather than as a Carrollian supertranslation.

The news smeared with test functions $F\in V$ generates the radiative part of the Weyl algebra $e^{iN(F)}=e^{i\phi(-\partial_u F)}$ that captures the finite-energy dynamics. Labeling the elements by $F\in V$ instead of by $f=-\partial_uF\in V_0$, the symplectic form takes a more standard form
\begin{align}
  \sigma^\mathrm{rad}(F,G) &= \sigma (-\partial_uF,-\partial_u G) \\
                           &= \frac{1}{2} \int du\,d^2z\,(F(u,z,\bar{z})\,\partial_u G(u,z,\bar{z}) - \partial_u F(u,z,\bar{z})\,G(u,z,\bar{z}))\,.
\end{align}
Since the charge commutes with the news
\begin{align}
  \label{eq:Q-central}
 [Q_{\epsilon},N(u,z,\bar z)]=0  \, ,
\end{align}
it is central in the radiative algebra.  So it is supported at null
infinity, commutes with the radiative algebra, and produces distinct
boundary vacua, hence~\eqref{eq:large-gauge} may be called a large
``gauge'' transformation. The name should be understood by analogy only, since the scalar
carries no gauge symmetry and $\epsilon$ parametrizes an asymptotic
shift symmetry rather than a residual gauge parameter. The whole algebra of observables is the
radiative algebra together with the soft algebra which is generated by
$U_{\epsilon}=e^{i Q_{\epsilon}}$ and $T_{h}=e^{i K_{\ell}(h)}$.

We also saw that the GNS representation of the vacuum $\omega$ factorizes into a radiative
part and a soft part, cf.,~\eqref{eq:hilbert-fac}. The memory $M(z,\bar z)$ acts trivially on the radiative part and is diagonal with respect to the standard basis, 
$M(z,\bar z) (v \otimes \delta_{m})=m(z,\bar z)(v \otimes
\delta_{m})$. The fact that the soft charge
$Q_{\epsilon}$ commutes with the radiative observables leads to
superselection sectors labeled by the eigenvalue profiles
$m(z,\bar z)$ of $M(z,\bar z)$ that radiative operators cannot change.

For S-matrix elements $\braket{\mathrm{out}|S|\mathrm{in}}$ and charge
$Q_{\epsilon,\mathrm{full}}^{\pm} = Q_{\epsilon}^{\pm} +
Q_{\epsilon,\mathrm{hard}}^{\pm}$, where the $\pm$ refers to the future and
past boundary, charge conservation is given by
\begin{align}
  \braket{\mathrm{out}|Q_{\epsilon,\mathrm{full}}^{+}S-SQ_{\epsilon,\mathrm{full}}^{-}|\mathrm{in}}=0 \,.
\end{align}
In a basis diagonalizing $M(z,\bar z)$, this leads to the soft Ward
identity (equivalent to the leading soft theorem)
\begin{align}
  \Big(\int d^2z \,\epsilon(z,\bar z) [m_{\mathrm{out}}(z,\bar z)-m_{\mathrm{in}}(z,\bar z)]\Big)
  \braket{\mathrm{out}|S|\mathrm{in}}
  =-\langle \mathrm{out}|\big(Q_{\epsilon,\mathrm{hard}}^{+}S - SQ_{\epsilon,\mathrm{hard}}^{-}\big)|\mathrm{in}\rangle \, .
\end{align}
Since this has to be true for any $\epsilon(z,\bar z)$ one obtains a
local version of this statement. In general this leads schematically
to a angle-by-angle balance law
\begin{align}
  m_{\mathrm{out}}(z,\bar z)-m_{\mathrm{in}}(z,\bar z) = -\int_{-\infty}^{+\infty}du \,J_{u}(u,z,\bar z) \, 
\end{align}
with a hard flux $J_{u}$. For the scalar case at hand, there is no hard
flux, so we find $ m_{\mathrm{out}}(z,\bar z)=m_{\mathrm{in}}(z,\bar z)$,
i.e., scattering is block-diagonal in the soft label.

In summary, the findings in Section~\ref{sec:quasifree-state}, align
with expectations for a (scalar toy) model for IR physics.  One could
have solely focused on the $u$-derivatives of the potentials $\phi$,
but this would miss the soft sector, including in principle measurable
memory effects.

\subsection{Conformal Carroll symmetries of the state}
\label{sec:conf-carr-symm}

The state $\omega$ defined in~\eqref{eq:state_evalv} is invariant under the extended conformal Carroll group which contains arbitrary supertranslations. To show this, we first represent a symmetry transformation $\alpha$ on the space of test functions,
\begin{equation}
    f\mapsto \rho_\alpha(f)\,.
\end{equation}
This induces a representation as automorphisms of the Weyl algebra,
\begin{equation}
    \pi(\alpha) \big(W(f)\big) = W\big(\rho_\alpha(f)\big)
\end{equation}
provided that the symplectic form is invariant,
\begin{equation}
    \sigma\big(\rho_\alpha(f),\rho_\alpha(g)\big) = \sigma (f,g)\,.
\end{equation}
The invariance of the state $\omega$ can then be formulated as
\begin{equation}
    \omega\big(W(\rho_\alpha(f))\big) = \omega(W(f))\, .
\end{equation}
We first consider the supertranslations that only affect the $u$-coordinate, 
\begin{equation}
    \alpha_b(u,z,\bar{z}) =(u+b(z,\bar{z}),z,\bar{z})\,,
\end{equation}
which include translations in $u$ (constant $b$), Carroll boosts (linear $b$), and the temporal part of the Carroll special conformal transformations (quadratic $b$). Under such a transformation, the symplectic form $\sigma$ in~\eqref{eq:sigma-flatholo} is invariant, 
and hence a representation
\begin{equation}
    \rho_{\alpha_b}(f)=f\circ \alpha_b^{-1}
\end{equation}
induces a representation as automorphisms of the Weyl algebra. Because $\big(\rho_{\alpha_b}(f)\big)_0=f_0$, and $\tilde{\mu}$ is invariant, we also confirm that the state $\omega$ is invariant. 
This symmetry is realized as a unitary operator $U_b$ on the Hilbert space $\mathcal{H}$. Let $f$ be a test function with $f_0=0$, and let $h$ be a test function on $S^2$. Denote the transformed function $f_b=\rho_{\alpha_b}(f)$, and similarly
\begin{equation}
    \ell_b (u)= \ell (u-b(z,\bar{z}))\,.
\end{equation}
Then the unitary operator $U_b$ acts as
\begin{equation}
    U_b \,e^{i\phi_{\mathrm{Fock}}(f)}\ket{0}\otimes \delta_h = e^{\frac{i}{2}\sigma(\ell_b h,\ell h)}\,e^{i\phi_{\mathrm{Fock}}((\ell_b-\ell)h)} \,e^{i\phi_{\mathrm{Fock}}(f_b)} \ket{0}\otimes \delta_h\, ,
\end{equation}
and the representation is strongly continuous.

Next we show invariance under spatial translations and rotations, dilatations and the spatial part of the Carrollian special conformal transformations. These can be summarized in the M{\"o}bius transformations
\begin{equation}
    \alpha_{A} (u,z,\bar{z}) = \bigg(\frac{1}{|cz+d|^2}u,\frac{az+b}{cz+d},\frac{\bar{a}\bar{z}+\bar{b}}{\bar{c}\bar{z}+\bar{d}}\bigg) = :(u',z',\bar{z}')\,,
\end{equation}
where
\begin{equation}
    A=\begin{pmatrix}
        a & b \\ c & d
    \end{pmatrix}
\end{equation}
is an $SL(2,\mathbb{C})$ matrix.
On test functions, we represent it as
\begin{equation}
    \rho_{\alpha_{A}}(f) (u',z',\bar{z}') =  |cz+d|^{4}\,f(u,z,\bar{z})\,.
\end{equation}
Then $\sigma$ is invariant because of the standard transformation of the measure coming from $\frac{dz'}{dz}=\frac{1}{(cz+d)^2}$, and the invariance of the sign function under (positive) rescaling. We further observe that $f_0=0 \Leftrightarrow \big(\rho_{\alpha_A}(f)\big)_0=0$. Also, $\tilde{\mu}$ is invariant because the shift in the integrand coming from 
\begin{equation}
   \log \bigg|\frac{1}{|cz+d|^2} (u_1-u_2)\bigg|= \log|u_1-u_2| - 2 \log |cz+d| 
\end{equation}
evaluates to zero since the test functions inserted into $\tilde{\mu}$ satisfy $f_0=0$. This shows that the state $\omega$ is invariant under M{\"o}bius transformations. The symmetries are realized as unitary operators $U_A$ on the Hilbert space $\mathcal{H}$. For a test function $f$ with $f_0=0$ and a test function $h$ on $S^2$, we introduce the notation
\begin{align}
    f_A(u',z',\bar{z}') &= |cz+d|^4 \,f(u,z,\bar{z})\,, & 
    h_A(z',\bar{z}') &= |cz+d|^2 \,h(z,\bar{z})\,.
\end{align}
Then for $f_0=0$ we have $(f_A)_0=0$, and $\big((\ell h)_A\big)_0 = h_A$.
The action of the unitary operator $U_A$ can then be written as
\begin{equation}
    U_A \,e^{i\phi_{\mathrm{Fock}}(f)} \ket{0}\otimes \delta_h = e^{-\frac{i}{2}\sigma((\ell h)_A,\ell h_A)}\,e^{i\phi_{\mathrm{Fock}}((\ell h)_A - \ell h_A)}\,e^{i\phi_{\mathrm{Fock}}(f_A)} \ket{0}\otimes \delta_{h_A}\,.
\end{equation}
Because of the nontrivial action in the nonseparable factor
($\delta_{h_A}$ generically depends on $A$), this representation is
not strongly continuous. Consequently, the
corresponding infinitesimal Lorentz or conformal Carroll generators
need not exist as self-adjoint operators in this representation. The
statement that $\omega$ is conformal Carroll-invariant should
therefore be understood at the level of the global group action by
automorphisms of the Weyl algebra, together with its unitary
implementation in the GNS representation leaving the vacuum vector~\eqref{eq:flat_Omega} invariant, rather than as the existence
of a strongly continuous representation of the full Lie algebra.

The M{\"o}bius transformations are the (spatial) conformal
transformations that are globally well-defined on $S^2$. Similarly,
one expects -- locally -- invariance under more general conformal
transformations (superrotations) which are however not well defined
globally.

There are more symmetries present in the theory. On the one hand, the invariance under M{\"o}bius transformations can be generalized to arbitrary spatial diffeomorphisms, replacing $|cz+d|^{-4}$ by the Jacobian $\big|\frac{\partial(z',\bar{z}')}{\partial(z,\bar{z})}\big|$. On the other hand, we have invariance under purely temporal generalized dilatations of the form $u\mapsto \lambda(z,\bar{z})u$ for a positive function $\lambda$ if the functions transform as
\begin{equation}
    f_\lambda(u',z,\bar{z}) = \lambda(z,\bar{z})^{-1}\,f(u,z,\bar{z})\,.
\end{equation}

\section{Discussion}
\label{sec:discussion}

We have analyzed free (conformal) Carrollian quantum field theories
and their implications for flat space holography from an algebraic
perspective, focusing in particular on Weyl algebras, states, and
their Hilbert space representations.

The massive electric Carrollian theory serves as the prototypical
example, admitting a well-defined Carroll limit both at the level of
the Weyl algebra and of the associated states. It possesses regular
ground and thermal states, the latter satisfying the KMS condition,
which is interpreted as placing the system in thermal equilibrium in
infinite volume.

The Carroll limit of the massless electric and magnetic theories is
more subtle.  To highlight the subtleties, let us review the limiting
procedure for the massless electric scalar from a different
viewpoint. For finite $\epsilon>0$, we can rewrite the smeared
Carrollian field in terms of the smeared Lorentzian field
(see~\eqref{eq:phieps} and~\eqref{smearedfields}),
\begin{equation}
    \phi_\epsilon(f) = \phi(f^{\epsilon})\,,
\end{equation}
where 
\begin{equation}
    f^{\epsilon}(t,x) = \epsilon^{-\frac{3}{2}}f(\epsilon^{-1} t,x)\,.
\end{equation}
In this sense, we can identify elements $W_\epsilon(f)$ of the Weyl
algebra of the rescaled fields with elements $W(f^\epsilon)$ of the
original Weyl algebra of the Lorentzian free scalar
($W_\epsilon(f)\mapsto W(f^\epsilon)$ defines an isomorphism). We then
obtain, for $\epsilon>0$, a one-parameter family of generating sets of
the Weyl algebra, such that the structure constants of the algebra
with respect to these sets -- encoded in $\sigma_\epsilon(f,g)$ --
have a well-defined limit for $\epsilon\to 0$. This is very similar to
the In{\"o}n{\"u}--Wigner contraction of Lie algebras, and the
resulting algebra -- defined by the limiting values of the structure
constants -- can be viewed as a type of contraction of the Weyl
algebra.\footnote{Note that even if the algebraic relations behave
  well in the limit, the $C^*$-properties are not necessarily under
  control. From the mathematical viewpoint, there are other notions of
  limiting procedures that are more appropriate in the $C^*$-context
  like asymptotic (homo-)morphisms \cite{Connes:1990} or
  (semi-)continuous fields of $C^*$-algebras (see, e.g.,
  \cite{Dixmier:1982}).} From this point of view, it is natural to
investigate whether the vacuum state of the scalar on Minkowski space
induces a well-defined state of the contracted algebra, by considering
the limiting behavior of
\begin{equation}
    \omega(W(f^\epsilon))=e^{-\frac{1}{2}G(f^\epsilon,f^\epsilon)} = e^{-\frac{1}{2}G_\epsilon(f,f)}\,.
\end{equation}
Indeed, as we have demonstrated, this limit is finite and describes a
well-defined state. In such a way, one obtains for the massless
electric scalar, as well as analogously for the magnetic scalar, 
well-defined vacuum (ground) states that minimize the energy and are
invariant under Carroll symmetries, yet these states fail to satisfy
regularity.  Whereas the obtained vacuum state for the massless
electric scalar can be interpreted as a state of sharp field momentum,
the corresponding vacuum state for the magnetic scalar is interpreted
as a state of sharp field values.  Conversely, we show that there are
no pure regular spacetime translation invariant quasifree states, therefore
no distinguished regular vacuum state exists. This is the standard
situation encountered in theories with zero modes or lacking
time-translation symmetry~\cite{Witten:2021jzq}.

A complementary route to analyze the Carrollian theories, in
particular their vacuum states, is to perform the quantization in the
functional Schrödinger representation (see,
e.g.,~\cite{Hatfield:2019sox}). For the electric case, this leads to
the functional Schrödinger equation
\begin{align}
  i \frac{\pd}{\pd t}\Psi[\phi,t] &= H \Psi[\phi,t]\,, &    H &= \frac{1}{2}\int d^{d}x 
  \left[
  -\frac{\vd^{2}}{\vd \phi(x)^{2}} +m^{2}\phi(x)^{2}
  \right] \, ,
\end{align}
whose ground-state solution takes the form
\begin{align}
  \Psi[\phi,t]=\mathcal{N} e^{-\frac{m}{2} \int d^{d}x \phi(x)^{2} - i Et} \, ,
\end{align}
where $\mathcal{N}$ is a normalization constant. This wave functional
must then still be regularized with respect to the natural inner
product
$\braket{\Psi_{2}|\Psi_{1}} = \int \prod_{x} d\phi(x)
\Psi_{2}^{*}[\phi]\Psi_{1}[\phi]$. Again we see that the $m \to 0$
limit is subtle.

We also discussed the implications of our results for flat space
holography. We showed that, even within the quasifree framework, the
definition of a vacuum state in the Carrollian setting is subtle. In
particular, guided by the singular vacuum, we constructed a
well-defined, albeit nonregular, state whose Hilbert space factorizes
into a standard Fock space and a nonseparable Hilbert space associated
with the zero modes, in agreement with expectations from infrared
physics. We further demonstrate that this state is invariant under
conformal Carroll symmetry.

This work opens various interesting areas for further exploration:
\begin{description}
\item[Algebraic (conformal) Carrollian QFT] A central
  open question left for future work is what a Carrollian quantum
  field theory is intrinsically, independent of its relativistic
  origin or limiting procedures. In a forthcoming
  work~\cite{Fredenhagen:2026xxx}, we develop a first intrinsic
  formulation within the framework of algebraic quantum field theory
  by adapting the Haag--Kastler axioms to Carrollian spacetimes,
  guided by their ultralocal causal structure. We further show that
  the Reeh--Schlieder property fails in this setting leading to a
  weakened entanglement structure, in agreement with expectations for
  putative flat space holographic duals.  These results provide a
  first step toward a systematic intrinsic understanding of Carrollian
  quantum field theory and its role in flat space holography.
  
\item[Curved space] One advantage of the algebraic approach to quantum
  field theory is that it can be naturally extended to general
  spacetime backgrounds, where the traditional Fock-space construction
  may be unavailable or ambiguous (see, e.g.,~\cite{Wald:1995yp}). The
  present work may therefore serve as a useful starting point for the
  analysis of (conformal) Carrollian quantum field theories on curved
  geometries, which are relevant for flat space holography from a
  Carrollian perspective. A natural question concerns the behaviour of
  the Carroll limit of Hadamard states, or of the Hadamard
  wavefront-set condition, which could be investigated using the
  methods developed here.

\item[Interacting theories] The tools employed in this work are not
  necessarily restricted to free scalar fields. It could be
  interesting to investigate higher spin and interacting theories from
  this perspective (see~\cite{Cotler:2024xhb} for some potential
  subtleties in the interacting case). In interacting theories where
  the dynamics plays a central role, the Weyl algebra alone may not
  capture the full structure of the limit. It would therefore be natural
  to investigate whether other operator-algebraic frameworks such as scaling
  algebras~\cite{Buchholz:1995gr} or resolvent algebras~\cite{Buchholz:2007gf}
  provide a suitable setting for extending the present contraction picture 
  beyond the free case. Roughly speaking, the scaling-algebra framework encodes
  renormalization-group limits by considering families of observables
  $(A_\epsilon)_{\epsilon\in (0,1]}$ localized in appropriately rescaled regions, and 
  extracting limit theories from the behavior of states on this larger algebra. 
  In this sense, the rescaled Weyl generators $W(f^\epsilon)$ appearing in our
  discussion may be viewed as simple examples of such scale-dependent observables. 
  It would be interesting to understand whether an anisotropic scaling adapted 
  to the Carroll limit can be formulated in this language.
  A related direction concerns Carrollian contractions of
  infinite-dimensional symmetry algebras, such as $W$-algebras, which
  appear in interacting theories~\cite{Fredenhagen:2025aqd}. Although the
  algebraic framework differs from the $C^*$-algebraic setting adopted
  here, both approaches are based on anisotropic scaling limits that
  induce a contraction of the underlying operator algebra. It would be
  interesting to clarify whether the contraction mechanism described
  here admits an analogue for more general interacting Carrollian
  theories.

\item[Flat space holography] It would be interesting to extend the
  analysis to higher-spin fields, in particular spins $1$ and $2$, to
  incorporate hard charges, and to go beyond the quasifree framework. Another natural direction is the study of more general
  states, such as KMS states. It may also be worthwhile to explore
  potential connections to black holes, especially in relation to
  their soft hair.
  
\item[Celestial holography] In this work, we have focused on states of
  Carrollian quantum field theory. Given its close connection to the
  celestial framework, it would be interesting to extend the present
  discussion to include a corresponding analysis in that context as
  well.

\item[von Neumann algebras] In this work we discussed states on unital
  $C^{*}$ algebras, which via the GNS construction gives rise to a
  Hilbert space and a natural von Neumann algebra. It might be
  interesting to determine their type and understand which role they
  and the nonregular states we discussed could play in recent advances
  of our understanding of holography (see \cite{Liu:2025krl} for a
  review).

  In particular, in strictly ultralocal theories spatial points
  decouple dynamically, so each point can be regarded as an
  independent quantum mechanical system. For the vacuum, the
  corresponding local operator algebra at a fixed point is therefore a
  type I von Neumann factor,
  $A_{x} \simeq \mathcal{B}(\mathcal{H}_{x})$. For a finite region,
  the full algebra is given by a tensor product of such factors
  $\mathcal{A}(\mathcal{O})=\bigotimes_{x\in \mathcal{O}}A_{x}$, and
  the global theory may be viewed (formally, in the continuum limit)
  as an infinite tensor product of quantum mechanical systems. In this
  sense, ultralocal field theories are structurally closer to ordinary
  quantum mechanics than to relativistic quantum field theories with
  type III local algebras. The absence of spatial coupling seems to
  lead to weakened ultraviolet entanglement
  structure~\cite{Bagchi:2014iea} (see our discussion above).  It may
  therefore be possible that a similar algebraic reduction occurs in a
  conformal Carrollian sector of flat space holography.
  
\item[$BMS_d$ invariant theories] It could be interesting to analyze,
  analogously to Section~\ref{sec:magnetic}, conformal Carrollian theories
  that derive as geometric actions of $BMS_{3}$ or
  $BMS_{4}$~\cite{Barnich:2017jgw,Merbis:2019wgk,Cotler:2024cia,Barnich:2022bni}
  from an algebraic perspective as well as contrast our work with
  \cite{Dappiaggi:2005ci} (see for further details
  \cite{Dappiaggi:2017kka}) which constructs a $BMS$ invariant state
  from an algebraic but complementary perspective. 
\end{description}

\section*{Acknowledgements}
\label{sec:acknowledgement}

We want to thank Klaus Fredenhagen, Daniel Grumiller, Florian Ecker, Alfredo Perez, Jakob Salzer and Tobias Sutter for
useful discussions.

\appendix

\section{Positivity}
\label{sec:positivity}

For completeness we show in Appendix~\ref{app:positivity_quasi_free}
that positivity of the state implies the
inequality~\eqref{eq:ineq_mu_sigma}. In
Appendix~\ref{app:positivity_thermal}, we furthermore present details
on the positivity of the massless electric thermal state.

\subsection{Conditions for positivity of quasifree states}
\label{app:positivity_quasi_free}

In this subsection we expand upon the positivity of quasifree states already mentioned in Section \ref{sec:quasifree-states}, i.e., states on a Weyl-algebra $\mathcal{W}(V,\sigma)$ of the form
\begin{align}
  \omega(W(f)) = e^{-\frac{1}{2}q(f)} \, ,
\end{align}
where $q:V\to[0,\infty]$ is a covariance as in
Section~\ref{sec:quasifree-states}, $V_q=\{f\in V:\,q(f)<\infty\}$, and
$\mu$ is the associated bilinear form on $V_q$. In the regular case
$V_q=V$, \cite{Kay:1988mu} showed that such a state is positive if and
only if $\mu$ satisfies
\begin{align}
    \mu(f,f)\mu(g,g) \geq \frac{1}{4} |\sigma(f,g) |^{2}  \, . \label{eq:ineq_mu_sigma}
\end{align}
Sufficiency follows there from the Fock space representation of
$\omega$; the converse direction is only sketched in
\cite{Kay:1988mu}. For completeness, and to make the argument
accessible to readers less familiar with the underlying constructions,
we provide a more detailed derivation in this appendix. It applies
without change to the nonregular case, since all elements entering it
lie in $V_q$, and yields~\eqref{eq:ineq_mu_sigma} for all $f,g\in V_q$.

As in \cite{Kay:1988mu} we consider the element 
\begin{equation}
    A:=A(f,g):= (W(f)-\mathbf{1})+i(W(g)-\mathbf{1})
\end{equation}
for arbitrary $f,g\in V_q$. A small computation shows
\begin{align}
    A^*A &= \big( (W(-f)-\mathbf{1})-i(W(-g)-\mathbf{1})\big) \big( (W(f)-\mathbf{1})+i(W(g)-\mathbf{1})\big)\\
    &=4-W(f)-W(-f)-W(g)-W(-g) 
    -i\left[e^{-i\sigma(f,g)/2}\,W(f-g)-W(-g)-W(f)\right. \nonumber\\  &\quad
    \left.-e^{i\sigma(f,g)/2}\,W(g-f)+W(-f)+W(g)\right]\,.
\end{align}
Evaluating $\omega$ at $A^*A$ results in
\begin{align}
    \omega(A^*A)
    &= 4-2e^{-\frac{1}{2}\mu(f,f)}-2e^{-\frac{1}{2}\mu(g,g)}-2\sin\left(\sigma(f,g)/2\right)\,e^{-\frac{1}{2}\mu(f-g,f-g)}\,.
\end{align}
Positivity of $\omega$ demands that this expression is positive for arbitrary $f,g\in V_q$ -- in particular one concludes that the smooth map
\begin{align}
    \lambda\mapsto\omega &(A(\lambda f,\lambda g)^*A(\lambda f, \lambda g))\\
    &= 4-2e^{-\frac{1}{2}\lambda^2\mu(f,f)}-2e^{-\frac{1}{2}\lambda^2\mu(g,g)}-2\sin\left(\lambda^2\sigma(f,g)/2\right)\,e^{-\frac{1}{2}\lambda^2\mu(f-g,f-g)}\,,
\end{align}
is nonnegative for any $\lambda \in \RR$. Expanding the above up to second order in powers of $\lambda$, we find
\begin{equation}
    \omega (A(\lambda f,\lambda g)^*A(\lambda f, \lambda g)) = \big(\mu(f,f)+\mu(g,g)-\sigma(f,g)\big)\lambda^2 + \mathcal{O}(\lambda^3)\,,
\end{equation}
and hence
\begin{equation}
    \mu(f,f)+\mu(g,g)\geq \sigma(f,g)
\end{equation}
for all $f,g\in V_q$. By replacing $g$ with $-g$ we can refine this to 
\begin{equation}
    \mu(f,f)+\mu(g,g)\geq |\sigma(f,g)|\,.\label{eq:intermediat_ineq}
\end{equation}
Now consider two cases:
First, assume that one of the functions, say $f=f_0$, satisfies
$\mu(f_0,f_0)=0$. Then \eqref{eq:intermediat_ineq} reduces to
\begin{equation}
    \mu(g,g)\geq |\sigma(f_0,g)|\,,
\end{equation}
which -- because the right hand side is linear under (positive)
scaling of $f_0$ -- can only hold if $\sigma(f_0,g)=0$.  Therefore,
\eqref{eq:ineq_mu_sigma} holds in that case since both sides vanish.
Next, assume that both functions give a nonzero value when inserted
into the quadratic form $\mu$. Because~\eqref{eq:ineq_mu_sigma} is
quadratic in $f$ and $g$ on both sides, it is enough to prove it for
$f=f_1$ and $g=g_1$ such that $\mu(f_1,f_1)=\mu(g_1,g_1)=1$. For such
elements \eqref{eq:intermediat_ineq} further collapses to
\begin{equation}
    2\geq |\sigma(f_1,g_1)|\,,
\end{equation}
which implies~\eqref{eq:ineq_mu_sigma} for $\mu(f_1,f_1)=\mu(g_1,g_1)=1$.

Finally, the converse direction extends to the nonregular case as
well. Since $q(f-g)=\infty$ whenever $f-g\notin V_q$, only pairs $f,g$
lying in the same coset of $V_q$ contribute to $\omega(C^*C)$ for
$C=\sum_f c_f\,W(f)$, so that the sum decomposes over $V/V_q$. Writing
$f=c+a$ and $g=c+b$ with $a,b\in V_q$, for $c$ a representative of a
given coset, bilinearity gives
$\sigma(c+a,c+b)=\sigma(c,b)-\sigma(c,a)+\sigma(a,b)$, so that the
phases involving $c$ can be absorbed into the coefficients
$\tilde{c}_a:=c_{c+a}\,e^{-\frac{i}{2}\sigma(c,a)}$ and the
contribution of the coset reduces to that of the regular quasifree
state with covariance $\mu$ on $\mathcal{W}(V_q,\sigma|_{V_q})$, which
is nonnegative by~\cite{Kay:1988mu}.
Appendix~\ref{app:positivity_thermal} carries this out explicitly in a
particular case.

\subsection{Positivity of the massless electric thermal state}
\label{app:positivity_thermal}

We show here that the functional~\eqref{masslessKMSstate} proposed as a thermal state for the massless electric scalar is positive, and hence -- as linearity and normalization are obvious -- defines a state. We need to check that for every $C=\sum_{f\in V_0}c_f W_0(f)$, with $c_f\in \CC$ and $c_f\neq 0$ for only finitely many $f$, we have
\begin{equation}
    \omega_0^\beta \big( C^*C\big)
    =\sum_{\underset{f^\phi-g^\phi=0}{f,g\in V_0}} c_f\overline{c_g}\,e^{-\frac{i}{2}\sigma_0(f,g)}e^{-\frac{1}{2\beta}\int d^dx\,(f^\pi(x)-g^\pi(x))^2}\geq 0\,.
\end{equation}
Since there are only finitely many $f$ for which $c_f$ is not $0$, the corresponding finitely many $f^\pi$ span a finite-dimensional subspace of $L_2(\RR^d)$. Let us say that this subspace is $n$-dimensional and take a basis $\{e^i(x)\}$ of this space, which satisfies $\int d^dx\, e^i(x)e^j(x)=\delta_{i,j}$. We can therefore write $f^\pi=\sum_{i=1}^n f^\pi_i e^i$, where $f^\pi_i$ are the corresponding real coefficients. We can use this to rewrite
\begin{equation}
    \int d^dx\,(f^\pi(x)-g^\pi(x))^2
    =\sum_{i=1}^n (f^\pi_i-g^\pi_i)^2\,,
\end{equation}
which implies
\begin{equation}
    e^{-\frac{1}{2\beta}\int d^dx\,(f^\pi(x)-g^\pi(x))^2}=\prod_{i=1}^n e^{-\frac{1}{2\beta}(f^\pi_i-g^\pi_i)^2}\,.
\end{equation}
Next, we  will use the identity
\begin{equation}
    e^{-(a-b)^2}=\frac{2}{\sqrt{\pi}}\int d\lambda\,e^{-2(a-\lambda)^2}e^{-2(b-\lambda)^2}\,.
\end{equation}
Applying this to every exponential in the product, we get
\begin{equation}
    e^{-\frac{1}{2\beta}\int d^dx\,(f^\pi(x)-g^\pi(x))^2}=\left(\frac{2}{\beta\pi}\right)^{\frac{n}{2}}\int d^n\lambda\,e^{-\frac{1}{\beta}\sum_{i=1}^n(f^\pi_i-\lambda_i)^2}e^{-\frac{1}{\beta}\sum_{i=1}^n(g^\pi_i-\lambda_i)^2}\,.
\end{equation}
Define now an equivalence relation $\sim$ on $V_0$ by $f\sim g\,:\iff f^\phi=g^\phi$. We get\footnote{Here we follow the procedure outlined in the section on Lagrangian subspaces in \cite{Fredenhagen:lecturenotes}.}
\begin{align}
    \omega_0^\beta \big( C^*C\big)=&\sum_{\underset{f^\phi-g^\phi=0}{f,g\in V_0}} c_f\overline{c_g}\,e^{-\frac{i}{2}\sigma_0(f,g)}e^{-\frac{1}{2\beta}\int d^dx\,(f^\pi(x)-g^\pi(x))^2} \\
    =& 
    \sum_{[h]\in V_0/\sim}\sum_{f,g\in [h]}c_f\overline{c_g}\,e^{-\frac{i}{2}\sigma_0(f,g)}\left(\frac{2}{\beta\pi}\right)^{\frac{n}{2}}\int d^n\lambda\,e^{-\frac{1}{\beta}\sum_{i=1}^n(f^\pi_i-\lambda_i)^2}e^{-\frac{1}{\beta}\sum_{i=1}^n(g^\pi_i-\lambda_i)^2}\,.
\end{align}
There is some abuse of notation present in the above expression since the decomposition $f^\pi=\sum_{i=1}^n f^\pi_i e^i$ only applies to those (finitely many) $f\in V_0$ that contribute to the sum. 
If $f,g\in [h]$, we have $f^\phi=g^\phi=h^\phi$, and we can decompose $\sigma_0(f,g)$ as
\begin{align}
    \sigma_0(f,g)&=- \int d^dx \, \big( f^\pi(x)g^\phi(x) - f^\phi(x) g^\pi(x) \big)\\
    &=
    - \int d^dx \, \big( f^\pi(x)h^\phi(x) - (f^\phi(x)-g^\phi(x))h^\pi(x) - h^\phi(x) g^\pi(x) \big)\\
    &=
    \sigma_0(f,h)-\sigma_0(g,h)\,.
\end{align}
For ease of notation, we define now the new coefficient functions
\begin{equation}
    \tilde{c}_{f,h}(\lambda):=c_f\,e^{-\frac{i}{2}\sigma_0(f,h)}e^{-\frac{1}{\beta}\sum_{i=1}^n(f^\pi_i-\lambda_i)^2}\,,
\end{equation}
which we can use to write
\begin{align}
    \omega_0^\beta \big( C^*C\big)=& 
    \sum_{[h]\in V_0/\sim}\sum_{f,g\in [h]}\left(\frac{2}{\beta\pi}\right)^{\frac{n}{2}}\int d^n\lambda\,\tilde{c}_{f,h}(\lambda)\overline{\tilde{c}_{g,h}(\lambda)}\\
    =& 
    \sum_{[h]\in V_0/\sim}\left(\frac{2}{\beta\pi}\right)^{\frac{n}{2}}\int d^n\lambda\,\left|\sum_{f\in [h]}\tilde{c}_{f,h}(\lambda)\right|^2\,.
\end{align}
The exchange of the integral and the sum is justified since we only sum over finitely many $f$. In this form, it is clear that this expression is nonnegative and therefore $\omega_0^\beta$ is indeed a state.

\section{States for flat holography}
\label{sec:stat-flat-hologr}

In this appendix we provide further technical details concerning the
analysis of states in flat space holography. In
Section~\ref{sec:positivity-non-zero-mode} we present the construction
of the one-particle Hilbert space. In
Section~\ref{sec:vac-rep-flatholo} we show that the representations
$\Pi_{\ell}$ of the Weyl algebra for the scalar field (defined
in~\eqref{eq:Pi-ell}) are well defined and unitarily equivalent for
different choices of $\ell$, and we establish their equivalence to the
GNS representation of the state~\eqref{eq:state_evalv}.

\subsection{One-particle Hilbert space for flat holography}
\label{sec:positivity-non-zero-mode}

We consider the two-point function~\eqref{eq:2pt-from-flat-holography}
obtained from flat space holography. On the space $V_0$ of test
functions with vanishing $u$-integral~\eqref{f0-integral}, $f_0=0$, it
can be used to define a bilinear form
(see~\eqref{scalarproduct-flatholo})
\begin{equation}
   \langle f,g\rangle =  \tilde{\mu}(f,g)+\frac{i}{2}\sigma(f,g)\,.
\end{equation}
We show here how $V_0$ can be completed to a Hilbert space $\mathcal{H}_0$ with $\langle \_,\_\rangle$ as the scalar product. For this we introduce the Fourier transform in the $u$-variable
\begin{equation}
    \mathring{f}(\xi,z,\bar{z}) = \int du\,f(u,z,\bar{z}) \,e^{i\xi u}\,.
\end{equation}
The logarithm occurring in $\tilde{\mu}$ (see~\eqref{mu-flatholo}) can be rewritten using the identity
\begin{equation}
    \log |a| = \int_0^\infty d\xi\,\frac{\cos \xi - \cos (a\xi)}{\xi} \,.
\end{equation}
We find (using that $\mathring{f}$ and $\mathring{g}$ vanish at $\xi=0$):
\begin{align}
    \tilde{\mu} (f,g) &= -\frac{1}{2\pi}\int du\,dv\,d^2z\,\left( \int_0^\infty d\xi\, \frac{\cos \xi- \cos (\xi (u-v))}{\xi}\right) \, f(u,z,\bar{z})\,g(v,z,\bar{z})\\
    &= \frac{1}{4\pi}\int_0^\infty d\xi \int du\,dv\,d^2z\,\frac{e^{i\xi(u-v)}+e^{-i\xi (u-v)}}{\xi} \, f(u,z,\bar{z})\,g(v,z,\bar{z})\\
    &= \frac{1}{4\pi}\int_0^\infty \frac{d\xi}{\xi} \int d^2z\,\Big(\mathring{f}(\xi,z,\bar{z})\,\overline{\mathring{g}(\xi,z,\bar{z})}+\overline{\mathring{f}(\xi,z,\bar{z})}\,\mathring{g}(\xi,z,\bar{z})  \Big)\, .
\end{align}
Similarly we can rewrite $\sigma$ given in~\eqref{eq:sigma-flatholo}. With the help of the identity
\begin{equation}
    \mathrm{sign}(t) = \frac{2}{\pi} \int_0^\infty d\xi \,\frac{\sin \xi t}{\xi}
\end{equation}
we find
\begin{align}
    \sigma(f,g)&=-\frac{1}{2}\int du\,dv\,d^2z\,\frac{2}{\pi} \int_0^\infty d\xi \,\frac{\sin \xi (u-v)}{\xi}f(u,z,\bar{z})\,g(v,z,\bar{z})\\
    &=-\frac{1}{\pi }\int_0^\infty d\xi \int du\,dv\,d^2z\,\frac{e^{i \xi (u-v)}-e^{-i\xi (u-v)}}{2i \xi}f(u,z,\bar{z})\,g(v,z,\bar{z})\\
    &= -\frac{1}{2\pi i }\int_0^\infty \frac{d\xi}{\xi} \int d^2z\,\Big(\mathring{f}(\xi,z,\bar{z})\,\overline{\mathring{g}(\xi,z,\bar{z})}-\overline{\mathring{f}(\xi,z,\bar{z})}\,\mathring{g}(\xi,z,\bar{z}))   \Big) \, .
\end{align}
In total we arrive at
\begin{equation}
    \langle f,g\rangle = \frac{1}{2\pi}\int_0^\infty \frac{d\xi}{\xi} \int d^2z\,\overline{\mathring{f}(\xi,z,\bar{z})}\,\mathring{g}(\xi,z,\bar{z})  \,.
\end{equation}
This is a scalar product on the complex Hilbert space $\mathcal{H}_0$
of complex functions on $\mathbb{R}_+\times S^2$ with
$\langle f,f\rangle <\infty$. The space $V_0$ of test functions is
embedded via $f\mapsto \mathring{f}$ as a dense subspace of
$\mathcal{H}_0$.

\subsection{Vacuum representation for flat holography}
\label{sec:vac-rep-flatholo}

In this subsection we provide some details about the representations
$\Pi_\ell$ (defined in~\eqref{eq:Pi-ell}) of the Weyl algebra
$\mathcal{W}(V,\sigma)$ of the scalar field from flat space
holography, based on the vector space
$V=\mathcal{C}_c(\mathbb{R}\times S^2)$ and the symplectic form
$\sigma$ given in~\eqref{eq:sigma-flatholo}. We first show
that~\eqref{eq:Pi-ell} indeed defines a representation, and then show
that representations $\Pi_\ell$ and $\Pi_{\ell'}$ are unitarily
equivalent. Moreover, we show that $\Pi_\ell$ is unitarily equivalent
to the GNS representation of the state~\eqref{eq:state_evalv}.

Using the definition of $\Pi_\ell$ in~\eqref{eq:Pi-ell}, we now show that
\begin{equation}\label{star-hom}
    \Pi_\ell\big(W(f)W(g)\big)= \Pi_\ell(W(f))\,\Pi_\ell(W(g))\,.
\end{equation}
As a preparation, we note that for any $h_1,h_2\in \mathcal{C}(S^2)$, the expression $\sigma(\ell h_1, \ell h_2 )$ vanishes:
\begin{align}
    \sigma(\ell h_1, \ell h_2 )&=-\frac{1}{2} \int du_1\,du_2\,d^2z\, \mathrm{sign}(u_1-u_2)\ell(u_1)\,h_1(z,\bar{z})\,\ell(u_2)h_2(z,\bar{z})=0\,.
\end{align}
Evaluating the left-hand side of~\eqref{star-hom} on a vector $v\otimes \delta_h$, we obtain
\begin{align}
    &\Pi_\ell(W(f)W(g))(v\otimes \delta_h)\\
    &\qquad =e^{-\frac{i}{2}\sigma(f, g)}\,\Pi_\ell(W(f+g))(v\otimes \delta_h)\\
    &\qquad = e^{-\frac{i}{2}\sigma(f, g)}e^{i\left(\phi_{\mathrm{Fock}}(P_\ell (f+g)) -\sigma(f+g,\ell h)-\frac{1}{2}\sigma(f+g,\ell (f_0+g_0))\right)}v\otimes \delta_{f_0+g_0+h} \,.
\end{align}
When evaluating the right-hand side of~\eqref{star-hom}, we find 
\begin{align}
   & \Pi_\ell(W(f))\Pi_\ell(W(g))(v\otimes \delta_h)\\
    &=e^{i\left(\phi_{\mathrm{Fock}}(P_\ell f) -\sigma(f,\ell h+\ell g_0)-\frac{1}{2}\sigma(f,\ell f_0)\right)}e^{i\left(\phi_{\mathrm{Fock}}(P_\ell g) -\sigma(g,\ell h)-\frac{1}{2}\sigma(g,\ell g_0)\right)}v\otimes \delta_{f_0+g_0+h}\\
    &=e^{i\left(\phi_{\mathrm{Fock}}(P_\ell f  +P_\ell g) -\sigma(f+g,\ell h)-\sigma(f,\ell g_0)-\frac{1}{2}\sigma(f,\ell f_0)-\frac{1}{2}\sigma(g,\ell g_0)-\frac{1}{2}\sigma(f-\ell f_0,g-\ell g_0)\right)}v\otimes \delta_{f_0+g_0+h}\\
    &=e^{i\left(\phi_{\mathrm{Fock}}(P_\ell (f  + g)) -\sigma(f+g,\ell h)-\frac{1}{2}\sigma(f,g)-\frac{1}{2}\sigma(f+g,\ell (f_0+g_0))\right)}v\otimes \delta_{f_0+g_0+h}\,.
\end{align}
Hence, \eqref{star-hom} holds, confirming that $\Pi_\ell$ is a representation of $\mathcal{W}(V,\sigma)$.

Two representations $\Pi_\ell$ and $\Pi_{\ell'}$ are unitarily equivalent. On $\mathcal{F}(\mathcal{H}_0)\otimes \ell^2(\mathcal{C}(S^2))$ we define the unitary operator $U_{\ell'\ell}$ by
\begin{equation}\label{U-ell-ell}
    U_{\ell' \ell}(v\otimes \delta_h) =
  e^{i\left(\phi_{\mathrm{Fock}}((\ell-\ell')h)+\frac{1}{2}\sigma(\ell'h,\ell
      h)\right)}v\otimes \delta_h\,.
\end{equation}
In the following, we show that 
\begin{equation}\label{intertwiner}
    U_{\ell' \ell} \,\Pi_\ell(W(f)) = \Pi_{\ell'}(W(f)) \,U_{\ell'
    \ell}\,.
\end{equation}

First, note that
\begin{equation}
    P_\ell f + (\ell-\ell')f_0=f-\ell'f_0=P_{\ell'} f\,.
\end{equation}
We now evaluate the left-hand side of~\eqref{intertwiner} on a vector $v\otimes \delta_h$:
\begin{align}
    &U_{\ell' \ell}\,\Pi_\ell(W(f))(v\otimes \delta_h)\nonumber \\
    &=e^{i\left(\phi_{\mathrm{Fock}}((\ell-\ell')(f_0+h))+\frac{1}{2}\sigma(\ell'(f_0+h),\ell(f_0+h))\right)}e^{i\left(\phi_{\mathrm{Fock}}(P_\ell f) -\sigma(f,\ell h)-\frac{1}{2}\sigma(f,\ell f_0)\right)}v\otimes \delta_{f_0+h}\\
    &=e^{i\left(\frac{1}{2}\sigma(\ell'f_0,\ell f_0)+\frac{1}{2}\sigma(\ell'f_0,\ell h)+\frac{1}{2}\sigma(\ell' h,\ell f_0) -\sigma(f,\ell h)-\frac{1}{2}\sigma(f,\ell f_0)-\frac{1}{2}\sigma((\ell-\ell')(f_0+h),P_\ell f)\right)}\nonumber \\
    &\quad\ \cdot e^{i\left(\phi_{\mathrm{Fock}}(P_{\ell'} f+(\ell-\ell')h)+\frac{1}{2}\sigma(\ell' h,\ell h)\right)}v\otimes \delta_{f_0+h}\\
    &=e^{i\left(\frac{1}{2}\sigma(\ell'f_0,\ell h)-\frac{1}{2}\sigma(f,\ell h)+\frac{1}{2}\sigma(\ell' f_0,f) +\frac{1}{2}\sigma(\ell' h,f)\right)} \, e^{i\left(\phi_{\mathrm{Fock}}(P_{\ell'} f+(\ell-\ell')h)+\frac{1}{2}\sigma(\ell' h,\ell h)\right)}v\otimes \delta_{f_0+h}\,.
    \label{result-LHS}
\end{align}
Evaluating the right-hand side of~\eqref{intertwiner} on a vector $v\otimes \delta_h$ results in
\begin{align}
    &\Pi_{\ell'}(W(f))\,U_{\ell' \ell}(v\otimes \delta_h)\nonumber \\
    &=e^{i\left(\phi_{\mathrm{Fock}}(P_{\ell'} f) -\sigma(f,\ell' h)-\frac{1}{2}\sigma(f,\ell' f_0)\right)}e^{i\left(\phi_{\mathrm{Fock}}((\ell-\ell')h)+\frac{1}{2}\sigma(\ell'h,\ell h)\right)}v\otimes \delta_{f_0+h}\\
    &=e^{i\left(\phi_{\mathrm{Fock}}(P_{\ell'} f+(\ell-\ell')h) -\sigma(f,\ell' h)-\frac{1}{2}\sigma(f,\ell' f_0)+\frac{1}{2}\sigma(\ell'h,\ell h)-\frac{1}{2}\sigma(P_{\ell'} f,(\ell-\ell')h)\right)}v\otimes \delta_{f_0+h}\\
    &=e^{i\left(-\frac{1}{2}\sigma(f,\ell'h)-\frac{1}{2}\sigma(f,\ell'f_0) -\frac{1}{2}\sigma(f,\ell h)+\frac{1}{2}\sigma(\ell'f_0,\ell h)\right)}\,e^{i\left(\phi_{\mathrm{Fock}}(P_{\ell'} f+(\ell-\ell')h) +\frac{1}{2}\sigma(\ell'h,\ell h)\right)}v\otimes \delta_{f_0+h}\,.
\end{align}
This coincides with~\eqref{result-LHS}, and hence~\eqref{intertwiner} holds.

Lastly, we show that $\Pi_\ell$ is unitarily equivalent to the GNS representation of the state $\omega$ given in~\eqref{eq:state_evalv}. Given that $\Pi_\ell$ already satisfies~\eqref{eq:state_exp_val}, i.e., $\langle\Omega|\Pi_\ell(W(f))|\Omega\rangle=\omega(W(f))$ with $|\Omega\rangle$ given in~\eqref{eq:flat_Omega}, we only need to show that $|\Omega\rangle$ is a cyclic vector for $\Pi_\ell$. The unitary equivalence is then provided by~\cite[Theorem 7.7]{Con00}.

From standard results on Fock space representations (see, e.g.,~\cite{PetzDenes1990Aitt}) we can conclude that the span of vectors $e^{i\phi_{\mathrm{Fock}}(f)}\ket{0}$ for $f_0=0$ is dense in $\mathcal{F}(\mathcal{H}_0)$. This implies that the vectors $e^{i\phi_{\mathrm{Fock}}(f)}\ket{0}\otimes \delta_h$ span a dense set in $\mathcal{H}$. Given any test function $f$ with $f_0=0$, as well as any test function $h$ on $S^2$, we obtain the vectors
\begin{equation}
    \Pi_\ell \big(W(\ell h)W(f)\big)|\Omega\rangle=e^{i\phi_{\mathrm{Fock}}(f)}|0\rangle\otimes \delta_{h}\,.
\end{equation}
As just argued, their span is dense in $\mathcal{H}$, and hence $\ket{\Omega}$ is cyclic.  

\subsection{Charge operator as a limit}
\label{sec:chargeoperator}

In this subsection, we show that for the family of test functions given by $f_\delta=-X_\delta'\,\epsilon$ (where $X_\delta$ is defined in~\eqref{defofXdelta}) the operators $\Pi_\ell(W(f_\delta))$ converge to the exponentiated charge operator $U_\epsilon$ for $\delta\to 0$ in the strong operator topology (as stated in~\eqref{Ueps-as-limit}). 

The test functions $X_\delta$ converge pointwise to the constant function $1$ such that for every compact interval, $X_\delta$ equals $1$ for sufficiently small $\delta$. This can be seen as follows. $\chi$ by assumption is a test function equal to $1$ on an interval around zero, say $[-a,a]$ with $a>0$. For $|u|\leq a/\delta$ and $s\geq 0$, we have $\delta\,e^{-s}\,|u|\leq a$, so every integrand in 
\begin{equation}
    X_\delta(u) = \frac{1}{\log \delta^{-1}}\int_0^{\log \delta^{-1}} ds\,\chi(\delta\,e^{-s}\,u)
\end{equation}
equals $1$, and hence $X_\delta(u)=1$ for $u \in [-a/\delta,a/\delta]$.

The test functions $f_\delta$ have $u$-integral zero, 
\begin{equation}
    (f_\delta)_0(z,\bar{z}) = \int du\, \big(-X_\delta '(u)\,\epsilon(z,\bar{z})\big) = 0\,.
\end{equation}
Using the definition~\eqref{eq:Pi-ell} of $\Pi_\ell$, we obtain
\begin{equation}\label{Piell_fdelta}
    \Pi_\ell(W(f_\delta)) (v\otimes \delta_h)=e^{i\phi_{\mathrm{Fock}}(f_\delta)} e^{-i\sigma(f_\delta,\ell h)}v\otimes \delta_{h}\,.
\end{equation}
Evaluating the exponent in the phase factor, one finds (using~\eqref{eq:sigma-flatholo})
\begin{align}
    -i\sigma(f_\delta,\ell h) &= -\frac{i}{2} \int du_1\,du_2\,d^2z\,\mathrm{sign}(u_1-u_2)\,X_\delta'(u_1)\,\epsilon(z,\bar{z})\, \ell(u_2)\,h(z,\bar{z})\\
    &= i\int du\,X_\delta(u)\,\ell(u) \int d^2z\, \epsilon(z,\bar{z})\,h(z,\bar{z})\,.
\end{align}
For $\delta$ sufficiently small, $X_\delta=1$ on the support of $\ell$ (see above), hence the integral over $u$ gives the factor 1, and we find
\begin{equation}\label{limit_of_phase}
    e^{-i\sigma(f_\delta,\ell h)} = e^{i \int d^2z\, \epsilon(z,\bar{z})\,h(z,\bar{z})}\quad \text{for sufficiently small}\ \delta\,.
\end{equation}
In the limit, the phase factor equals the phase appearing in the action of $U_{\epsilon}$ (see~\eqref{eq:soft-exp}). 

The remaining task is to show that $e^{i\phi_{\mathrm{Fock}}(f_\delta)}$ strongly converges to the identity. As $f\mapsto e^{i\phi_{\mathrm{Fock}}(f)}$ is strongly continuous in the Fock representation, it is enough to show that $f_\delta \to 0$ in $\mathcal{H}_0$. 

Using the scalar product~\eqref{scalarproduct-flatholo} on $\mathcal{H}_0$ and the definition~\eqref{mu-flatholo}, we find
\begin{align}
    ||f_\delta||_{\mathcal{H}_0}^2 &= \tilde{\mu}(f_\delta,f_\delta)\\
    &= -\frac{1}{2\pi} \int du_1\,du_2\,d^2z\,\log |u_1-u_2|\,f_\delta(u_1,z,\bar{z})\,f_\delta(u_2,z,\bar{z})\\
    &= \frac{||\epsilon||^2}{(\log \delta^{-1})^2} \int_0^{\log \delta^{-1}}\!\!\! ds_1\int_0^{\log \delta^{-1}}\!\!\!ds_2\, C(s_1,s_2) 
    \label{integraloverC}
\end{align}
with 
\begin{equation}
    ||\epsilon||^2 = \int d^2z\, |\epsilon(z,\bar{z})|^2
\end{equation}
and
\begin{align}
    C(s_1,s_2) &= -\frac{\delta^2\,e^{-s_1-s_2}}{2\pi} \int du_1\,du_2\,\log|u_1-u_2|\,\chi'(\delta\,e^{-s_1}\,u_1)\,\chi'(\delta\,e^{-s_2}\,u_2)\\
    &= -\frac{1}{2\pi} \int dx_1\,dx_2\,\log|\delta^{-1}\,e^{s_1}\,x_1-\delta^{-1}\,e^{s_2}\,x_2|\,\chi'(x_1)\,\chi'(x_2)\\
    &= -\frac{1}{2\pi} \int dx_1\,dx_2\,\log|x_1-e^{s_2-s_1}\,x_2|\,\chi'(x_1)\,\chi'(x_2)\,,
\end{align}
where in the last step we have used the fact the constant factors in the logarithm drop out because $\int \chi'=0$. As $C(s_1,s_2)=C(s_2,s_1)$, we have
\begin{equation}\label{defofc}
    C(s_1,s_2) = -\frac{1}{2\pi} \int dx_1\,dx_2\,\log\big|x_1-e^{-|s_1-s_2|}\,x_2\big|\,\chi'(x_1)\,\chi'(x_2)=:c(|s_1-s_2|)\,.
\end{equation}
We now show that $c(s)$ is bounded by a constant independent of $s\geq 0$.
By assumption, $\chi$ is $1$ on an interval $[-a,a]$ and $0$ outside an interval $[-b,b]$, therefore $\chi'(u)$ can be nonzero only for $a\leq |u|\leq b$. In the integral, we estimate $\chi'(x_1)$ by its supremum $||\chi'||_\infty$,
\begin{align}
    |c(s)| \leq \frac{||\chi'||_\infty}{2\pi} \int_{-b}^b dx_1 \int_{-b}^b dx_2\, \big| \log|x_1 - e^{-s}\,x_2|\big| \,|\chi'(x_2)|\,.
\end{align}
As $|x_1-e^{-s}x_2|\leq 2\,b$, we have
\begin{equation}
    |c(s)| \leq \frac{||\chi'||_\infty}{2\pi} \int_{-2b}^{2b} dx \, \big| \log|x|\big| \int_{-b}^b dx_2\,|\chi'(x_2)| =:K_1 < \infty.
\end{equation}
We next argue that for large differences of $s$, $c(s)$ is exponentially suppressed by the factor $e^{-s}$. We first observe that we can rewrite the logarithm in the integrand of~\eqref{defofc} as 
\begin{equation}
    \log \big| x_1-e^{-s}x_2\big| = \log |x_1| + \log \bigg|1 - e^{-s} \frac{x_2}{x_1}\bigg|\,.
\end{equation}
The first summand drops out when integrating over $x_2$ together with $\chi'(x_2)$. The second is suppressed (by a term proportional to $e^{-s}$) when the exponential factor becomes small because $|\frac{x_2}{x_1}|\leq \frac{b}{a}$ is bounded on the support of $\chi'(x_1)\chi'(x_2)$. In total, one finds that there are positive constants $S_0$ and $K_2$ such that
\begin{equation}
    |c(s)|\leq K_2\,e^{-s}\quad \text{for}\ s \geq S_0 \,.
\end{equation}
This now allows to estimate the integral appearing in~\eqref{integraloverC}:
\begin{align}
    \int_0^{\log \delta^{-1}} \!\!\!ds_1\int_0^{\log \delta^{-1}}\!\!\!ds_2\, C(s_1,s_2) &= 2 \int_0^{\log \delta^{-1}} \!\!\!ds_1 \int_{s_1}^{\log \delta^{-1}}\!\!\!ds_2\,C(s_1,s_2) \\
    & \leq 2 \int_0^{\log \delta^{-1}} \!\!\!ds_1 \int_{s_1}^{\infty} ds_2\,|c(s_2-s_1)|\\
    &= 2 \log \delta^{-1} \int_{0}^{\infty} ds\,|c(s)| \,.
\end{align}
The integral appearing in the last expression is finite,
\begin{align}
    \int_{0}^{\infty} ds\,|c(s)| &= \int_0^{S_0} ds\,|c(s)| + \int_{S_0}^\infty ds\, |c(s)|\\
    & \leq S_0\,K_1 + K_2 \int_{S_0}^\infty ds\, e^{-s} =: K_3 < \infty \,.
\end{align}
Finally, we obtain for the norm of $f_\delta$ the estimate
\begin{align}
    ||f_\delta||^2_{\mathcal{H}_0} \leq \frac{2\,||\epsilon||^2}{\log \delta^{-1}} K_3
\end{align}
which goes to zero for $\delta \to 0$. 

For sufficiently small $\delta$, the phase identity above (see~\eqref{Piell_fdelta} and~\eqref{limit_of_phase}) gives
\begin{equation}
\Pi_\ell(W(f_\delta))
=\bigl(e^{i\phi_{\mathrm{Fock}}(f_\delta)}\otimes I\bigr)U_\epsilon.
\end{equation}
Since $\|f_\delta\|_{\mathcal H_0}\to0$, strong continuity of the Fock Weyl operators implies
$\Pi_\ell(W(f_\delta))\to U_\epsilon$ strongly on $\mathcal H$.

\bibliographystyle{utphys} 
\bibliography{bibl_local} 

\end{document}